\documentclass{aa}
\usepackage[utf8]{inputenc}
\usepackage{graphicx}
\usepackage[english]{babel}
\usepackage{amsmath}
\usepackage{mathtools}
\usepackage{amssymb}
\usepackage{txfonts}
\usepackage{hyperref} 
\usepackage{color}
\usepackage{rotating}
\usepackage{enumitem}
\usepackage[labelfont=bf]{caption}
\usepackage{subcaption}
\usepackage{ragged2e}
\usepackage{threeparttable}
\usepackage{placeins}
\usepackage{url}
\hypersetup{colorlinks=true,citecolor=blue}

\begin{document}

\title{Water Condensation Zones around Main Sequence Stars}

\author{Martin Turbet\inst{1,2,3}, Thomas J. Fauchez\inst{4,5,6}, Jeremy Leconte\inst{3}, Emeline Bolmont\inst{2,7}, Guillaume Chaverot\inst{2,7}, Francois Forget\inst{1}, Ehouarn Millour\inst{1}, Franck Selsis\inst{3}, Benjamin Charnay\inst{8}, Elsa Ducrot\inst{9}, Micha\"el Gillon\inst{10}, Alice Maurel\inst{11,1}, Geronimo L. Villanueva\inst{4,6}}

\institute{
$^1$ Laboratoire de M\'et\'eorologie Dynamique/IPSL, CNRS, Sorbonne Universit\'e, Ecole Normale Sup\'erieure, Universit\'e PSL, Ecole Polytechnique, Institut Polytechnique de Paris, 75005 Paris, France\\
$^2$ D\'epartement d'astronomie de l'Universit\'e de Gen\`eve, Chemin Pegasi 51, 1290 Sauverny, Switzerland\\
$^3$ Laboratoire d'astrophysique de Bordeaux, Univ. Bordeaux, CNRS, B18N, allée Geoffroy Saint-Hilaire, 33615 Pessac, France\\
$^4$ NASA Goddard Space Flight Center, 8800 Greenbelt Road, Greenbelt, MD 20771, USA\\
$^5$ Integrated Space Science and Technology Institute, Department of Physics, American University, Washington DC\\
$^6$ NASA GSFC Sellers Exoplanet Environments Collaboration\\
$^7$ Centre pour la Vie dans l'Univers, Universit\'e de Gen\`eve, Switzerland\\
$^8$ LESIA, Observatoire de Paris, Universit\'e PSL, CNRS, Sorbonne Universit\'e, Universit\'e Paris Cit\'e, 5 place Jules Janssen, 92195 Meudon, France\\
$^9$ Universit\'e Paris-Saclay, Universit\'e Paris-Cit\'e, CEA, CNRS, AIM, Gif-sur-Yvette, 91191, France\\
$^{10}$ Astrobiology Research Unit, University of Li\`ege, All\'ee du 6 Ao\^ut, 19, 4000 Li\`ege, Belgium\\
$^{11}$ Sorbonne Universit\'es, UPMC Universit\'e Paris 6 et CNRS, UMR 7095, Institut d’Astrophysique de Paris, 98 bis bd Arago, 75014 Paris, France}

  \date{Submitted to A\&A}
  \abstract
   {Understanding the set of conditions that allow rocky planets to have liquid water on their surface -- in the form of lakes, seas or oceans -- is a major scientific step to determine the fraction of planets potentially suitable for the emergence and development of life as we know it on Earth. This effort is also necessary to define and refine the so-called "Habitable Zone" (HZ) in order to guide the search for exoplanets likely to harbor remotely detectable life forms. Until now, most numerical climate studies on this topic have focused on the conditions necessary to maintain oceans, but not to form them in the first place. Here we use the three-dimensional Generic Planetary Climate Model, historically known as the LMD Generic Global Climate Model (GCM), to simulate water-dominated planetary atmospheres around different types of Main-Sequence stars. The simulations are designed to reproduce the conditions of early ocean formation on rocky planets due to the condensation of the primordial water reservoir at the end of the magma ocean phase. We show that the incoming stellar radiation (ISR) required to form oceans by condensation is always drastically lower than that required to vaporize oceans. We introduce a Water Condensation Limit, which lies at significantly lower ISR than the inner edge of the HZ calculated with three-dimensional numerical climate simulations. This difference is due to a behavior change of water clouds, from low-altitude dayside convective clouds to high-altitude nightside stratospheric clouds. Finally, we calculated transit spectra, emission spectra and thermal phase curves of TRAPPIST-1b, c and d with H$_2$O-rich atmospheres, and compared them to CO$_2$ atmospheres and bare rock simulations. We show using these observables that JWST has the capability to probe steam atmospheres on low-mass planets, and could possibly test the existence of nightside water clouds. }
  

\titlerunning{Water Condensation Zones around Main Sequence Stars}
\authorrunning{M. Turbet et al.} 

\maketitle

\section{Introduction}
\label{section_introduction}

Understanding the set of conditions that allow rocky planets to have liquid water on their surface -- in the form of lakes, seas or oceans -- is a major scientific step to determine the fraction of planets potentially suitable for the emergence and development of life as we know it on Earth. Based on our experience on Earth, liquid water as a solvent is thought to be a necessary (but not sufficient) condition for the existence of life, along with the presence of an energy source and nutrients C,H,N,O,P,S \citep{Forget:2013,Shields:2016}. While it is now known to exist in the subsurface of a large sample of planetary bodies (icy moons, Pluto, Triton; see \citet{Lunine:2017} and references therein), planets that have liquid water on their surface have two additional favorable properties \citep{TurbetSelsis2021}: (1) liquid water is exposed to stellar radiation ; and (2) liquid water is in interaction with the atmosphere, which facilitates the remote search for tracers of life (using the exoplanetary atmosphere to search for biomarkers).

This has motivated numerous theoretical studies to determine the range of parameters (e.g. type of star, incoming stellar radiation) required for a planet to sustain surface liquid water reservoirs, and is the main driver for the definition of the Habitable Zone (HZ) \citep{Kasting:1993,Kopparapu:2013,Kopparapu:2014}. The HZ is now a widely used, first-order criterion to determine a zone to search for life-compatible planets, as well as to evaluate the potential habitability of detected planets (see e.g., \citealt{Borucki:2011}).

So far, most numerical climate studies in the field focused on evaluating the conditions needed for a planet to stabilize surface liquid water, whether using one-dimensional (1D) climate models \citep{Kasting:1993,Kopparapu:2013,Kopparapu:2014,Ramirez:2014c,Ramirez:2016,Yang:2016,Ramirez:2017b,Ramirez:2018,Koll:2019,Chaverot:2022} or three-dimensional (3D) Global Climate Models (GCMs) \citep{Wordsworth:2011ajl,Yang:2013,Shields:2013,Yang:2014b,Bolmont:2016aa,Kopparapu:2016,Turbet:2016,Turbet:2017epsl,Boutle:2017,Wolf:2017,Wolf:2017b,Kopparapu:2017,Turbet:2018,Way:2018,Delgenio2019,Kodama:2019,Yang:2019a,Yang:2019b,Boutle:2020,Wolf:2020,Colose:2021}. 
However, this approach implicitly assumes surface liquid water was available in the first place. This assumption was recently challenged in \citet{Turbet2021} that explored the conditions of surface ocean formation by condensation of the primordial water vapor reservoir.
Terrestrial planets are indeed believed to form hot due to their initial accretion energy, and thus to cross a magma ocean stage \citep{Zahnle:1988,Abe:1997,Hamano:2013,Lebrun:2013,Salvador:2017,Lichtenberg:2023} – where superficial water is present only in the form of vapor – before evolving towards their final state. For liquid water oceans to ever appear on a planetary surface, water initially present in the young and warm planetary atmosphere must be able to condense on the surface in the first place \citep{Hamano:2013,Lebrun:2013,Turbet2021}.

The conditions leading to the condensation of a water ocean after the magma ocean phase have first been studied with 1D numerical climate models \citep{Abe:1988,Hamano:2013,Lebrun:2013,Salvador:2017}, which neglect the effects of atmospheric dynamics and clouds. Recently, \citet{Turbet2021} simulated water condensation for the first time in a GCM, specifically for the cases of Earth and Venus, but with potentially far-reaching consequences for a broader range of planets \citep{Kasting2021}. They showed that -- in water-rich atmospheres -- water clouds tend to preferentially form on the nightside and at the poles, owing to the strong subsolar water vapor absorption. Clouds have thus a strong net warming effect that inhibits surface water condensation even at modest insolations, with strong implications for the early habitability of Earth and Venus.

Here we seek to extend the work of \citet{Turbet2021} by extending GCM simulations to the cases of exoplanets, and especially those orbiting stars cooler than the Sun. Terrestrial planets orbiting low-mass, cool, red stars (a.k.a. M-stars) are indeed the best targets we currently have for remote sensing of their atmospheres, and thus to search for tracers of habitability and life \citep{Morley:2017,Kaltenegger:2017,Lustig-Yaeger:2019,Wunderlich2019,Fauchez:2019}. These planets are thus potentially unique natural laboratories to be able to confront our theories on planetary habitability with observations, in the event that they were able to preserve their atmospheres \citep{Zahnle:2017}.

\section{Method}
\label{section_method}

All the numerical experiments presented in this study are based on the Generic Planetary Climate Model (PCM), which is historically known as the Laboratoire de Météorologie Dynamique (LMD) Generic GCM.

\subsection{The Generic PCM}

The Generic PCM is a sophisticated 3D Global Climate Model (GCM) that has previously been developed and used to simulate a wide range of exoplanetary atmospheres ranging from that of temperate rocky planets to warm mini-Neptunes \citep{Wordsworth:2011ajl,Leconte:2013aa,Charnay:2015a,Charnay:2015b,Turbet:2016,Turbet:2018,Fauchez:2019,Charnay:2021}. Specifically, the model has been adapted to simulate water-rich planetary atmospheres \citep{Leconte:2013nat} with numerous applications: studying runaway greenhouse on Earth \citep{Leconte:2013nat}, post-impact atmospheres on Mars \citep{Turbet:2020impact}, water cloud formation on temperate mini-Neptunes \citep{Charnay:2021}, water condensation on early Earth and Venus \citep{Turbet2021}.

We summarize the main features of the model below. The GCM includes a complete radiative transfer that takes into account the absorption and scattering by the atmosphere, the clouds and the surface. The radiative transfer calculations, which are based on the correlated-k approach \citep{Fu:1992}, use 55 spectral bands in the thermal infrared (from 0.65 to 100$\mu$m ; designed to capture the thermal emission of the planet) and 45 in the visible domain (from 0.3 to 6.5$\mu$m ; designed to capture the incoming stellar radiation). 
Opacity tables were computed as in \citet{Turbet2021} using the HITRAN~2008 database for wavelength above 1$\mu$m and the HITEMP~2010 database below. We made this choice because most of the differences between HITRAN and HITEMP are below 1$\mu$m (see comparisons done in \citealt{Goldblatt:2013}), and that this significantly reduces the computation time of the opacity tables.
The opacity tables also account for the continuum absorptions (from collision-induced absorptions, dimer and far-wings) of N$_2$-N$_2$, H$_2$O-H$_2$O and H$_2$O-N$_2$, using the HITRAN collision-induced absorption (CIA) database \citep{Karman:2019} and the MT$\_$CKD v3.5 database \citep{Mlawer:2012}. Subgrid-scale dynamical processes (turbulent mixing and convection) were parameterized as described in \citet{Leconte:2013nat}. Moist convection was taken into account following a moist convective adjustment scheme that originally derives from \citet{Manabe:1967} and was later generalized for water-dominated atmospheres in \citet{Leconte:2013nat}. Relative humidity evolves freely and is limited to 100$\%$ (no supersaturation). In practice, when an atmospheric grid cell reaches 100$\%$ saturation and the corresponding atmospheric column has an unstable temperature vertical profile, the moist convective adjustment scheme is performed to get a stable moist adiabatic lapse rate. When condensing, water vapour forms liquid water droplets and/or water ice particles, depending on the atmospheric temperature and pressure, forming clouds. 
We used a fixed number of activated cloud condensation nuclei (CCNs) per unit mass of air to determine  -- based on the amount of condensed water -- the local, effective radii of H$_2$O cloud particles \citep{Turbet2021}. 
The effective radius is then used to compute the radiative properties and the sedimentation velocity of cloud particles. Water precipitation is divided into rainfall and snowfall, depending on the nature of the cloud particles, determined solely by the atmospheric temperature. Rainfall is parameterised to account for the conversion of cloud liquid droplets to raindrops by coalescence with other droplets \citep{Boucher:1995}. Rainfall is considered to be instantaneous, that is it goes directly to the surface, but can evaporate while falling through subsaturated layers \citep{Gregory:1995}. In hot and steamy simulations, re-evaporation of precipitation is always complete, that is, precipitation always fully evaporates in the dry lower atmosphere before it reaches the surface. The snowfall rate is calculated using the sedimentation velocity of particles, assumed to be equal to the terminal velocity that we approximate by a Stokes law modified with a ’slip-flow’ correction factor \citep{Rossow:1978}. 

The simulations presented in this paper were performed at a spatial resolution of 64~$\times$~48 in longitude~$\times$~latitude. In the vertical direction, the model is composed of 40 distinct atmospheric layers for a 10~bar surface pressure atmosphere, ranging from the surface up to a few Pascals. The exact number of atmospheric layers and model top pressure slightly vary depending on the surface pressure considered. The dynamical time step of the simulations is $\sim$~15~s, but can slightly vary from one simulation to another. The radiative transfer and the physical parameterisations (such as condensation, convection, etc.) are calculated every 100 and 5 dynamical time steps, respectively.

The main input parameters of our simulations are summarized in Table~\ref{table:method_param}. 

\begin{table*}
\centering
\begin{tabular}{lr}
   \hline
   Physical parameters & Values \\
   \hline
   Obliquity ($^\circ$) & 0 \\
   Orbital eccentricity & 0  \\
   Bare ground albedo & 0.2 \\
   Ground thermal inertia (J~m$^{-2}$~s$^{-1/2}$~K$^{-1}$) & 2000 \\
   Surface Topography & flat \\
   Surface roughness coefficient (m) & 0.01 \\
   No. of H$_2$O cloud condensation nuclei (CCN) & 10$^5$ \\
   for ice and liquid (kg$^{-1}$) & \\
   \hline
\end{tabular}
\caption{Summary of the fixed parameters used in all the new GCM simulations performed in this study.}
\label{table:method_param}
\end{table*}

More details on the model and the parameterizations used in this study can be found in \citet{Turbet2021}, and all the references therein. Compared to \citet{Turbet2021}, we have proceeded to several changes listed below:
\begin{itemize}
    \item We varied the type of host star.
    \item We varied the rotation period, and rotation mode 
    into synchronous rotation. We also fixed the obliquity to 0$^{\circ}$.
    \item We varied the values of the masses/radii of planets, depending on the numerical experiments, either to match real planets (in particular, the TRAPPIST-1 inner planets) or to make comparisons with previous studies (in particular, with \citealt{Yang:2013}).
    \item We varied the internal heat flux. For this, we simply added a constant (through time, and at all latitudes and longitudes) heat flux at the bottom of the model.
\end{itemize}
These changes are detailed in Tables~\ref{table_simu_1} and \ref{table_simu_2}. 

\subsection{Strategy of the numerical experiments}

Using the Generic PCM, we set up a series of 3D GCM simulations, assuming a N$_2$+H$_2$O-rich atmospheric composition, designed to identify the conditions required to trigger water condensation from the atmosphere onto the surface. 

For this, we followed a similar strategy than in \citet{Turbet2021}. Given a host star (T$_{\text{eff}}$ between 2600 and 5780~K), we first start with a GCM simulation with partial pressures P$_{\rm N_2}$~=~1~bar and P$_{\rm H_2O}$~=~10~bar (in our baseline scenario), with a high incoming stellar radiation (500~W~m$^{-2}$ , corresponding to 1.47 the insolation on Earth). P$_{\rm H_2O}$~=~10~bar is much below the possible content of a primordial steam atmosphere (Earth oceans would correspond to $\sim$~300~bar). Using 10~bar ensures the feasibility of the GCM simulations, and should be sufficient to reach conclusions valid for higher contents, as demonstrated in Section~\ref{subsection_results_water_content}. For this baseline simulation, we assume an isothermal cloud-free atmospheric temperature profile at 10$^3$~K for the initial state. While the upper atmospheric layers evolve very rapidly towards equilibrium, we used a strategy proposed in \citet{Turbet2021} to converge the deepest layers of the atmosphere, and detailed below. We run the GCM for typically 10 Earth years starting from the initial state described earlier and then we proceed to:
\begin{enumerate}
 \item run the GCM for 2 Earth years.
 \item extrapolate the evolution of the temperature field over $n_{\text{Earth years}}$ (the number of Earth years on which we decide to extrapolate the temperature field) using:
\begin{equation}
T_{i,j,k}(t+n_{\text{Earth years}})=T_{i,j,k}(t)+n_{\text{Earth years}}~\times~\Delta T_{mean,k},
\label{convergence_iteration_scheme}
\end{equation}
with $T_{i,j,k}$ the temperature at the $i$,$j$,$k$ spatial coordinates (corresponding respectively to longitude, latitude and altitude coordinates) and $\Delta T_{mean,k}$ the change of the mean horizontal (averaged over all longitudes and latitudes) temperature field at the altitude layer $k$ calculated over the second Earth year of step~1. $n_{\text{Earth years}}$ was first set arbitrarily to 100, then 50, 20, 10 and eventually 0, when the planetary atmosphere is close to convergence. The transition from one $n_{\text{Earth years}}$ value to another was chosen manually to minimize the convergence time, while avoiding numerical instabilities.
\item repeat the two previous steps, until the top of the atmosphere (TOA) radiative imbalance is lower than $\sim$~1~W/m$^2$. Moreover, we also check that surface temperatures are stable.
\end{enumerate}
We repeated the same procedure for simulations at lower incoming stellar radiations (ISR) until runaway water condensation is reached, except that we used the final equilibrium state of simulations at (previous) higher ISR as the initial state.
In comparison to \citet{Turbet2021}, the convergence time of the simulations is much longer, in particular for simulations around the coolest stars (TRAPPIST-1 and Proxima Cen). This is due to the fact that for cooler stars, a very small fraction of the stellar flux reaches the low atmosphere and the surface, due to the near-IR absorption by water vapor. For instance, for a GCM simulation of 10~bar of water vapor around TRAPPIST-1 near the water condensation limit, it takes about 150~Earth years of GCM simulations to converge the simulation. This is equivalent to around 2000~Earth years of evolution, taking extrapolation into account. We discuss this aspect in more details in the results section, as it also has an impact on the thermal profiles \citep{Selsis:2023}. We note that \citet{Turbet2021} did some sensitivity tests to ensure that the accelerated convergence scheme is working effectively. We thus did not perform any additional test here.

On the whole, and taking advantage of the convergence scheme presented above, we have managed to converge all the simulations presented in this work using a total of 1300~kh CPU on the French supercomputer OCCIGEN. This corresponds to a total of 70 simulations listed in this work, in addition to numerous preliminary and sensitivity simulations (included in the total CPU used and in the carbon budget provided in the Acknowledgments Section).

Finally, we note that for each combination of ISR and star type, we have calculated the rotation period $T$ self consistently with the assumption of synchronous rotation, using the following formula: 

\begin{equation}
    T~=~2\pi~\frac{\Bigl(d^2_{\odot-\oplus}~(\frac{L_{\star}}{L_{\odot}})~(\frac{F_{\oplus}}{F_p})\Bigr)^{\frac{3}{4}}}{\sqrt{GM_{\star}}}
\end{equation}
with $d^2_{\odot-\oplus}$ the Sun-Earth distance (in m); $L_{\star}$ and $L_{\odot}$ the stellar and solar luminosity, respectively; $F_p$ and $F_{\oplus}$ the stellar flux received at the TOA of the planet and on Earth, respectively; G the gravitational constant (in m$^3$~kg$^{-1}$~s$^{-2}$); $M_{\star}$ the stellar mass (in kg). The formula is simply a reformulation of Kepler's third law of planetary motion and assumes that the rotation period is equal to the orbital period (i.e. assumes the synchronous rotation). The luminosity of the star is taken equal to that at the ZAMS (Zero Age Main Sequence), i.e. at the beginning of the Main Sequence Phase (which can be much lower than during the Pre Main Sequence Star, in particular for M stars). This is done because here we are interested in determining the minimal ISR at which a planet has condensed its water reservoir at the start of the Main Sequence Star. This roughly corresponds to the minimum luminosity during the lifetime of stars. For each star type, we used the following mass-luminosity properties:
\begin{itemize}
\item For the K5 star, we chose L$_{\text{K5}}$~=~0.17~L$_{\odot}$, M$_{\text{K5}}$~=~0.7~M$_{\odot}$.
\item For the M3 star, we chose L$_{\text{M3}}$~=~1.2~$\times$~10$^{-2}$~L$_{\odot}$, M$_{\text{M3}}$~=~0.31~M$_{\odot}$. We used the stellar properties of L98-59 \citep{Kostov:2019}.
\item For the M5.5 star, we chose L$_{\text{M5.5}}$~=~1.5~$\times$~10$^{-3}$~L$_{\odot}$, M$_{\text{M5.5}}$~=~0.12~M$_{\odot}$. We used the stellar properties of Proxima Centauri \citep{Faria2022}.
\item For the M8 star, we chose L$_{\text{M8}}$~=~5.5~$\times$~10$^{-4}$~L$_{\odot}$, M$_{\text{M8}}$~=~0.09~M$_{\odot}$. We used the stellar properties of TRAPPIST-1 \citep{Agol2021}.
\end{itemize}

\subsection{Ensemble of simulations}

We set up a grid of GCM simulations (summarized in Tables~\ref{table_simu_1} and \ref{table_simu_2}) designed to explore the conditions of primordial ocean condensation for various types of terrestrial exoplanets, listed below:
\begin{itemize}
    \item SUN (taken from \citealt{Turbet2021}), K5, M3, Pcen and T1 are designed to explore surface water condensation on an Earth-like planet (1M$_{\oplus}$, 1R$_{\oplus}$) synchronously rotating (except for the Sun case) around its host star. The simulations assume respectively a G2V star (Sun spectrum), a K5 star (4400~K blackbody spectrum), a M3 star (3400~K blackbody spectrum), a M5.5 star (Proxima Cen PHOENIX BT-Settl synthetic spectrum) and a M8 star (TRAPPIST-1 PHOENIX BT-Settl synthetic spectrum). All the simulations assume a H$_2$O partial pressure of 10~bar, and a N$_2$ partial pressure of 1~bar.
    \item M3-YANG is designed to reproduce the same planetary parameters (2~R$_{\oplus}$, 60~Earth days rotation period) as used in the standard simulations of \citet{Yang:2013}, except the simulations assume a warm start (1~bar of N$_2$, 10~bar of H$_2$O).
    \item M3-3bar, M3-1bar, M3-0.3bar are designed to explore the impact of water content on the surface water condensation.
    \item T1b, T1c, T1d, T1e, Pcen-b are designed to explore surface water condensation (and observability) on known exoplanets TRAPPIST-1b, TRAPPIST-1c, TRAPPIST-1d, TRAPPIST-1e and Proxima b, respectively.
    \item T1b-3bar, T1b-1bar are designed to explore the role of water content on the climate (and observability) of TRAPPIST-1b.
    \item T1b-Fgeo1, T1b-Fgeo5 and T1b-Fgeo25 are designed to explore the role of internal heat flux on the climate (and observability) of TRAPPIST-1b. The internal heat flux is possibly high on TRAPPIST-1b due to tidal heating \citep{Turbet:2018aa}
\end{itemize}

\subsection{Computation of observables}

Transmission and emission spectra were simulated using the Planetary Spectrum Generator \citep[PSG\footnote{\url{https://psg.gsfc.nasa.gov/}}, ][]{Villanueva:2018,Villanueva:2022}. PSG is an online radiative transfer model integrating up-to-date radiative transfer methods and spectroscopic parameterizations, including a realistic treatment of multiple scattering in layer-by-layer spherical geometry. Multiple scattering from aerosols in the atmosphere is computed using the discrete ordinates method, which employs the plane-parallel atmosphere in which the radiation field is approximated by a discrete number of streams angularly distributed. The angular dependence of the aerosol scattering phase function is represented with Legendre polynomials for which the number of expansion terms equal to the number of stream pairs. For each available aerosol type \citep[e.g.,][]{Massie:2023}, look-up-tables of Legendre expansion coefficients are pre-computed using an assumed particle size distribution. PSG contains billions of spectral lines of thousands of species from several constantly updated spectroscopic repositories such as HITRAN, ExoMol, JPL, CDMS, GSFC-Fluor. In this work, we used the latest HITRAN database \citep{Gordon:2022}.
The PSG noise model for NIRISS, NIRSpec and MIRI instruments have been benchmarked 
with the JWST ETC\footnote{\url{https://jwst.etc.stsci.edu/}}. The PSG noise model was benchmarked with current post JWST launch sensitivities by using the JWST Exposure-Time-Calculator (ETC) version~1.7. We used the NASA JWST website\footnote{\url{https://webb.nasa.gov/content/webbLaunch/whereIsWebb.html}} to establish the optics temperature post commissioning ($\sim$~40K). For each instrument, we computed the unsaturated signal-to-noise ratio (SNR) from the ETC by using the appropriate filter, subarray, readout pattern, group per integration and integrations per exposure. We adjusted the number of exposures, exposure time, readout noise and dark current in PSG to match these numbers. The total background level for the observations (e-/s) was retrieved from the ETC, and we used a consistent background level in PSG. To estimate the effective instrument throughput efficiencies needed by PSG, we compared the extracted flux from the source (e-/sec) in the ETC and we adjusted the throughput parameter in PSG to match these levels. Finally, we checked that the simulated SNR from the ETC matched the ones we computed with PSG for a set of representative cases.

To ingest the 3D atmospheric fields (temperature, pressure, volume mixing ratios of molecular species, and mass mixing ratios and particle sizes for liquid and water ice clouds) produced by the Generic PCM and create synthetic spectra, we used the PSG module named GlobES\footnote{\url{https://psg.gsfc.nasa.gov/apps/globes.php}} (Global Exoplanetary Spectra, \citealt{Villanueva:2022}). 
For transmission spectroscopy, GlobES computes the radiative transfer at each GCM latitude chunk across the terminator to create a local spectrum. Each of them are then averaged to create an average terminator spectrum for the planet. 
For emission spectra, GLobES performs radiative
transfer simulations across the whole observable disk, and the individual spectra are integrated considering the projected area of each bin. 
Because radiative transfer calculations across the full observable disk are very expensive, we have horizontally binned GCM data by a factor of 3, as typically done to speed up the calculation, leading to 3~$\times$~3 (latitude~$\times$~longitude = 9) binning, and a speed up of 9 times while sacrificing a little accuracy. Thermal phase curves have been computed at all phase angles (with a step of 5 degrees) along the orbit. Broadband phase curves have been integrated over the MIRI 5--25 $\mu m$ range (see Fig. \ref{fig:phase_curves_T1bA} to \ref{fig:phase_curves_T1dB}). To compare with MIRI observations in photometric bands, we have integrated the high-resolution emission spectra using the spectral transmission function of each MIRI band (for F1280W and F1500W bands, in this work).

\section{Results}
\label{section_results}

\subsection{Preferential nightside cloud formation and climate multi-stability}
\label{subsection_results_yang2013}

\begin{figure*}
\centering 
\includegraphics[width=\linewidth]{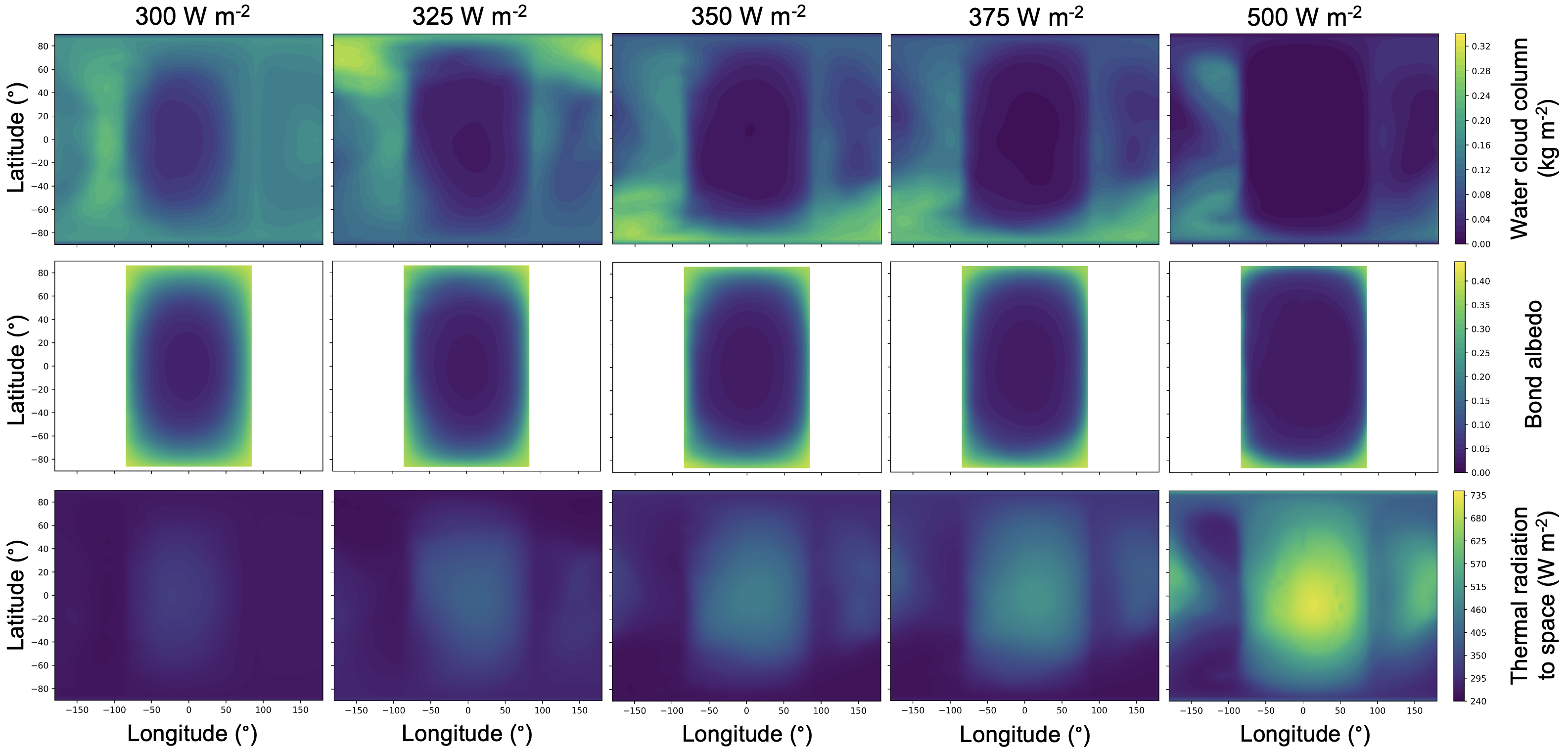}
\caption{Water cloud column (top), albedo (middle) and thermal emission to space (bottom) horizontal maps, for an initially hot and steamy 2~R$_{\oplus}$ planet (with a 10~bar of H$_2$O + 1~bar of N$_2$ atmosphere) synchronously rotating (fixed rotation period of 60 Earth days) around a M3 star. The simulations shown here were forced to ISR (defined here and throughout the manuscript as the global mean average value, i.e. a factor of 4 lower than the stellar flux at the substellar point) ranging from 500~W~m$^{-2}$ or 1.47~F$_{\oplus}$ (right) down to 300~W~m$^{-2}$ or 0.88~F$_{\oplus}$ (left), and the outputs were averaged over 250~Earth days. The associated simulations names are M3-YANG-1, 2, 3, 4 and 5. For reference, insolation is 340.5~W~m$^{-2}$ on present-day Earth.}
\label{fig_yang_P60days_2Rearth_cloud_alb_otr}
\end{figure*}

Our most striking result is that in all the simulations we did (listed in Tables~\ref{table_simu_1} and \ref{table_simu_2}) water clouds form preferentially on the nightside and at the poles. This is illustrated in Fig.~\ref{fig_yang_P60days_2Rearth_cloud_alb_otr} (top row) that shows the column-integrated water cloud distribution for a tidally-locked planet (2~R$_{\oplus}$, 1.4~g$_{\oplus}$) around a M3 star, for a fixed rotation period of 60 Earth days, and for various ISR ('M3-YANG' cases in Table~\ref{table_simu_2}). The parameters used here for the star and the planet closely match those used in the standard simulations of \cite{Yang:2013}, except that it is assumed here that the entire water reservoir is initially vaporized in the atmosphere. While 'cold start' GCM simulations in \cite{Yang:2013} -- with a global surface liquid water ocean -- predict the formation of low-altitude, convective dayside clouds, our 'hot start' GCM simulations -- with a steam atmosphere and a dry surface -- predict the formation of high-altitude, stratospheric nightside clouds. 

\begin{figure*}
\centering 
\includegraphics[width=\linewidth]{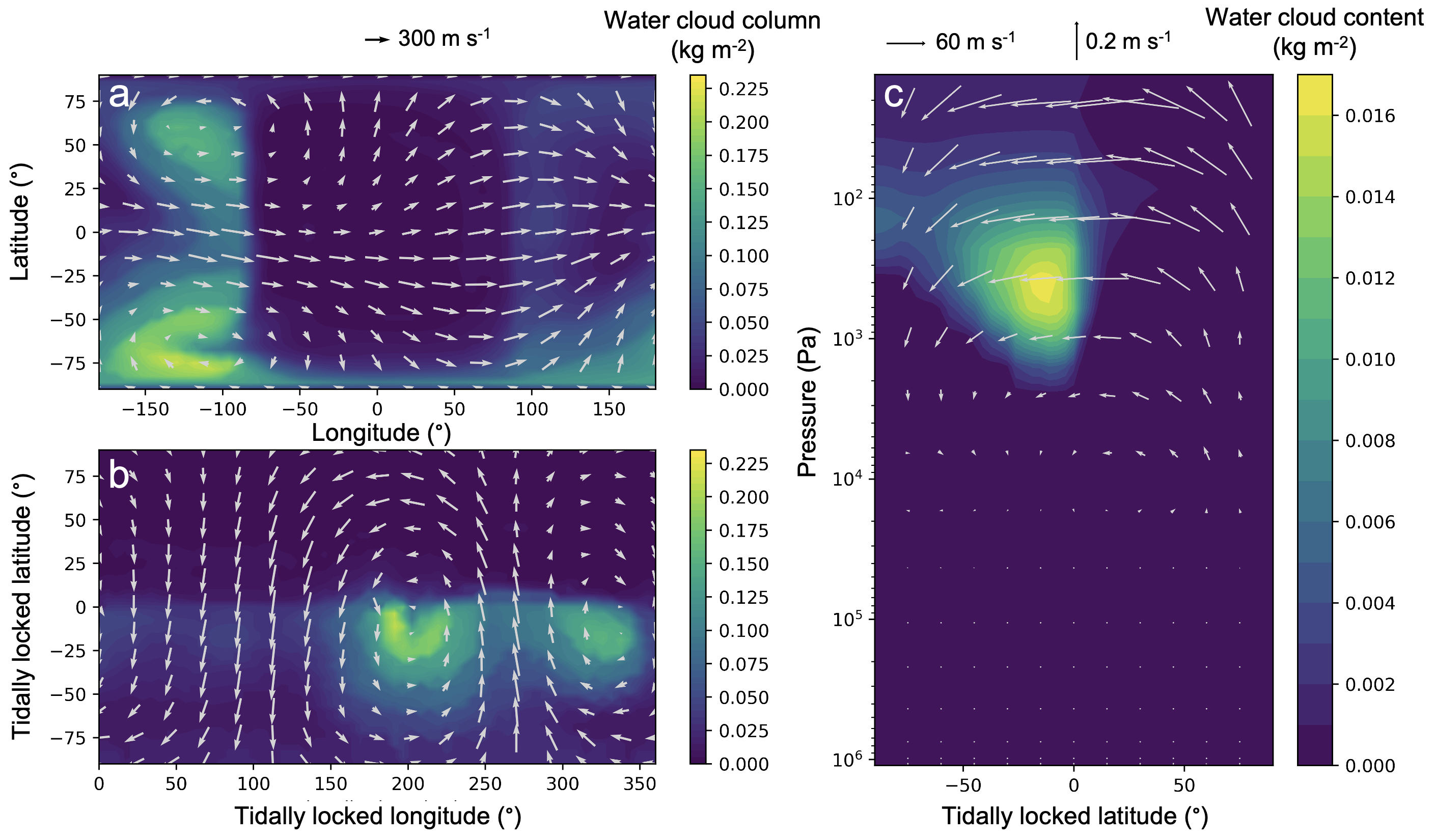}
\caption{Water cloud horizontal and vertical distributions. Water cloud column (a) in latitude vs longitude coordinates and (b) tidally-locked coordinates (b). Water cloud content vertical distribution (c) as a function of tidally-locked latitude (averaged over all tidally-locked longitudes). The plots were computed for an initially hot and steamy 2~R$_{\oplus}$ planet (with a 10~bar of H$_2$O + 1~bar of N$_2$ atmosphere) synchronously rotating (fixed rotation period of 60 Earth days) around a M3 star, and forced to an ISR of 500~W~m$^{-2}$ (simulation M3-YANG-1). The outputs were averaged over 250~Earth days. For reference, insolation is 340.5~W~m$^{-2}$ on present-day Earth. Wind fields are superimposed on the plots. Winds are plotted at a pressure of $\sim$~50~Pa in panels a and b, and averaged over all tidally-locked (TL) longitudes in panel c.}
\label{fig_TL_plots_yang2013}
\end{figure*}

Fig.~\ref{fig_TL_plots_yang2013}b shows the horizontal distribution of clouds and winds in the tidally-locked (TL) coordinates \citep{Koll:2015}. In the TL coordinate system, the TL latitude is equal to 90$^{\circ}$ at the substellar point and -90$^{\circ}$ at the antisubstellar point. Fig.~\ref{fig_TL_plots_yang2013}b was derived by converting Fig.~\ref{fig_TL_plots_yang2013}a (which is similar to Fig.~\ref{fig_yang_P60days_2Rearth_cloud_alb_otr}, top row, right column) using the python package of \citet{Koll:2020}. The TL coordinates highlight well the strong dichotomy in cloud distribution between the dayside and the nightside of the planet, with two maxima located near the two wind gyre structures (near -120$^\circ$ in longitude, $\pm$60$^\circ$ in latitude).

The cloud formation mechanism at play has already been identified and discussed for the cases of early Earth and Venus in \citet{Turbet2021}, and we find that the same mechanism also applies here to tidally-locked exoplanets. In short, water vapor which is the main component of the atmosphere efficiently absorbs the incident stellar flux. This shortwave heating breaks moist convection on the dayside and thus prevents the formation of convective clouds. This shortwave heating also drives strong winds in the stratosphere that transport air parcels from the dayside to the nightside, which is qualitatively similar to the results of \cite{Fujii:2017}. As the air parcels move to the nightside, they are cooled by longwave emission (i.e., thermal emission to space) which reduces their temperature down to the saturation temperature of water vapor. This leads to large-scale condensation, and thus to the formation of stratospheric clouds preferentially located on the nightside and at the poles. The 3D distribution of water clouds and winds is depicted in Fig.~\ref{fig_TL_plots_yang2013}bc using the TL coordinates. We note that we observe in Fig.~\ref{fig_TL_plots_yang2013}abc a slight decrease of the cloud cover at the antisubstellar point. This is particularly visible in the TL coordinates, with a decrease of clouds between -60$^{\circ}$ and -90$^{\circ}$ TL latitudes. This reduction in cloud content is associated with a downward movement of air parcels near the antisubstellar point (see the wind field in Fig.~\ref{fig_TL_plots_yang2013}c). This subsidence produces adiabatic warming thus reducing the amount of clouds in this region, which is qualitatively similar to the results of \cite{Charnay:2021}. We also note that the asymmetry between North and South hemispheres in the cloud distribution (see e.g., Fig.~\ref{fig_TL_plots_yang2013}a) is due to temporal variability. The asymmetry in fact vanishes when averaging over a longer period of time (e.g. 5 Earth years, instead of 250 Earth days, as illustrated in Fig.~\ref{fig_map_CLOUD_ALB_OTR_stars}).


The preferential accumulation of clouds on the nightside and lack of clouds on the dayside has two consequences: (1) lowering the bond albedo (see Fig.~\ref{fig_yang_P60days_2Rearth_cloud_alb_otr}, middle row); and (2) producing a strong greenhouse effect on the nightside (see Fig.~\ref{fig_yang_P60days_2Rearth_cloud_alb_otr}, bottom row). Fig.~\ref{fig_yang_P60days_2Rearth_cloud_alb_otr} (bottom row) illustrates that the planetary thermal radiation displays a strong day-to-night amplitude. This is first due to the greenhouse effect of nightside clouds ; and second, this is due to the fact that a significant fraction of the ISR is absorbed high in the atmosphere, which is qualitatively similar to the results of \cite{Wolf:2019}. This shortwave heating heats up the atmosphere on the dayside, which thermally re-radiates to space, further increasing the day-night Outgoing Thermal Radiation (OTR) contrast. This 3D effect is the direct consequence of the distribution of the stellar flux, combined with the strong, direct absorption of this flux by water vapor.

\begin{figure*}
\centering 
\includegraphics[width=1.0\linewidth]{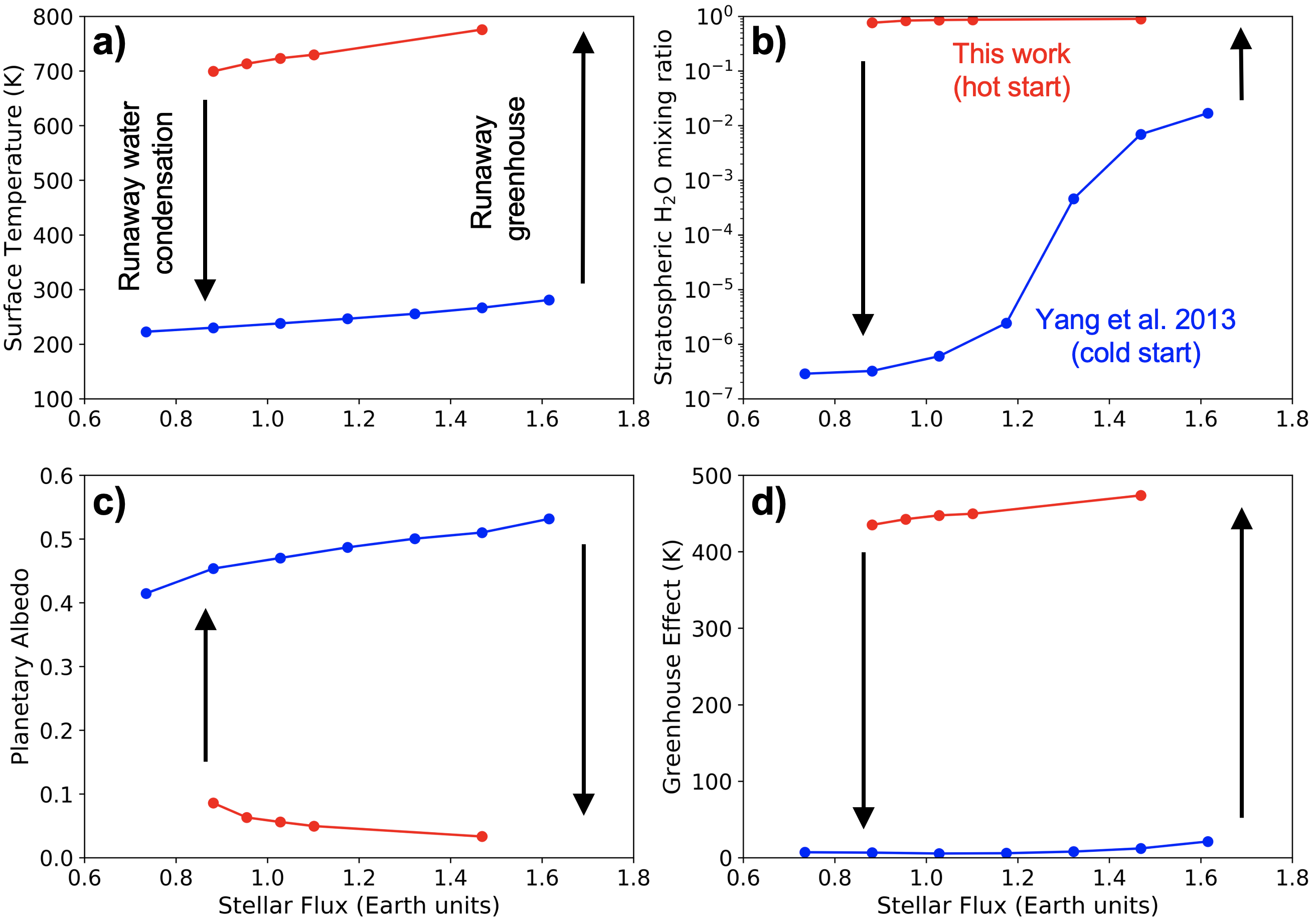}
\caption{Hysteresis loops and conditions of ocean formation for an initially hot and steamy 2~R$_{\oplus}$ planet (with a 10~bar of H$_2$O + 1~bar of N$_2$ atmosphere) synchronously rotating (fixed rotation period of 60 Earth days) around a M3 star (simulation M3-YANG-1). The figure depicts the evolution of surface temperature (top left), stratospheric water mixing ratio (top right), bond albedo (bottom left) and greenhouse effect (bottom right) as a function of the incoming stellar flux. The blue branches correspond to simulation results \citep{Yang:2013} that assume water is initially condensed on the surface. The red branches (this study) correspond to initially hot and steamy simulation results, that assume water is initially in vapor form in the atmosphere.}
\label{fig_comparison_fig2_yang2013}
\end{figure*}

As a consequence, we find the same result than in \cite{Turbet2021}, that depending on the initial state (all water initially in the form of vapor in the atmosphere ; versus all water initially condensed at the surface), there is a wide range of ISR for which climate end states are different. This hysteresis is illustrated in Fig.~\ref{fig_comparison_fig2_yang2013} for the surface temperature (Fig.~\ref{fig_comparison_fig2_yang2013}a), planetary albedo (Fig.~\ref{fig_comparison_fig2_yang2013}c), upper atmosphere water vapor mixing ratio (Fig.~\ref{fig_comparison_fig2_yang2013}b) and greenhouse effect ((Fig.~\ref{fig_comparison_fig2_yang2013}d). The greenhouse effect G is defined as the difference between the surface temperature and the equilibrium temperature of the planet:
\begin{equation}
    G~=~T_{\text{surf}}-\Bigl(\frac{OTR}{\sigma}\Bigr)^{\frac{1}{4}}
\end{equation}
with T$_{\text{surf}}$ the global mean surface temperature, OTR the global mean thermal radiation to space and $\sigma$ the Stefan–Boltzmann constant. Our new simulations (Fig.~\ref{fig_comparison_fig2_yang2013}, red branches) are compared to that of the standard simulations of \cite{Yang:2013} (Fig.~\ref{fig_comparison_fig2_yang2013}, blue branches) in the standard case of a 2~R$_{\oplus}$, 1.4~g$_{\oplus}$ planet synchronously rotating (fixed rotation period of 60 Earth days) around a M3 star.

Fig.~\ref{fig_comparison_fig2_yang2013}a shows the evolution of the surface temperature as a function of ISR, for two types of initial states. The arrows indicate an estimate of the ISR threshold limit at which the radiative TOA budget becomes unstable, leading to a runaway greenhouse (for the 'cold start' sequence) and a runaway water condensation (for the 'hot start' sequence). For the hot start sequence, the arrows are positioned halfway between the last simulation (or lowest ISR) that does not lead to surface water condensation and the first simulation (or higher ISR) that does lead to surface water condensation (i.e., the simulation for which rainwater hits the surface). For the cold start sequence, the arrows are positioned halfway between the last simulation (or highest ISR) that avoid runaway greenhouse and the first simulation (or lowest ISR) that goes into runaway greenhouse \citep{Yang:2013}. As already identified in \cite{Turbet2021}, near the water condensation limit the amount of dayside clouds grows when reducing the ISR, leading to an increase in bond albedo (see Fig.~\ref{fig_comparison_fig2_yang2013}c, red branch). As soon as the water condensation limit is attained, the dayside becomes largely covered by clouds, which makes the bond albedo jump. This produces a strong TOA radiative desequilibrium, which inevitably leads to the condensation of water vapor on the surface. This imbalance is maintained until almost all the water reservoir has condensed onto the surface. This cloud-driven positive feedback is what triggers the runaway water condensation.

\subsection{Water condensation around different types of stars}
\label{subsection_results_various_stars}

\begin{figure*}
\centering 
\includegraphics[width=\linewidth]{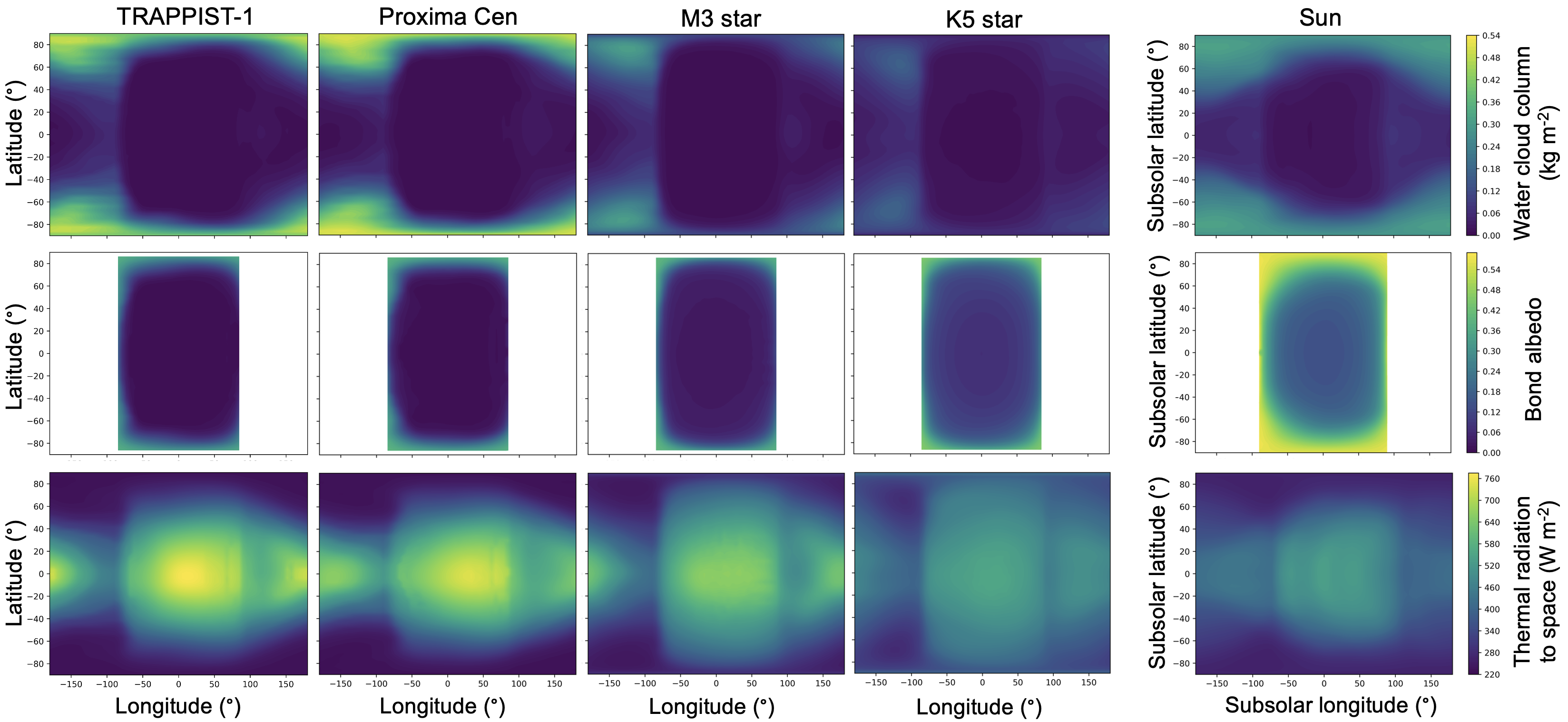}
\caption{Water cloud column (top), albedo (middle) and thermal emission to space (bottom) horizontal maps, for an initially hot and steamy 1~R$_{\oplus}$ planet (with a 10~bar of H$_2$O + 1~bar of N$_2$ atmosphere) orbiting around different types of stars: TRAPPIST-1 (T$_{\star}$~$\sim$~2600~K), Proxima Centauri (T$_{\star}$~$\sim$~3050~K), a M3 star (T$_{\star}$~$\sim$~3400~K), a K5 star (T$_{\star}$~$\sim$~4400~K) and the Sun (T$_{\star}$~$\sim$~5780~K). The associated simulations names are T1-1, Pcen-1, M3-1, K5-1 and SUN-1, respectively. The simulations shown here were forced to an ISR of 500~W~m$^{-2}$ (i.e. 1.47~F$_{\oplus}$), and the outputs were averaged over 100~Earth days. We made the assumption of a synchronous rotation for all the simulations, except for those of the Sun (where we took the results directly from \cite{Turbet2021} in which rotation period is equal to 1 Earth day, orbital period is equal to 365 Earth days, and obliquity is equal to 23.5$^{\circ}$). For this simulation only, the maps were calculated in the heliocentric frame (that is, keeping the subsolar point at 0$^{\circ}$ longitude and 0$^{\circ}$ latitude), and using an average of five Earth years as in \cite{Turbet2021}.}
\label{fig_map_CLOUD_ALB_OTR_stars}
\end{figure*}

We conducted a grid of simulations to explore the conditions for water condensation depending on the type of host star (from Sun-like stars to late M stars ; see cases 'SUN', 'K5', 'M3', 'Pcen' and 'T1' in Table~\ref{table_simu_1}), for an Earth-like planet (1M$_{\oplus}$, 1R$_{\oplus}$) synchronously rotating (except for the Sun case) around its host star. Again, we find that in all the simulations, water clouds preferentially form on the nightside and at the poles. This is illustrated in Fig.~\ref{fig_map_CLOUD_ALB_OTR_stars} (top row) that shows the column-integrated water cloud distribution for the planet depending on the type of host star. Here again, the albedo (Fig.~\ref{fig_map_CLOUD_ALB_OTR_stars}, middle row) and the OTR (Fig.~\ref{fig_map_CLOUD_ALB_OTR_stars}, bottom row) maps are correlated and anti-correlated with the distribution of clouds, respectively.

As discussed in the previous section, we observe again a slight decrease of the cloud cover at the antisubstellar point (Fig.~\ref{fig_map_CLOUD_ALB_OTR_stars}, top row), associated with subsidence warming. We note that the cloud patterns are broadly similar between all types of host star, in particular all those assuming a synchronous rotation. The cloud content is maximum near the two wind gyre structures – the coldest parts of the atmospheres – in all simulations, although their exact location and extension is affected by the parameters of the simulations (in particular, the rotation rate ; see \citealt{Carone:2014,Carone:2015,Carone:2016}).

While the thermal emission maps (Fig.~\ref{fig_map_CLOUD_ALB_OTR_stars}, bottom row) are qualitatively similar between all simulations, we observe a much higher dayside emission for the lowest mass stars (TRAPPIST-1, Proxima Cen) than the other cases. This is first due to the bond albedo being lower (see Fig.~\ref{fig_map_CLOUD_ALB_OTR_stars}, middle row ; see also discussions below), but also, and above all, to the fact that low-mass stars emit more near-IR flux which gets absorbed higher up in the atmosphere \citep{Fujii:2017} and directly re-emitted to space on the dayside \citep{Wolf:2019}.

\begin{figure}
\centering 
\includegraphics[width=\linewidth]{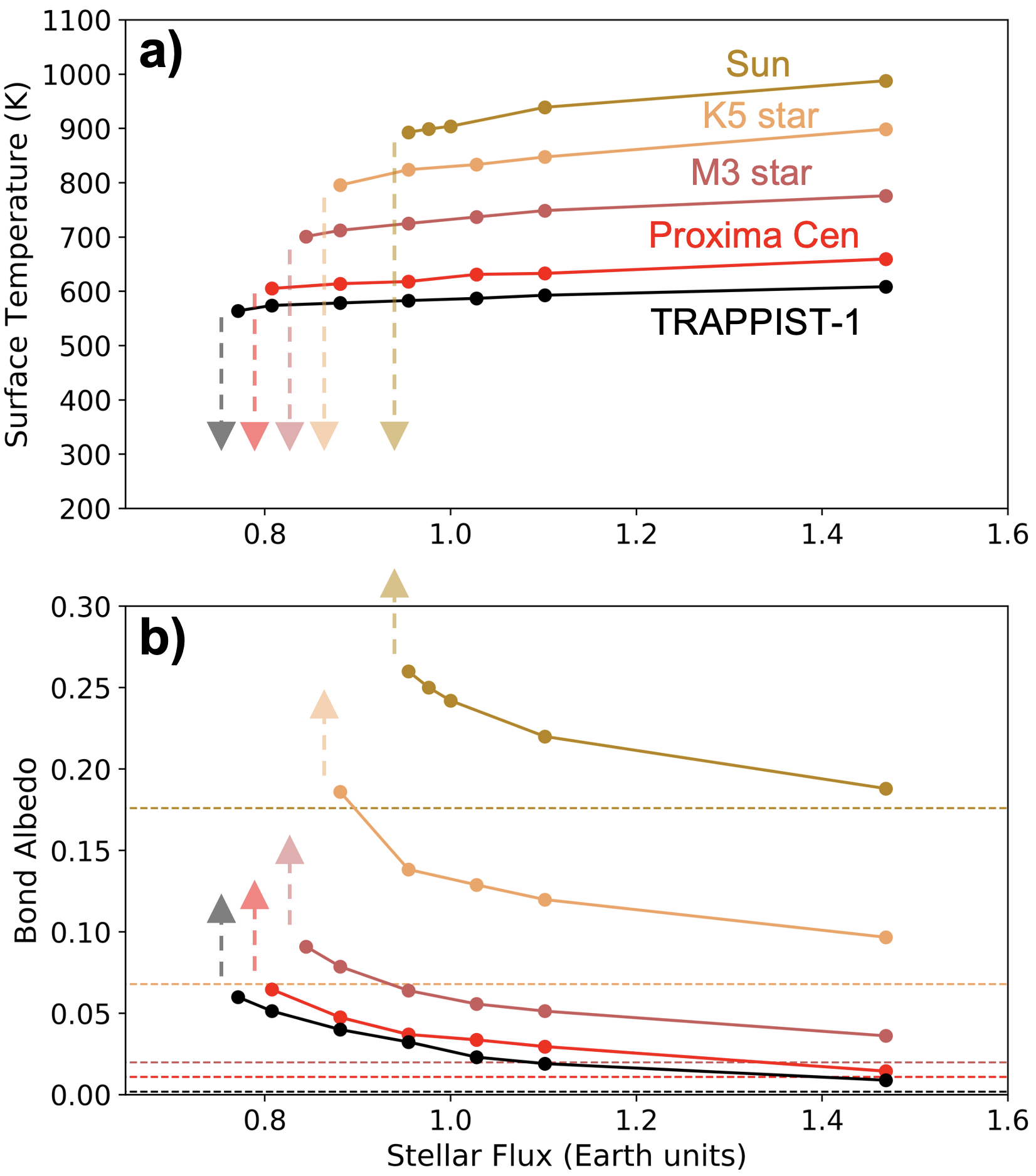}
\caption{Conditions of ocean formation for initially hot and steamy 1~R$_{\oplus}$ planets (with a 10~bar of H$_2$O + 1~bar of N$_2$ atmosphere) orbiting around different types of stars: TRAPPIST-1 (T$_{\star}$~$\sim$~2600~K), Proxima Centauri (T$_{\star}$~$\sim$~3050~K), a M3 star (T$_{\star}$~$\sim$~3400~K), a K5 star (T$_{\star}$~$\sim$~4400~K) and the Sun (T$_{\star}$~$\sim$~5780~K). The associated simulations names are T1-1 to 8, Pcen-1 to 7, M3-1 to 7, K5-1 to 6 and SUN-1 to 6, respectively. The figure depicts the evolution of surface temperature (top) and bond albedo (bottom) as a function of the incoming stellar flux. We made the assumption of a synchronous rotation for all the simulations, except for those of the Sun (where we took the results directly from \cite{Turbet2021} in which rotation period is equal to 1 Earth day, orbital period is equal to 365 Earth days, and obliquity is equal to 23.5$^{\circ}$). Horizontal dashed lines indicate the bond albedos of planets in runaway greenhouse, obtained from 1D cloud-free calculations \citep{Kopparapu:2013}. The dashed arrows indicate the ISR threshold limit at which the radiative TOA budget becomes imbalanced, leading to a runaway water condensation.}
\label{fig_comparison_stars}
\end{figure}

For each type of stars, we used the GCM simulations to build their water condensation sequences, illustrated in Fig.~\ref{fig_comparison_stars}. Fig.~\ref{fig_comparison_stars}a shows the evolution of the surface temperature as a function of ISR, for all the types of host stars. The dashed arrows indicate the ISR threshold limit at which the radiative TOA budget becomes imbalanced, leading to a runaway water condensation. 

There are several noticeable points in this Fig.~\ref{fig_comparison_stars}:
\begin{enumerate}
    \item Firstly, we notice that the surface temperatures are lower for cooler stars despite that more ISR is absorbed by the planet (as illustrated in Fig.~\ref{fig_comparison_stars}b showing the bond albedos). This is somewhat counter-intuitive, as cooler stars have been shown to produce consistently warmer climates on planets with Earth-like atmospheres \citep{Kopparapu:2013,Shields:2013,Godolt:2015,Wolf:2019}. We discuss this effect in more details below.
    \item Secondly, the bond albedo decreases for cooler stars, which was expected (see e.g., \citealt{Kopparapu:2013,Shields:2013}). Moreover, the bond albedo increases when the ISR decreases. This effect, which is discussed in Section~\ref{subsection_results_yang2013}, is what drives the sharp radiative budget deficit at the TOA, and thus what defines the runaway water condensation for all types of stars. Finally, as the ISR reaches the highest values we have explored, Fig.~\ref{fig_comparison_stars}b shows that the bond albedos seem to converge asymptotically towards the results from 1D cloud-free calculations (here taken from \citealt{Kopparapu:2013}). This is due to the fact that the dayside of the planet is getting warmer and therefore has fewer clouds.
    \item Thirdly, the water condensation limit ISR threshold decreases in a monotonic way when decreasing the stellar effective temperature. In other words, this indicates that for a given ISR, water condensation is more difficult to attain for planets around M stars than around Sun-like stars. Implications are discussed in Section~\ref{subsection_results_HZ}.
\end{enumerate}

\begin{figure*}
\centering 
\includegraphics[width=\linewidth]{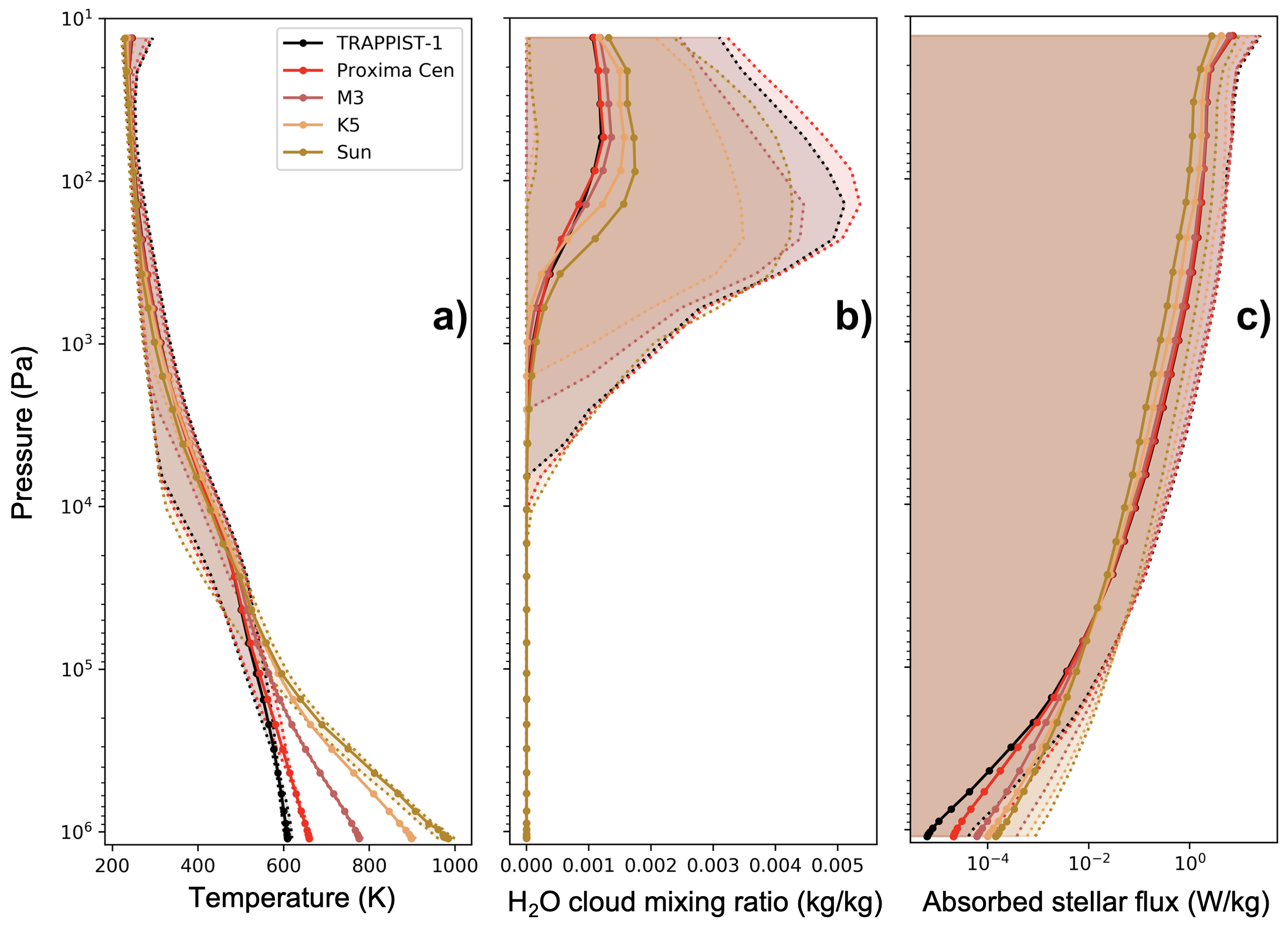}
\caption{Vertical profiles (global mean, with an envelope defined by global maximum and minimum values) of atmospheric temperatures (left), water cloud mixing ratio (middle) and absorbed stellar fluxes (right). The profiles were computed for an initially hot and steamy 1~R$_{\oplus}$ planet (with a 10~bar of H$_2$O + 1~bar of N$_2$ atmosphere) orbiting around different types of stars: TRAPPIST-1 (T$_{\star}$~$\sim$~2600~K), Proxima Centauri (T$_{\star}$~$\sim$~3050~K), a M3 star (T$_{\star}$~$\sim$~3400~K), a K5 star (T$_{\star}$~$\sim$~4400~K) and the Sun (T$_{\star}$~$\sim$~5780~K). The curves were drawn from the same GCM simulations as in Fig. \ref{fig_map_CLOUD_ALB_OTR_stars}. The associated simulations names are T1-1, Pcen-1, M3-1, K5-1 and SUN-1, respectively.}
\label{fig_vertical_stars}
\end{figure*}

To understand in more details why surface temperatures decrease (in thick H$_2$O-rich atmospheres) for cooler stars, we show in Fig.~\ref{fig_vertical_stars}abc vertical profiles of the atmospheric temperatures, cloud mixing ratios and absorbed stellar flux. While atmospheric temperatures are higher for cooler stars in the upper atmosphere (in the stratosphere, typically above 10$^4$~Pa), this is reversed below in the troposphere where atmospheric temperatures become lower for cooler stars. Following \citet{Selsis:2023}, we found that the absorption of the ISR by water bands -- combined with Rayleigh scattering of H$_2$O molecules -- is so efficient, that not enough shortwave flux reaches the lower atmosphere to sustain convection. This is illustrated in Fig.~\ref{fig_vertical_stars}c, which shows indeed that the absorbed stellar radiation drops in the low atmosphere due to most of the flux being absorbed higher in the atmosphere. \citet{Selsis:2023} explored and discussed this effect in details, and we verify here that the same effect appear in all our GCM simulations of thick H$_2$O-rich atmospheres. The clouds have little effect on the temperature structure in the low atmosphere, given that the dayside is almost entirely depleted in clouds (Fig.~\ref{fig_map_CLOUD_ALB_OTR_stars}, top row). In other words, from the point of view of the shortwave radiative transfer calculations, the atmosphere behaves closely to the cloud-free case thus validating the approach of \citet{Selsis:2023}.

\subsection{Impact of the water content}
\label{subsection_results_water_content}

\begin{figure}
\centering 
\includegraphics[width=\linewidth]{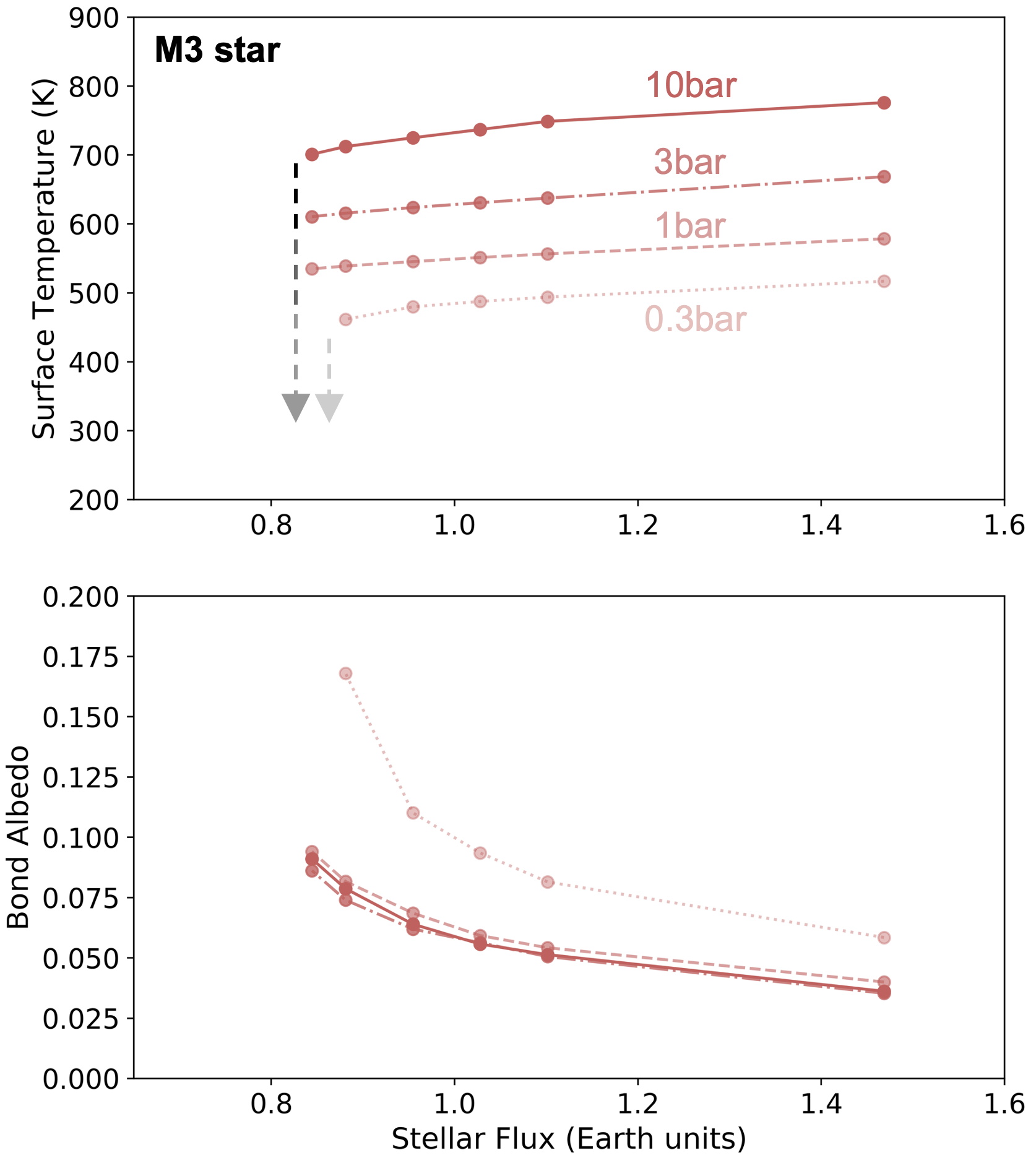}
\caption{Conditions of ocean formation for initially hot and steamy 1~R$_{\oplus}$ planets synchronously rotating around a M3 star (T$_{\star}$~$\sim$~3400~K). The figure depicts the evolution of surface temperature (top) and bond albedo (bottom) as a function of the incoming stellar flux, for four different bulk atmospheric compositions: 10~bar of H$_2$O + 1~bar of N$_2$ (M3), 3~bar of H$_2$O + 1~bar of N$_2$ (M3-3bar), 1~bar of H$_2$O + 1~bar of N$_2$ (M3-1bar) and 0.3~bar of H$_2$O + 1~bar of N$_2$ (M3-0.3bar).}
\label{fig_comparison_water_contents}
\end{figure}

We conducted a series of simulations to explore the effect of the water mass fraction in our simulations (M3 from Table~\ref{table_simu_1} and M3-3bar, M3-1bar, M3-0.3bar from Table~\ref{table_simu_2}), and how it may potentially affect the water condensation limit. Fig.~\ref{fig_comparison_water_contents} shows the water condensation sequences for H$_2$O atmospheric pressures from 0.3 to 10~bar (all simulations assume 1~bar of N$_2$ as a background gas).

We first notice that the surface temperatures increase with water content, as already seen in \citet{Turbet2021}, and as expected from the increasing greenhouse effect of water vapor. Note however that for planets orbiting very cool stars, the surface temperature increases very little with H$_2$O partial pressure, because the stellar flux in the lower atmosphere is too low to maintain convection and thus the thermal profile reaches an isotherm (see Section~\ref{subsection_results_TRAPPIST1} where this is discussed in the context of TRAPPIST-1b).

The bond albedo sequence looks very similar for all H$_2$O partial pressures explored, except the one at P$_{\rm H_2O}$~=~0.3~bar. For this case, the water vapor content is low enough to reduce the fraction of ISR absorbed by the atmosphere, which leads to an increase in bond albedo (as a reminder, the surface albedo is fixed at 0.2 in these simulations). A similar behavior is seen in the 1D cloud-free simulations of \citet{Kopparapu:2013} (see e.g., their Fig.~6a) where the bond albedo for an Earth-like planet around a M-type star decreases as surface temperature increases, due to the increasing water vapor content. Moreover, given that less ISR is absorbed on the dayside by water vapor, the dayside cloud positive feedback (producing the runaway water condensation introduced in previous subsections) is even more efficient. This is visible in Fig.~\ref{fig_comparison_water_contents}b revealing that the bond albedo increases more rapidly for the P$_{\rm H_2O}$~=~0.3~bar simulation when decreasing ISR.
As a result, the water condensation limit occurs at slightly higher ISR for the P$_{\rm H_2O}$~=~0.3~bar simulation (see Fig.~\ref{fig_comparison_water_contents}a) than for the other H$_2$O partial pressures explored. It is not clear if this trend would remain for even lower H$_2$O atmospheric mixing ratios and/or H$_2$O partial pressures, given that the saturation temperature also decreases with partial pressure. This deserves to be explored in more details, with possibly some interesting implications for the case of early Venus, and more generally for the case of land exoplanets \citep{Kodama:2015,Ding:2020}.

The fact that the evolutions of the bond albedo as a function of ISR (see Fig.~\ref{fig_comparison_water_contents}b) look very similar for all GCM simulations with P$_{\rm H_2O}$~$\geq$~1~bar is a good indication that we are capturing well here the climate regime in which H$_2$O is a dominant gas (starting from P$_{\rm H_2O}$~=~1~bar, up to pressures as high as desired). Although the amount of water has an impact on the surface temperature (see Fig.~\ref{fig_comparison_water_contents}a), for planets more irradiated than the water condensation limit the surface temperatures are always higher than that of the saturation temperature of the assumed H$_2$O partial pressure. The fact that the P$_{\rm H_2O}$~=~10~bar simulation produces surface temperatures even higher than the critical temperature of water (T$_{\text{crit}}$~=~647~K) indicates that for P$_{\rm H_2O}$~$\ge$~10~bar, the water condensation limit would remain unchanged. More specifically, H$_2$O-rich atmospheric simulations at different total pressures have thermal profiles that overlap well (see Fig.~\ref{vertical_profiles_T1bcd_2}c, in the context of TRAPPIST-1b). Increasing the surface pressure to higher values would likely produce an extension of the thermal profile, while keeping a similar profile at low atmospheric pressures. 

For planets orbiting ultra-cool stars, \citet{Selsis:2023} showed that the thermal profile in the lower atmosphere can converge to an isotherm at a temperature that is lower than that of the critical point of water, potentially permitting the co-existence of a steam atmosphere and liquid water oceans for planets more irradiated than the runaway greenhouse limit. Although we did not explore this possibility in detail, our simulations for TRAPPIST-1 inner planets (see Figs.~\ref{vertical_profiles_T1bcd_1} and \ref{vertical_profiles_T1bcd_2}) support this result. This result might however be restricted to a purely theoretical experiment, since a tiny internal heat flux is enough to increase the surface temperatures above the critical temperature of water (\citealt{Selsis:2023} ; see also Fig.~\ref{vertical_profiles_T1bcd_2}b in the context of TRAPPIST-1b).

To summarize, the sensitivity experiment described in this subsection shows that the choice of P$_{\rm H_2O}$~=~10~bar for our GCM calculations is a relevant choice to represent all the cases where water is a major component of the atmosphere. Additional radiative active gases (e.g. CO$_2$) could change the TOA radiative budget, likely producing more warming (see \citealt{Turbet2021}), but this deserves to be explored quantitatively (see discussions in Section~\ref{section_discussions}).

\subsection{Consequences for the Habitable Zone}
\label{subsection_results_HZ}

\begin{figure*}
\centering 
\includegraphics[width=17cm]{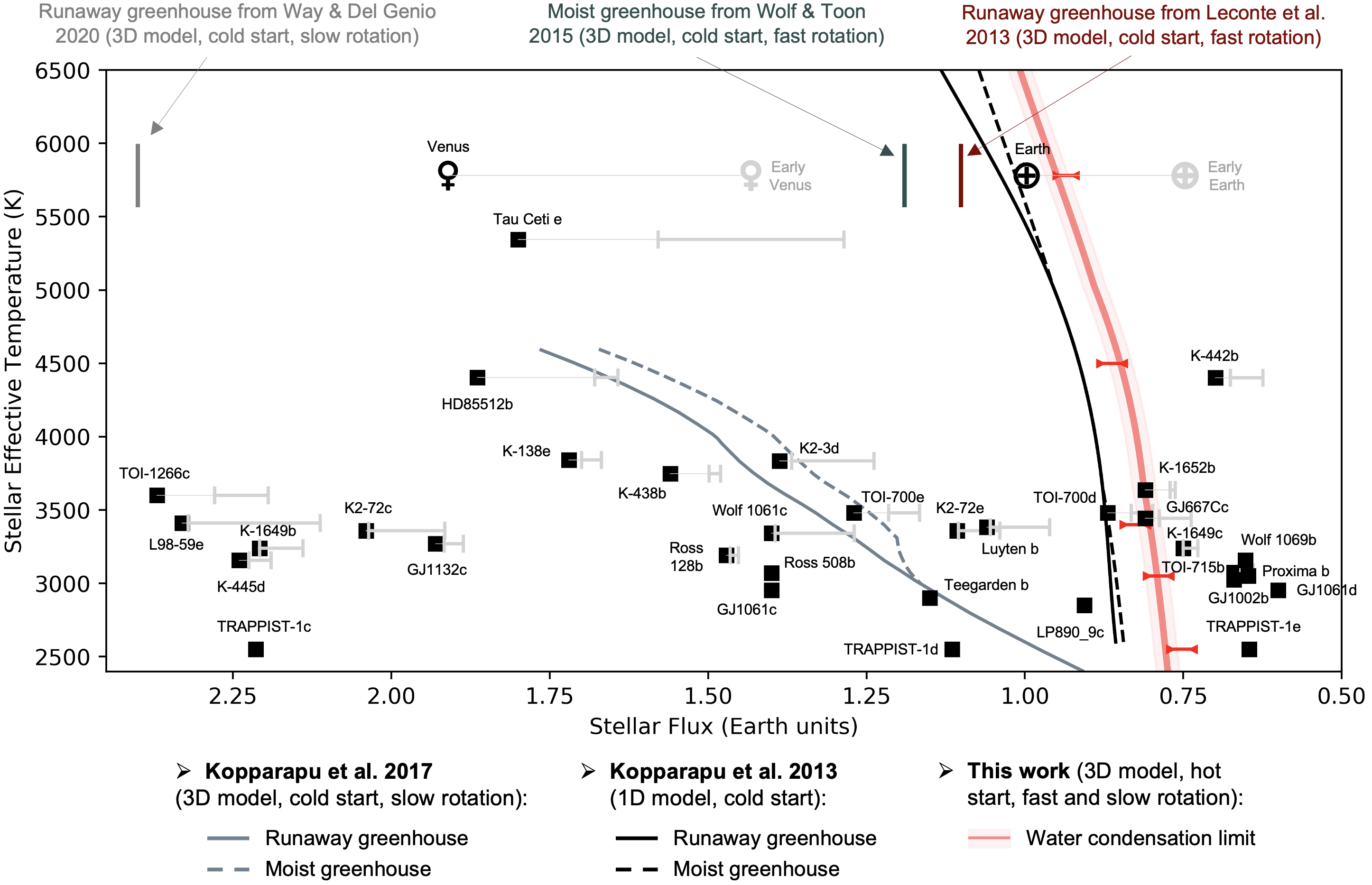} 
\caption{Various inner edge boundaries of the Habitable Zone (HZ) for different types of stars, based on the results of 1D calculations \citep{Kopparapu:2013} and 3D calculations for slow \citep{Kopparapu:2017,Way:2020} and fast rotators \citep{Leconte:2013nat,Wolf:2015}. The black solid and dashed lines were calculated based on 1D climate calculations, for the runaway greenhouse and the moist greenhouse, respectively \citep{Kopparapu:2013}. The grey solid and dashed lines were calculated based on 3D climate calculations of tidally-locked planets, for the runaway greenhouse and the moist greenhouse, respectively \citep{Kopparapu:2017}. The red bold line (and red brackets) indicates the newly calculated water condensation limit. All the currently known planets and exoplanets (with a mass $\leq$ 5~M$_{\oplus}$ and/or a radius $\leq$ 1.6~R$_{\oplus}$) in or near the inner boundary of the HZ are shown (black squares). We also added (grey brackets) the incoming stellar radiation (ISR) they received at the ZAMS (Zero Age Main Sequence) based on their age estimates, using the grid of stellar models of \citet{Baraffe2015}. The water condensation limit (red bold line) should be compared with the ISR that planets received at the ZAMS (grey brackets).}
\label{fig_revised_HZ}
\end{figure*}

\begin{figure*}
\centering 
\sidecaption
\includegraphics[width=12cm]{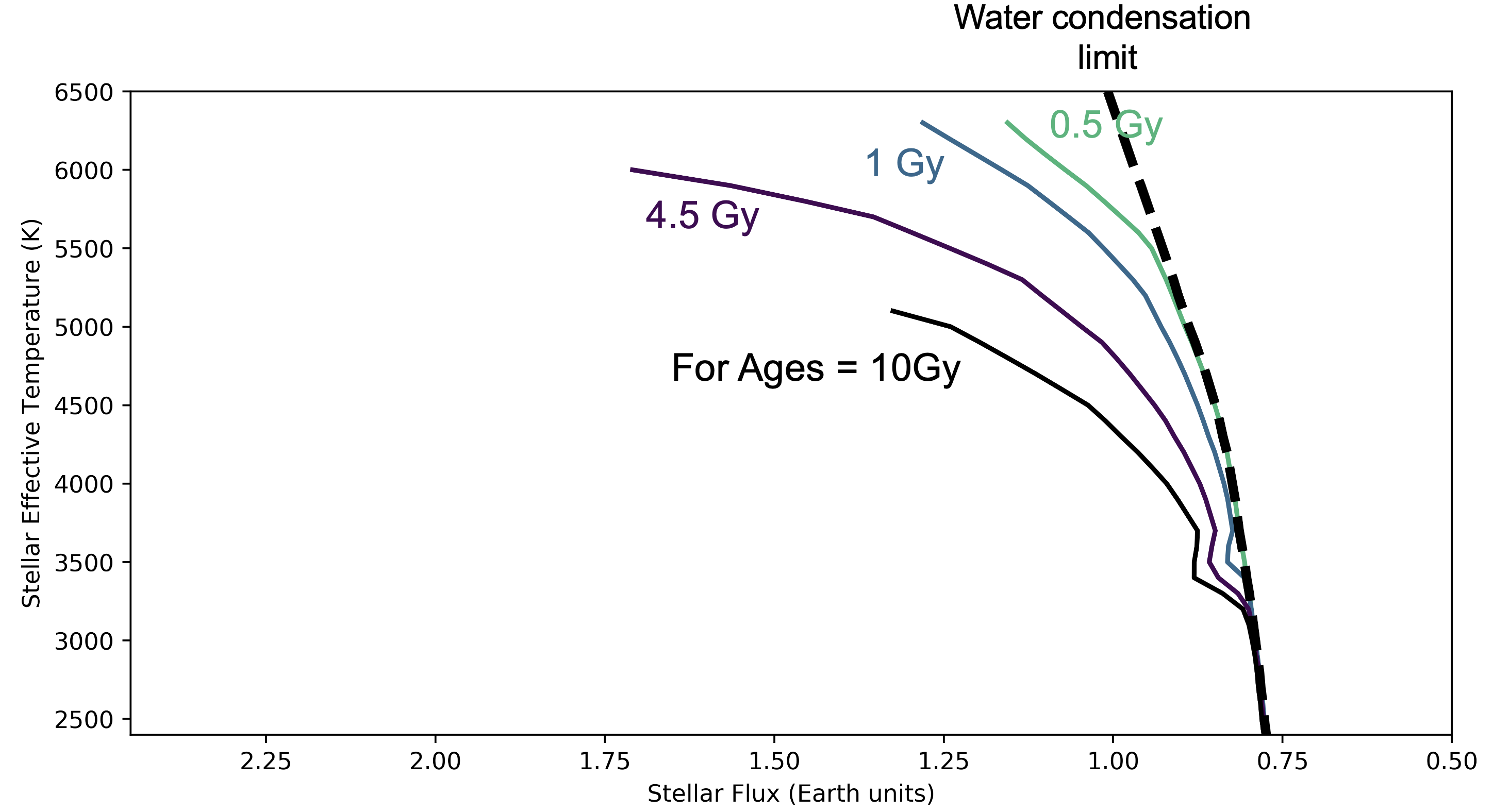} 
\caption{Various positions of the water condensation limit, depending on the age of the planetary system considered. The standard water condensation limit (black dashed line) is valid at the Zero Age Main Sequence (ZAMS). It can be used as is to evaluate the condensation of water on a planet, provided that the insolation of the planet is taken at the ZAMS (see e.g., grey brackets in Fig~\ref{fig_revised_HZ}). The other water condensation limits (solid colored lines) indicate, for a planet of age X (with X = 0.5, 1, 4.5, and 10~Gyr here), that if that planet is to the left of the age X curve, then it has never been able to, or will never be able to, condense its water at the surface. These curves were calculated from the standard water condensation curve (black dashed line), and shifted using the grid of stellar models of \citet{Baraffe2015}. Numerical values of all the curves are provided in Table~\ref{table:fits_IHZ}.}
\label{fig_revised_HZ2}
\end{figure*}

We synthesized our results to define a Water Condensation Zone (WCZ) which we propose to use to redefine the inner boundary of the Habitable Zone (HZ). We remind that the ISR at which the surface water condensation starts is necessarily less than or equal to the ISR required to trigger the runaway greenhouse. It is therefore a more restrictive criterion. Fig.~\ref{fig_revised_HZ} highlights in bold red the water condensation limit calculated with GCM simulations for five types of host stars (Sun-like, K5, M3, M5.5, M8 stars). Planets receiving lower ISR than this limit are within the WCZ of their host star, meaning ocean formation by condensation is permitted. To put this result in context, we have added various inner edge boundaries of the HZ, based on the results of 1D calculations \citep{Kopparapu:2013} and 3D calculations for slow \citep{Kopparapu:2017,Way:2020} and fast rotators \citep{Leconte:2013nat,Wolf:2015}. We note that the inner edge of the HZ is at much higher ISR for slowly rotating planets, because GCM simulations predict for these planets the formation of highly reflective, dayside convective clouds \citep{Yang:2013,Yang:2014b} that significantly expands the HZ towards higher ISR.

\begin{figure}
\centering 
\includegraphics[width=\linewidth]{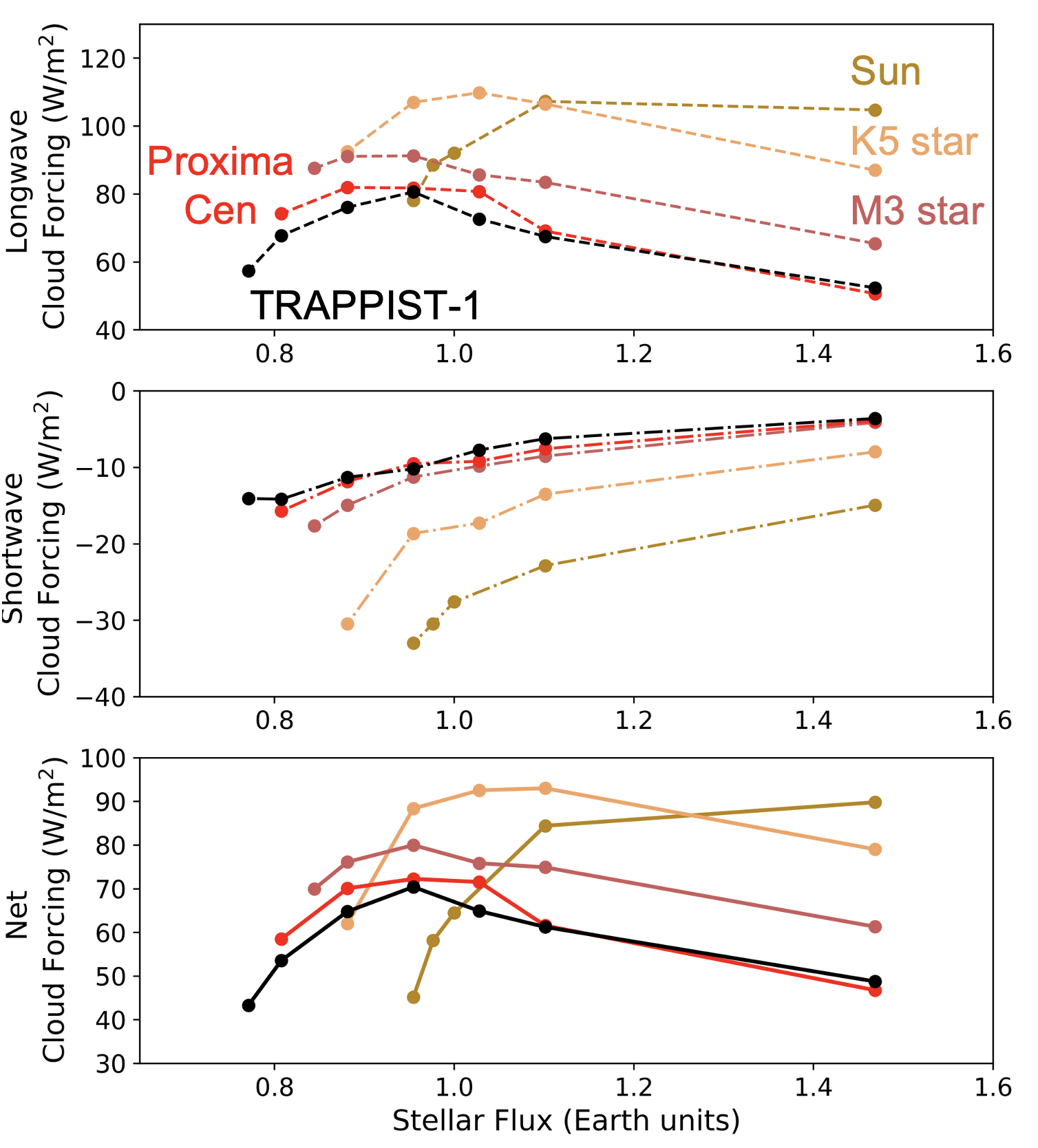}
\caption{Cloud forcings for initially hot and steamy 1~R$_{\oplus}$ planets (with a 10~bar of H$_2$O + 1~bar of N$_2$ atmosphere) orbiting around different types of stars: TRAPPIST-1 (T$_{\star}$~$\sim$~2600~K), Proxima Centauri (T$_{\star}$~$\sim$~3050~K), a M3 star (T$_{\star}$~$\sim$~3400~K), a K5 star (T$_{\star}$~$\sim$~4400~K) and the Sun (T$_{\star}$~$\sim$~5780~K). The associated simulations names are T1-1 to 8, Pcen-1 to 7, M3-1 to 7, K5-1 to 6 and SUN-1 to 6, respectively. The figure depicts the greenhouse effect (top) of clouds, the amount of incident radiation (middle) reflected back by the clouds (the more negative the value, the greater the reflected flux), and the net radiative effect (bottom) of clouds (positive values mean warming), as a function of the incoming stellar flux. In all these simulations, clouds produce a strong atmospheric warming.}
\label{fig_RT_clouds}
\end{figure}

To derive the water condensation limit, we have followed the two steps listed below:
\begin{itemize}
    \item Based on the results presented in Section~\ref{subsection_results_various_stars}, we estimated for the 5 types of host stars (Sun-like, K5, M3, M5.5, M8 stars) 5 estimates of the water condensation limit (Fig.~\ref{fig_revised_HZ}, red brackets) bracketed with the ISR of two GCM simulations (the ISR of the first GCM simulation reaching runaway water condensation ; and the ISR of the last GCM simulation not reaching runaway water condensation).
    \item We fitted the 5 brackets with the empirical HZ inner edge of \citet{Kopparapu:2013} shifted by 0.065~F$_{\oplus}$ (i.e. 0.065 the insolation on Earth). This corresponds roughly to a difference of 22~W~m$^{-2}$ between the inner edge of the HZ calculated in \citet{Kopparapu:2013} and our water condensation limit. We did not attempt to fit the points with a more sophisticated curve, since the \citet{Kopparapu:2013} fit with a shift is sufficient to satisfactorily match the results of our GCMs simulations. The condensation curve looks broadly similar to the HZ inner edge of \citet{Kopparapu:2013} because low-mass stars emit more flux in the near-IR, which produces more shortwave warming that delays water condensation. This is the same effect that makes the runaway greenhouse occur at lower ISR for planets around low-mass stars \citep{Kopparapu:2013}. Compared to 1D cloud-free calculations, nightside clouds add an additional greenhouse effect, which is mainly responsible for the shift of the HZ inner edge toward lower ISR. The net effect of clouds is roughly similar across the various types of host stars (see Fig.~\ref{fig_RT_clouds}) .
\end{itemize}

It is important to note that the ISR that matters to determine if a planet can condense its primordial water vapour reservoir into oceans is the ISR the planet received at the beginning of the Main Sequence Phase of its host star. This is commonly referred to as the Zero Age Main Sequence (ZAMS). For the types of stars we have studied here, the ZAMS corresponds to the moment when (1) the stellar luminosity is minimal while (2) planets are already fully formed. In Fig.~\ref{fig_revised_HZ}, we indicated the ISR of all the currently known planets and exoplanets (with a mass $\leq$ 5~M$_{\oplus}$ and/or a radius $\leq$ 1.6~R$_{\oplus}$) in or near the inner boundary of the HZ are shown (black squares), along with the ISR they received at the ZAMS (grey brackets). The planets are based on the following publications: HD85512b \citep{Pepe:2011}, GJ667Cc \citep{Bonfils:2013,Anglada-Escude:2013}, $\tau$ Ceti e \citep{Tuomi:2013}, Kepler-138e \citep{Rowe:2014,Piaulet:2023}, Kepler-442b and Kepler-438b \citep{Torres:2015}, K2-3d \citep{Crossfield:2015}, Kepler-445d \citep{Muirhead:2015}, Proxima b \citep{Anglada:2016}, K2-72c and e \citep{Dressing:2017}, TRAPPIST-1c, d and e \citep{Gillon:2017}, Kepler-1652b \citep{Torres:2017}, Wolf 1061c, Luyten b \citep{Astudillo-Defru:2017}, Kepler-1649b and c \citep{Angelo:2017,Vanderburg:2020}, Ross 128b \citep{Bonfils:2018}, GJ 1132c \citep{Bonfils:2018b}, Teegarden b \citep{Zechmeister:2019}, L98-59e \citep{Kostov:2019}, TOI-700d and e \cite{Gilbert:2020,Gilbert:2023}, TOI-1266c \citep{Demory:2020}, GJ 1061c and d \citep{Dreizler:2020}, LP-890-9c \citep{Delrez:2022}, Ross 508b \citep{Harakawa:2022}, GJ 1002b \citep{Suarez-Mascareno:2023}, Wolf 1069b \citep{Kossakowski:2023}, TOI-715b \citep{Dransfield:2023}.
While the previous limits of the inner edge of the HZ \citep{Kopparapu:2013,Leconte:2013nat,Wolf:2015,Kopparapu:2017,Way:2020} should be compared with the ISR that planets receive today, the water condensation limit (Fig.~\ref{fig_revised_HZ}, bold red line) should be compared with the ISR that planets received at the ZAMS.

The ISR of planets at the ZAMS were calculated by (1) using stellar properties, including age estimates and associated uncertainties based on up-to-date publications ; and by (2) calculating the variation in the luminosity of the star between the ZAMS and its current measured age. The grey brackets in Fig.~\ref{fig_revised_HZ} correspond to the $\pm$1$\sigma$ age estimates. For planets for which age constraints are either very loose or non-existent (mostly for low-mass stars), we arbitrarily used age estimates between 1 and 10~Gyr, which is the maximum value in the grid of stellar models of \citealt{Baraffe2015}). While ISR change significantly with age (between now and the ZAMS) for planets around solar-type and K-type stars, they change very little around low-mass (M-type) stars. Sun-like stars for instance have a luminosity that increases by about 40$\%$ during the first 4.5~Gigayears of evolution in their Main Sequence. Ultra-cool stars like TRAPPIST-1 have an extended Pre-Main-Sequence (PMS) phase during which their luminosity can decrease by more than 2 orders of magnitude within the first Gigayear of evolution \citep{Bolmont:2017}. However, once they reach their minimum luminosity (at the ZAMS), then their luminosity increases very little on a time scale less than or equal to the age of the universe.

Fig.~\ref{fig_revised_HZ} illustrates well the findings of \citet{Turbet2021} that (1) Venus never received a low enough insolation to condense its oceans, and (2) that the Earth was able to condense its oceans thanks to the fact that the Sun's luminosity was fainter in the past than it is today (a.k.a. the Faint Young Sun Opportunity). It also shows that the standard water condensation curve (bold red line in Fig.~\ref{fig_revised_HZ}) can be compared directly to the ISR received today by planets orbiting around M stars, because their luminosity evolves very little on the main sequence phase. We predict that several planets (K2-3d, TOI-700d, TOI-700e, K2-72e, Luyten b, LP890-9c, etc.) which are located well inside the Habitable Zone calculated using 3D models for tidally-locked planets \citep{Kopparapu:2017}, may in fact never have been able to form surface oceans. Interestingly, for TOI-700d this prediction depends on the age of the planet, which controls the minimum ISR it received at the ZAMS. This result highlights the importance of accurately measuring the age of planetary systems (at least, for planets around early-M and more massive stars) to assess their ability to form oceans in the first place.

In parallel, and to facilitate the usability of our results by the community, we have produced in Fig.~\ref{fig_revised_HZ2} a series of water condensation curves at different ages. The curves indicate, for a planet of age X (with X = 0.5, 1, 4.5, and 10~Gyr here), that if that planet is to the left of the age X curve, then it has never been able to, or will never be able to, condense its water at the surface. The numerical values of the curves are provided in Table~\ref{table:fits_IHZ}, for further uses. These curves were (1) calculated from the standard water condensation curve (black dashed line in Fig.~\ref{fig_revised_HZ2} ; which is the same than the bold red line in Fig.~\ref{fig_revised_HZ}) and then (2) multiplied by the ratio $\frac{L_{\star, \text{at age = X}}}{L_{\star, \text{at the ZAMS}}}$. This ratio was determined using the grid of stellar models of \citet{Baraffe2015}. We note that the most massive stars in Fig.~\ref{fig_revised_HZ2} are too short-lived for the 4.5 and 10~Gigayears limits to be defined. We also note that to evaluate the potential habitability of a planet, the curves in Fig.~\ref{fig_revised_HZ2} should be used in tandem with the traditional HZ \citep{Kopparapu:2013,Kopparapu:2017}. In the case of planets orbiting around highly evolved stars (for example, the Earth in more than 1 billion year), the inner edge of the Habitable Zone may be more restrictive than the Water Condensation Limit. In other words, the planet may have formed its first oceans by condensation, but then lost them several billion years later when a runaway greenhouse was triggered by the increase in the star's luminosity \citep{Leconte:2013nat,Wolf:2015}.

\begin{table*}
\centering
\resizebox{0.62\linewidth}{!}{%
\begin{tabular}{lccccr}
   \hline
   Star Type & Water Condensation & WCL at 0.5~Gy & at 1~Gy & at 4.5~Gy & at 10~Gy\\
   & Limit (WCL) & & & & \\
   (T$_{\text{eff}}$) & (F$_{\oplus}$) & (F$_{\oplus}$) & (F$_{\oplus}$)& (F$_{\oplus}$) & (F$_{\oplus}$)  \\
   \hline
2300  &  0.772  &         &  0.772  &  0.772  &  0.775  \\
2400  &  0.774  &         &  0.774  &  0.774  &  0.778  \\
2500  &  0.777  &         &  0.777  &  0.778  &  0.781  \\
2600  &  0.779  &         &  0.779  &  0.781  &  0.783  \\
2700  &  0.782  &         &  0.782  &  0.783  &  0.786  \\
2800  &  0.784  &         &  0.784  &  0.784  &  0.788  \\
2900  &  0.787  &  0.787  &  0.787  &  0.789  &  0.791  \\
3000  &  0.790  &  0.790  &  0.790  &  0.792  &  0.795  \\
3100  &  0.793  &  0.793  &  0.793  &  0.795  &  0.800  \\
3200  &  0.796  &  0.796  &  0.796  &  0.800  &  0.808  \\
3300  &  0.799  &  0.799  &  0.800  &  0.816  &  0.838  \\
3400  &  0.803  &  0.803  &  0.806  &  0.845  &  0.881  \\
3500  &  0.806  &  0.806  &  0.831  &  0.858  &  0.880  \\
3600  &  0.810  &  0.810  &  0.830  &  0.854  &  0.876  \\
3700  &  0.814  &  0.814  &  0.824  &  0.849  &  0.875  \\
3800  &  0.817  &  0.817  &  0.827  &  0.856  &  0.890  \\
3900  &  0.820  &  0.820  &  0.831  &  0.863  &  0.905  \\
4000  &  0.824  &  0.824  &  0.836  &  0.872  &  0.921  \\
4100  &  0.828  &  0.828  &  0.843  &  0.884  &  0.942  \\
4200  &  0.832  &  0.832  &  0.850  &  0.896  &  0.964  \\
4300  &  0.838  &  0.838  &  0.859  &  0.910  &  0.988  \\
4400  &  0.843  &  0.843  &  0.867  &  0.923  &  1.011  \\
4500  &  0.849  &  0.850  &  0.875  &  0.939  &  1.037  \\
4600  &  0.856  &  0.858  &  0.885  &  0.957  &  1.075  \\
4700  &  0.863  &  0.866  &  0.895  &  0.976  &  1.114  \\
4800  &  0.871  &  0.875  &  0.905  &  0.995  &  1.155  \\
4900  &  0.879  &  0.884  &  0.916  &  1.016  &  1.196  \\
5000  &  0.888  &  0.894  &  0.929  &  1.046  &  1.240  \\
5100  &  0.896  &  0.903  &  0.941  &  1.076  &  1.325  \\
5200  &  0.904  &  0.912  &  0.953  &  1.105  &  1.412  \\
5300  &  0.911  &  0.922  &  0.971  &  1.134  &  1.499  \\
5400  &  0.919  &  0.932  &  0.992  &  1.186  &  1.588  \\
5500  &  0.927  &  0.943  &  1.014  &  1.241  &         \\
5600  &  0.935  &  0.962  &  1.036  &  1.297  &         \\
5700  &  0.943  &  0.988  &  1.066  &  1.354  &         \\
5800  &  0.951  &  1.014  &  1.096  &  1.455  &         \\
5900  &  0.959  &  1.040  &  1.126  &  1.565  &         \\
6000  &  0.968  &  1.071  &  1.164  &  1.709  &         \\
6100  &  0.976  &  1.101  &  1.203  &         &         \\
6200  &  0.984  &  1.130  &  1.242  &         &         \\
6300  &  0.992  &  1.156  &  1.281  &         &         \\
6400  &  1.000  &         &         &         &         \\
6500  &  1.008  &         &         &         &         \\
\hline
\end{tabular}
}
\caption{Numerical values of the water condensation curves at different ages shown on Fig.~\ref{fig_revised_HZ2}. We note that the values have been interpolated between the different GCM simulations using the same empirical fit as in \citet{Kopparapu:2013} (see main text for more detailed explanations).}
\label{table:fits_IHZ}
\end{table*}

In summary, using a grid of GCM simulations, we have defined for the first time a Water Condensation Zone (WCZ), that we propose to use as a new standard to define the inner edge of the HZ. It is fortunate that this limit lies very close -- albeit at lower ISR -- to the 1D cloud-free calculations of \citet{Kopparapu:2013}, which are the most widely used definition in the exoplanet community. This is mainly because (1) water vapor profiles are similar between 1D and 3D climate models, and (2) the absence of dayside clouds makes the atmosphere behave as if it were cloud-free from the point of view of shortwave heating. There is also a significant impact (1) of the greenhouse effect of nightside clouds, and (2) of the temporal evolution of stellar luminosity, which both quantitatively affect the water condensation limit (e.g., compared to \citealt{Kopparapu:2013}).

An important point to note is that the question of the ability of planets to condense their primordial atmospheric water vapor reservoir is in fact even more constraining for planets orbiting M-stars (than Sun-like stars), since the level of ISR they receive is much higher during their PMS, which can last up to 1 gigayear \citep{Baraffe2015} for the lowest mass stars. Indeed, to evaluate the ability of a planet to condense its primordial water reservoir, the water condensation limit should be compared to the minimum ISR received by the planet over the course of its evolution (in general, at the beginning of the Main Sequence Phase, a.k.a. the ZAMS), and not to its currently observed ISR. As a consequence, unlike planets orbiting solar stars, planets around M stars cannot benefit from a lower ISR period at the beginning of their evolution \citep{Turbet2021} favorable for water condensation. This is well illustrated in Fig.~\ref{fig_revised_HZ2} by looking at the difference between the water condensation limit at the ZAMS (dashed black lines), and the water condensation limit at several ages (solid colored lines).

\subsection{Application to TRAPPIST-1 planets and their observability with JWST}
\label{subsection_results_TRAPPIST1}

\subsubsection{Water condensation and clouds}

\begin{figure*}
\centering 
\includegraphics[width=\linewidth]{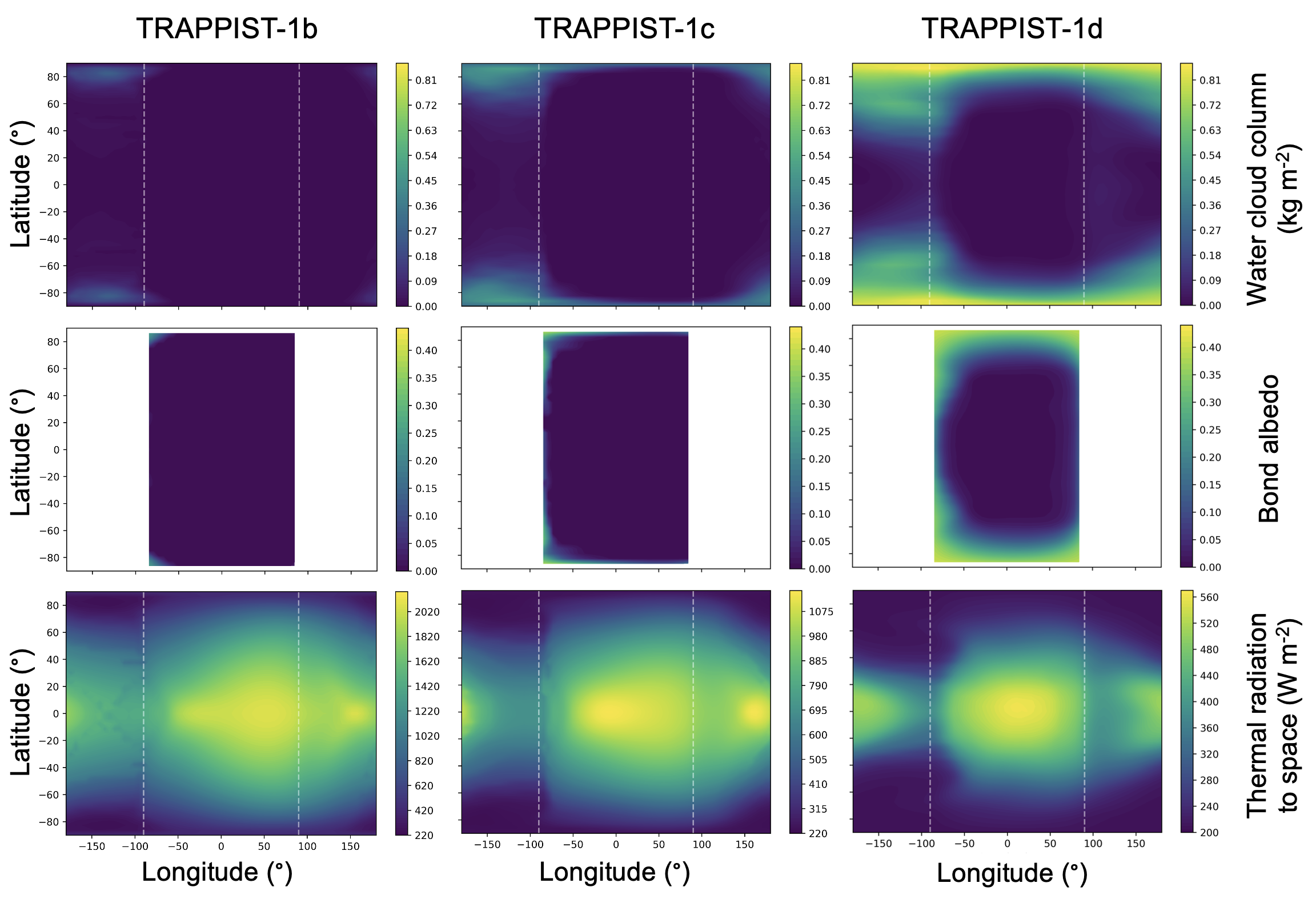}
\caption{Water cloud column (top), albedo (middle) and thermal emission to space (bottom) horizontal maps, for TRAPPIST-1b (left), TRAPPIST-1c (middle) and TRAPPIST-1d (right). The associated simulations names are T1b, T1c and T1d, respectively. The simulations assume initially hot and steamy tidally-locked planets (with a 10~bar of H$_2$O + 1~bar of N$_2$ atmosphere), and the outputs were averaged over 100~Earth days. The dashed white lines indicate the location of the terminator.}
\label{fig_map_TRAPPIST-1bcd}
\end{figure*}

We have extended our GCM simulations of H$_2$O-rich atmospheres to known exoplanets, in particular to TRAPPIST-1 planets (see Table~\ref{table_simu_2} for a detailed list). Fig.~\ref{fig_map_TRAPPIST-1bcd} shows the cloud distribution, bond albedo and OTR for TRAPPIST-1 inner planets (based on the stellar and planet properties provided in \citealt{Agol2021}, following \citealt{Mann:2019} and \citealt{Ducrot:2020}). Again, clouds are located mostly in the nightside, high in the atmosphere. It is clearly visible that the cloud horizontal extension reduces with increasing ISR. The extension of clouds has impacts on the observables -- in particular transit spectra -- which we discuss in the following subsections.

For the three TRAPPIST-1 inner planets, in none of the cases that we have explored the water vapor is able to condense on the surface, and thus to form oceans. TRAPPIST-1d, which lies very close to the runaway greenhouse limit (and could be either in or out, depending on subtle cloud feedbacks ; see \citealt{Wolf:2017,Turbet:2018aa}), is unable to reach the runaway water condensation and should thus lie outside the HZ. Simulations of TRAPPIST-1e (and also Proxima b, for which we also performed a simulation, as it is a prime target for the JWST and the ELTs) systematically lead to the quasi-complete condensation of the water vapor reservoir onto the surface. This confirms what had been calculated in Section~\ref{subsection_results_various_stars}, in the case of an Earth-like planet around TRAPPIST-1, that the water condensation limit is located somewhere between the orbit of TRAPPIST-1d and e. Given that TRAPPIST-1b, c and d are outside of the Water Condensation Zone of their star, they are unlikely to be habitable planets. More generally, there is a significant number of small exoplanets, which had been previously put inside or close to the HZ calculated with 3D models \citep{Kopparapu:2017}, and that may never have condensed their primordial water reservoir. This includes TOI-700d \citep{Gilbert:2020}, TOI-700e \citep{Gilbert:2023}, LP-890-9c \citep{Delrez:2022}, Teegarden b \citep{Zechmeister:2019}, K2-72e \citep{Dressing:2017}, K2-3d \citep{Crossfield:2015}, etc.

\begin{figure*}
\centering 
\includegraphics[width=17cm]{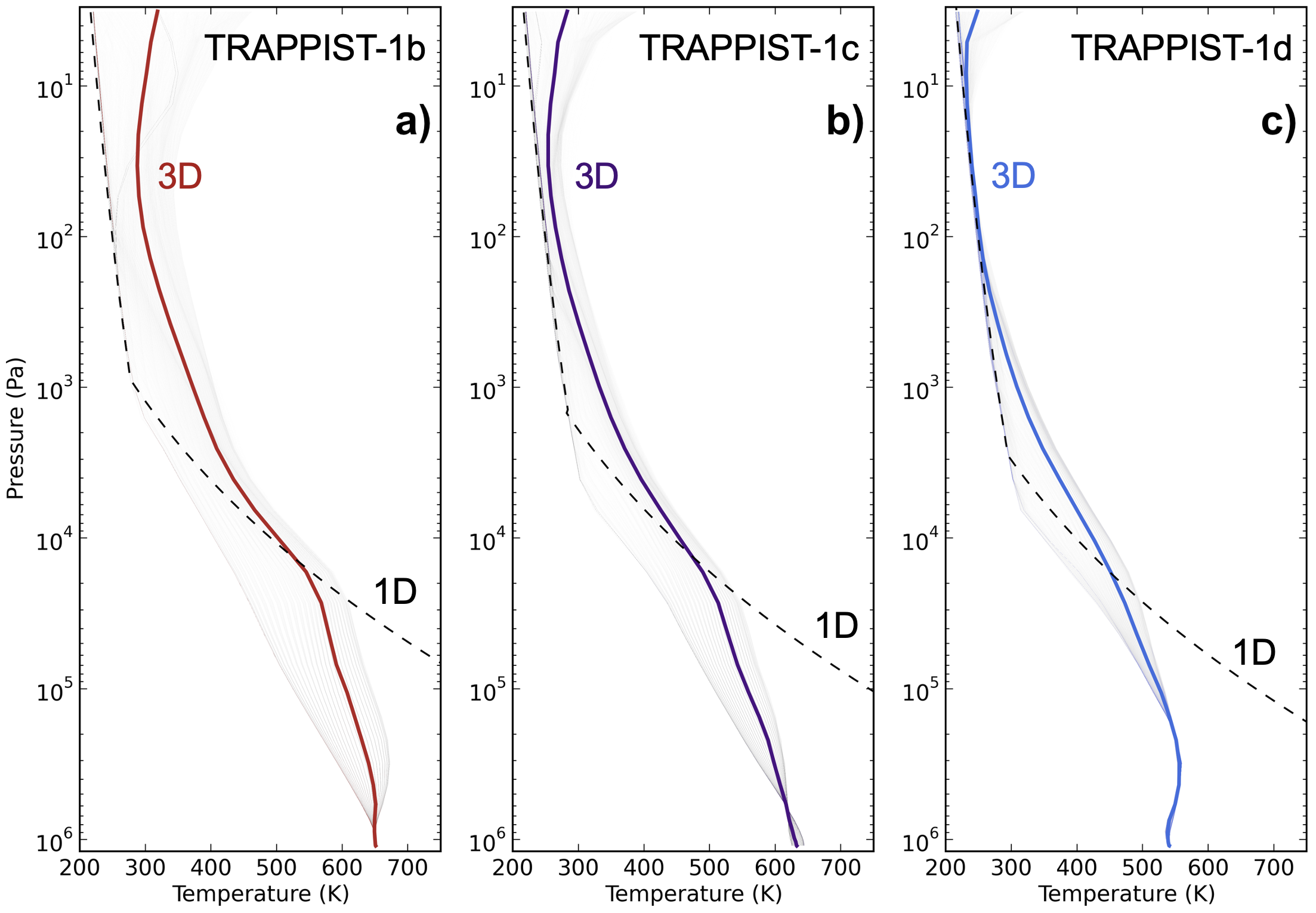}
\caption{Temperature vertical profiles for TRAPPIST-1b (a), TRAPPIST-1c (b) and TRAPPIST-1d (d). The associated simulations names are T1b, T1c and T1d, respectively. The solid lines are global mean, temporal average (over 100~Earth days) temperature profiles calculated from GCM simulations. The shaded lines are snapshots taken at different latitude and longitude points to show the range of profiles within a simulation. The dashed lines are calculated with a 1D inverse radiative-convective model \citep{Turbet2019aa,Turbet2020aa}, similar to those used in \citet{Kopparapu:2013} to calculate the inner edge of the HZ.}
\label{vertical_profiles_T1bcd_1}
\end{figure*}

\begin{figure*}
\centering 
\includegraphics[width=17cm]{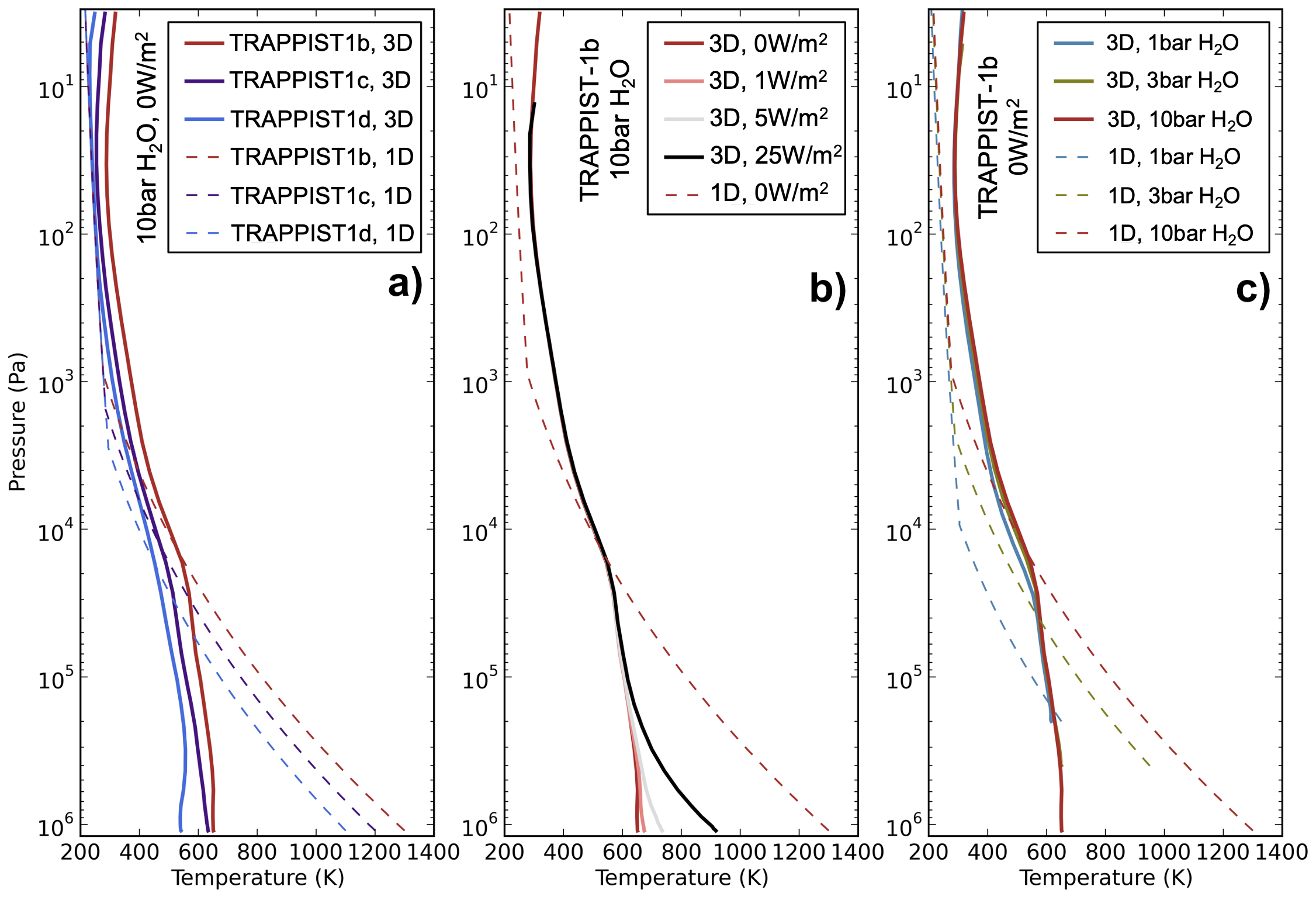}
\caption{Temperature vertical profiles (a) comparing TRAPPIST-1b, c and d (simulations T1b, T1c and T1d), (b) several internal heat fluxes for TRAPPIST-1b (simulations T1b, T1b-Fgeo1, T1b-Fgeo5 and T1b-Fgeo25), (c) several water partial pressures for TRAPPIST-1b (simulations T1b, T1b-3bar, T1b-1bar). The solid lines are global mean, temporal average (over 100~Earth days) temperature profiles calculated from GCM simulations. The dashed lines are calculated with a 1D inverse radiative-convective model \citep{Turbet2019aa,Turbet2020aa}, similar to those used in \citet{Kopparapu:2013} to calculate the inner edge of the HZ.}
\label{vertical_profiles_T1bcd_2}
\end{figure*}

We have performed several sensitivity simulations to explore the thermal structure of the TRAPPIST-1 planetary atmospheres, summarized in Figs.~\ref{vertical_profiles_T1bcd_1} and \ref{vertical_profiles_T1bcd_2}. Again, our results confirm the 1D accelerated time-marching model results of \citet{Selsis:2023} (see a comparison in Extended Data Fig.~3 of \citealt{Selsis:2023}) that the thermal profiles strongly diverge from that calculated from 1D inverse radiative-convective models. In particular, due to the lack of shortwave flux reaching the lower atmosphere and surface, the low atmosphere is much colder than predicted from the adiabatic profile. The water vapor content also has a limited impact on the thermal structure (see Fig.\ref{vertical_profiles_T1bcd_2}c). The internal heat flux, which is possibly high on TRAPPIST-1b due to tidal heating \citep{Turbet:2018aa}, has a strong impact on the lower part of the atmosphere (see Fig.\ref{vertical_profiles_T1bcd_2}b), by transporting energy from the surface to the upper part of the atmosphere. This brings an extra heating term warming the atmospheric layers near the surface. This confirms with a 3D GCM the results of \citet{Selsis:2023}. Both the water content and internal heat fluxes (in the range of simulations explored here) have almost no impact on the temperature structure above 0.1~bar as well as on the location and amount of water clouds.

The fact that the thermal profiles calculated with the GCM are very different from the 1D inverse climate calculations has several important consequences \citep{Selsis:2023}: reduction of the lifetime of a magma ocean ; modification of the mass-radius relationships, etc.

All the vertical profiles discussed in this Section are used as input of radiative transfer calculations to compute various types of observables (transit spectra, eclipses, phases curves) described in following subsections.

\subsubsection{Transit Spectra}
\label{subsubsection_transit_spectra}

\begin{figure*}
\sidecaption
\includegraphics[width=12cm]{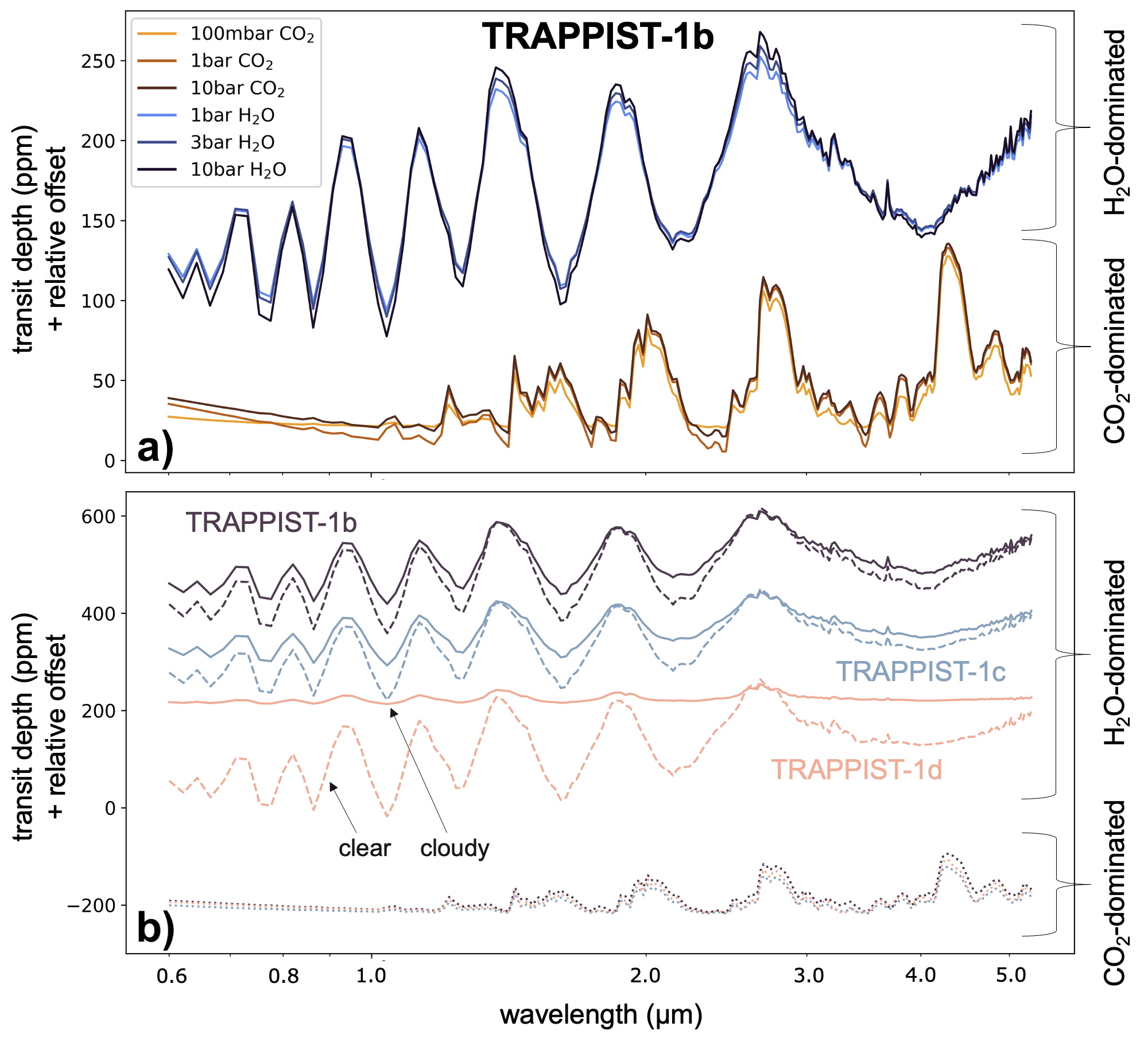}
\caption{Various transit spectra computed with PSG for the three innermost planets of the TRAPPIST-1 system calculated from 3D GCM simulations (assuming H$_2$O-dominated composition ; and also CO$_2$-dominated, for comparison). In the upper panel (a), we compare transit spectra of TRAPPIST-1b for various compositions and pressures ; in the lower panel (b), we compare transit spectra of TRAPPIST-1b, c and d for two compositions only (10~bar H$_2$O + 1~bar N$_2$ ; 10~bar CO$_2$). Dashed lines indicate H$_2$O cloud-free atmospheres, while solid lines assume H$_2$O cloudy ones. Dotted lines indicate CO$_2$ atmospheres (including the effect of CO$_2$ ice clouds, which is however extremely limited here ; note also that CO$_2$ condensation was cut off for TRAPPIST-1b and c simulations with 100~mbar of CO$_2$ as this would cause the atmosphere to collapse on the nightside.). We note that these are relative transit depths (we shifted the curves with arbitrary offsets to compare them more easily).}
\label{transit_spectra_T1bcd}
\end{figure*}

We used the various GCM simulations of TRAPPIST-1b, c and d to compute their expected signature in transit spectroscopy. Fig.~\ref{transit_spectra_T1bcd} summarizes our results for H$_2$O-dominated atmospheres that we compare -- for the context -- to spectra for CO$_2$-dominated atmospheres, calculated also from dedicated GCM simulations (not shown).

Firstly, we observe that the amplitude of water bands in the transit spectra is quite strong, up to 150~ppm for TRAPPIST-1b. We compared our spectra with those of \citet{Koll:2019b} (calculated between 1.75-3~$\mu$m) for cloud-free H$_2$O-dominated atmospheres. Our cloud-free spectra (see Fig.~\ref{transit_spectra_T1bcd}b, dashed lines) look qualitatively similar. When correcting the amplitude of the transit depth features by the scale height ratio (about 35$\%$ change) using the latest mass measurement of \citet{Agol2021} (compared to that of \citealt{Grimm:2018}), the cloud-free spectra of \citet{Koll:2019b} are also quantitatively similar to ours. The amplitude of water bands is significantly stronger than that of CO$_2$ bands (up to 75~ppm for TRAPPIST-1b), due mainly to a larger scale height (with a lower mean molecular weight, and a higher temperature for H$_2$O than CO$_2$). We note that the 10~bar CO$_2$ spectra look qualitatively similar to those of \citet{Lustig-Yaeger:2019} and \citet{Koll:2019b} (calculated between 1.75-3~$\mu$m), although the amplitudes of CO$_2$ absorption bands are lower. This is again most likely because we used planet properties of \citet{Agol2021}, while \citet{Lustig-Yaeger:2019} and \citet{Koll:2019b} used old values (with a higher mass for TRAPPIST-1b) from \citet{Grimm:2018}. Fig.~\ref{transit_spectra_T1bcd}a illustrates that the relative transit depths (i.e. the transit depth corrected from an arbitrary offset) are very similar for the range of atmospheric pressures simulated (from 1 to 10~bar for H$_2$O, which confirms the results of \citealt{Koll:2019b}; from 0.1 to 10~bar for CO$_2$, which confirms the results of \citealt{Morley:2017}). This is simply due to the fact that the atmospheric structure in the upper atmosphere (i.e. typically for pressures lower than 0.1~bar, which is also the radiative part of the atmosphere, to which transit spectroscopy is most sensitive to) is weakly impacted by the total pressure (see e.g. Fig.~\ref{vertical_profiles_T1bcd_2}c). This result differs for CO$_2$ atmospheres from the predictions of \citet{Koll:2019b}, which used a parameterization of the heat redistribution that produces day-to-night temperature variations significantly larger than what our GCM simulations predict. This impacts the temperature at the terminator, the atmospheric height scale, and thus the amplitude of the molecular bands in the transit spectra as a function of surface pressure.

Secondly, we observe that the impact of clouds varies strongly from one planet to another. This is illustrated in Fig.~\ref{transit_spectra_T1bcd}b comparing the transit spectra of TRAPPIST-1b, c, d with and without the effect of clouds (solid versus dashed lines). In fact, the horizontal extension of the clouds (towards the dayside, and thus the terminator regions) increases when decreasing the ISR (see Fig.~\ref{fig_map_TRAPPIST-1bcd}, top row). This is simply because the planet is warmer which reduces the area of cloud stability. The contrast is particularly striking between TRAPPIST-1d -- with a significant amount of water clouds reaching the terminator -- and the two inner planets, with a lack of clouds. For the TRAPPIST-1 system, we thus find that the ISR required for the water clouds not to significantly blur the transit spectra is somewhere between that of TRAPPIST-1c and d. More generally, we find -- using the grid of simulations performed for this study -- that removing most water clouds of the terminator requires an ISR typically at least twice that of the Earth.

TRAPPIST-1b, c, d are transit spectroscopy targets of the JWST space-based observatory, starting from Cycle 1, with both NIRSpec and NIRISS instruments. To prepare and anticipate the outcome of these observations, we computed error bars on the synthetic spectra by cumulating all transit observations of JWST Cycle 1, for the three TRAPPIST-1 inner planets (see Figs.~\ref{transit_spectra_T1bcd_H2O} and \ref{transit_spectra_T1bcd_CO2}). This includes a total of 2 transits of TRAPPIST-1b with NIRISS \citep{Lim:2021}, 4 transits of TRAPPIST-1c with NIRSpec \citep{Rathcke:2021} and 2 with NIRISS \citep{Lim:2021}, and 2 transits of TRAPPIST-1d with NIRSpec \citep{Lafreniere:2017}. Our spectra reveal that water vapor dominated (i.e., without H$_2$/He as the dominant gases) atmospheres are theoretically within reach of JWST Cycle 1 observations using transit spectroscopy, for TRAPPIST-1b and c only. For TRAPPIST-1d, the H$_2$O bands are much weaker due to the effect of clouds forming at the terminator, and a detection would require a larger amount of transits. Detection thresholds are provided in Table~\ref{table:num_transits}, for each of the atmospheric scenarios (H$_2$O and CO$_2$-dominated atmospheres) and observations planned for the first cycle of JWST. We note that here we did not account for the transit stellar contamination effect \citep{Rackham:2018,Rackham:2023}, which will be an issue (see review in \citealt{Turbet:2020review} and references therein).

\begin{table*}
\centering
\begin{tabular}{lccr}
   \hline
  Case & SNR Cycle 1 transits & Number of transits at 5$~\sigma$ & Instrument \\
   \hline
T1b 0.1 bar CO$_2$ & 2.9 & 7 & NIRISS \\
T1b 1 bar CO$_2$ & 5.1 & 2 & NIRISS \\
T1b 10 bar CO$_2$ & 4.7 & 3 & NIRISS \\
T1b 1 bar H$_2$O & 19.2 & 1 & NIRISS \\
T1b 3 bar H$_2$O & 21.3 & 1 & NIRISS \\
T1b 10 bar H$_2$O & 36.8 & 1 & NIRISS \\
   \hline
T1c 0.1 bar CO$_2$ & 1.8 & 15 & NIRISS \\
T1c 0.1 bar CO$_2$ & 2.8 & 13 & NIRSpec \\
T1c 1 bar CO$_2$ & 3.3 &  5 & NIRISS \\
T1c 1 bar CO$_2$ & 4.6 & 5 & NIRSpec \\
T1c 10 bar CO$_2$ & 2.4 & 9 & NIRISS \\
T1c 10 bar CO$_2$ & 3.5 & 9 & NIRSpec \\
T1c 10 bar H$_2$O & 14.9 & 1 & NIRISS \\
T1c 10 bar H$_2$O & 17.4 & 1 & NIRSpec \\
\hline
T1d 0.1 bar CO$_2$ & 3.4 & 5 & NIRSpec \\
T1d 1 bar CO$_2$ & 5.4 & 2 & NIRSpec \\
T1d 10 bar CO$_2$ & 4.4 & 3 &  NIRSpec \\
T1d 10 bar H$_2$O & 4.5 & 3 & NIRSpec \\
 \hline
\end{tabular}
\caption{SNR values to detect an atmosphere assuming Cycle~1 observations and number of transits to achieve a 5$~\sigma$ detection. }
\label{table:num_transits}
\end{table*}

\begin{table*}
\centering
\begin{tabular}{lcr}
   \hline
  Case & SNR Cycle 1 eclipses & SNR Cycle 1 eclipses \\
   & (MIRI F1280W) & (MIRI F1500W) \\
   \hline
1b airless 0 albedo & 8.7 & 9.0 \\
1b airless 0.2 albedo & 7.6 & 7.9 \\
1b 0.1 bar CO$_2$ & 4.3 & 2.3 \\
1b 1 bar CO$_2$ & 2.7 & 2.1 \\
1b 10 bar CO$_2$ & 2.2 & 1.9 \\
1b 1 bar H$_2$O & 5.6 & 5.0 \\
1b 3 bar H$_2$O & 5.5 & 5.0 \\
1b 10 bar H$_2$O & 5.5 & 4.9 \\
   \hline
1c airless 0.2 albedo & & 5.5 \\
1c 0.1 bar CO$_2$ & & 1.3 \\
1c 1 bar CO$_2$ & & 0.9 \\
1c 10 bar CO$_2$ & & 1.1 \\
1c 10 bar H$_2$O & & 3.2 \\
 \hline
\end{tabular}
\caption{SNR values to detect a secondary eclipse assuming Cycle~1 observations. Here we corrected the SNR to adjust the Kurucz stellar flux of TRAPPIST-1 in the F1500W MIRI filter to the observed value \citep{Greene:2023,Zieba:2023}.}
\label{table:num_eclipses}
\end{table*}

\subsubsection{Secondary Eclipses}
\label{subsubsection_emission_spectra}

\begin{figure*}
\centering 
\subfloat[TRAPPIST-1b, eclipse depths]{
\label{subfig:eclipse_T1b}
\includegraphics[width=17cm]{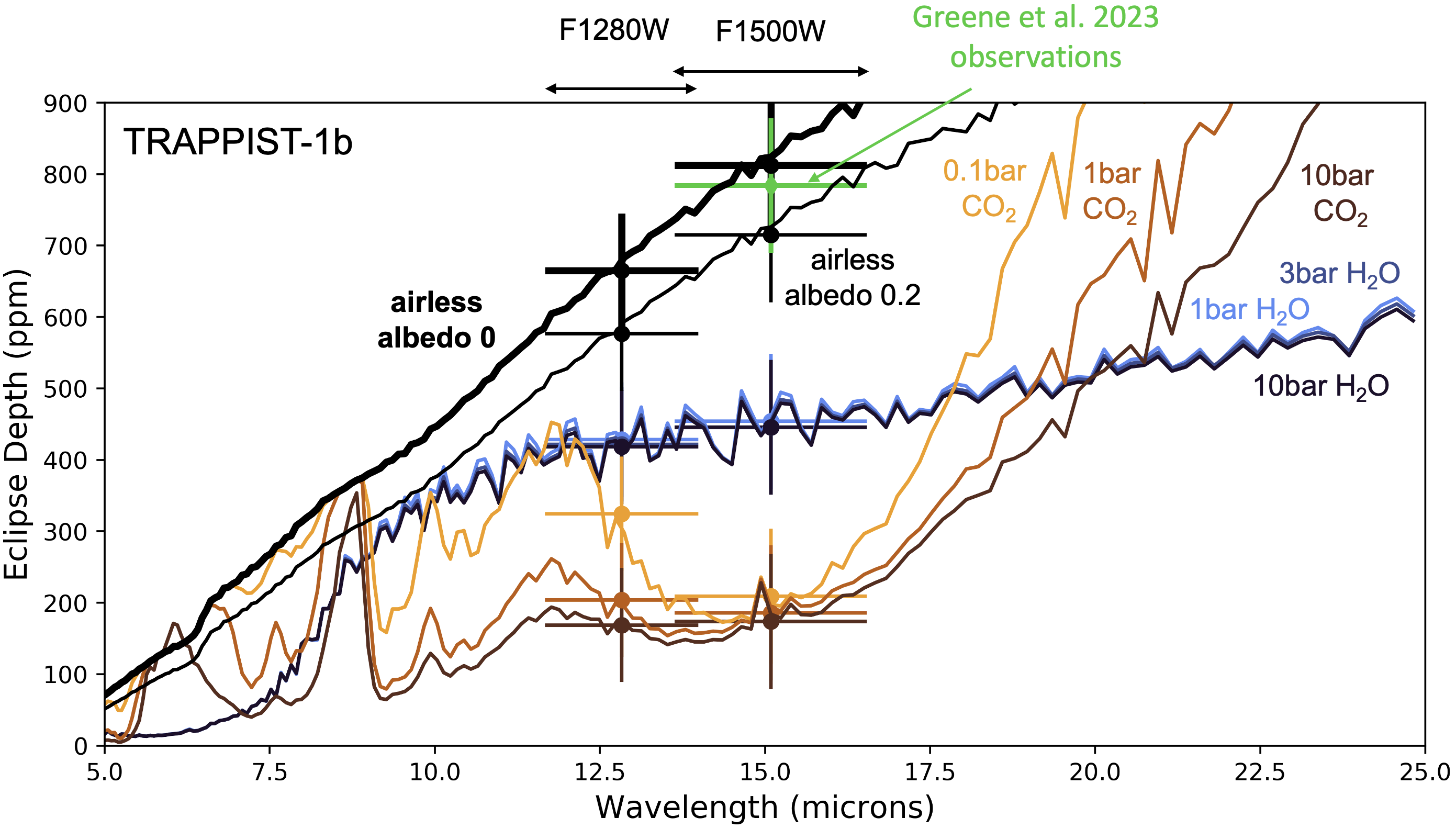}}
\\
\subfloat[TRAPPIST-1c, eclipse depths]{
\label{subfig:eclipse_T1c}
\includegraphics[width=17cm]{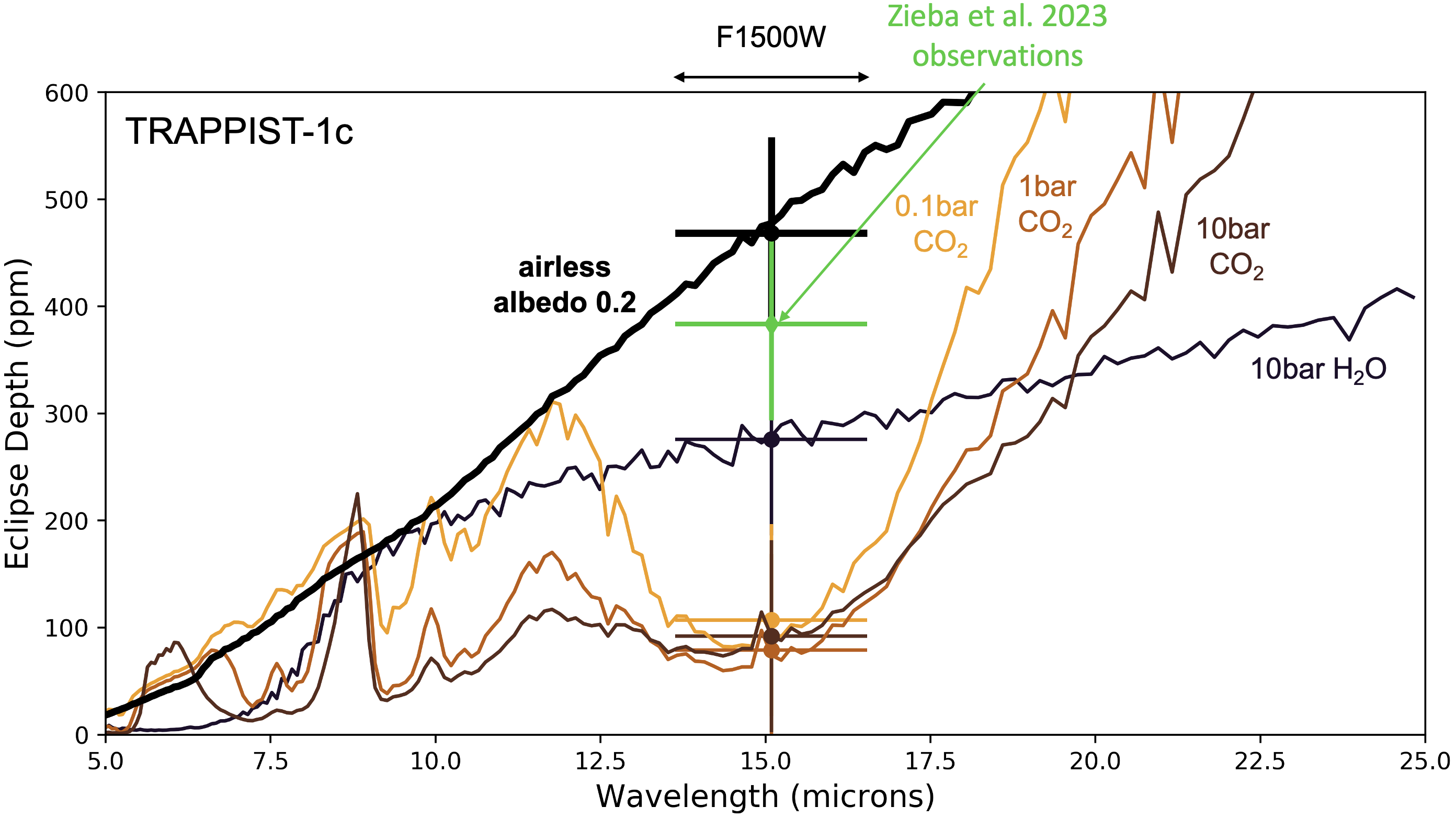}}
\caption{Emission spectra of TRAPPIST-1b (upper panel) and c (lower panel), calculated at the secondary eclipse. The spectra are calculated using GCM simulations of H$_2$O-rich (blue to grey colors) and CO$_2$-rich atmospheres (red to orange colors). Emission spectra are also computed for airless planets (in bold black). 1-$\sigma$ error bars are calculated using the JWST Cycle 1 observation program (more details in Section-\ref{subsubsection_emission_spectra}).
\label{fig:emission_spectra_T1bc}}
\end{figure*}

Using the same GCM simulations, we computed the expected signals in thermal emission of TRAPPIST-1b and c at the secondary eclipse. Fig.~\ref{fig:emission_spectra_T1bc} summarizes our results by showing the emission spectra of H$_2$O-dominated atmospheres compared -- for the context -- with CO$_2$-dominated atmospheres, as well as airless planets. TRAPPIST-1b and c are secondary eclipse targets of the JWST, starting from Cycle 1, with the MIRI instrument. This includes a total of 10 eclipses for TRAPPIST-1b (5 in the F1280W filter in \citealt{Lagage:2017}, 5 in the F1500W filter in \citealt{Greene:2017}) and 4 eclipses for TRAPPIST-1c in the F1500W filter \citep{Kreidberg:2021}. We have added in Fig.~\ref{fig:emission_spectra_T1bc} the first results of this observational campaign in the F1500W filter for TRAPPIST-1b \citep{Greene:2023} and TRAPPIST-1c \citep{Zieba:2023}. We note that in Fig.~\ref{fig:emission_spectra_T1bc} we have artificially reduced the observed secondary eclipses of TRAPPIST-1b and c by 9.8$\%$ to correct for the difference between the observed stellar flux in the MIRI F1500W filter (equal to 2.595~mJy, based on \citealt{Greene:2023} and \citealt{Zieba:2023} which provide an absolute stellar flux in very good agreement) and that predicted by the stellar spectra used for TRAPPIST-1 (using Kurucz 2005 stellar templates). We computed error bars near 15~$\mu$m (F1500W MIRI filter) on the synthetic spectra by using the observed values from \citet{Greene:2023} for TRAPPIST-1b ($\pm$~99~ppm) and from \citet{Zieba:2023} for TRAPPIST-1c ($\pm$~94~ppm), divided by the square root of the ratio between the synthetic Kurucz stellar spectrum of TRAPPIST-1 and the observed stellar flux in the MIRI F1500W filter \citep{Greene:2023,Zieba:2023}. For TRAPPIST-1b, we evaluated the error bars near 12.8~$\mu$m (F1280W MIRI filter) by using the error bars near 15~$\mu$m (F1500W MIRI filter) rescaled by the square root of the ratio of synthetic Kurucz stellar fluxes between F1280W and F1500W filters.

We observe that the eclipse depth is strong for H$_2$O-rich atmospheres (up to 400~ppm and 250~ppm for TRAPPIST-1b and c, respectively, near 15~$\mu$m). We note that the spectra are very similar for the explored atmospheric pressure range (from 1 to 10~bar of H$_2$O, which confirms the results of \citealt{Koll:2019b} for TRAPPIST-1b, calculated between 5-12~$\mu$m). This is because most of the planetary thermal emission comes from atmospheric layers at or above 1~bar \citep{Boukrouche:2021}. In fact, we are not sensitive here to the contribution of the surface thermal emission \citep{Hamano:2015} because the surface is not hot enough to emit in the visible. This result is likely to persist even for water contents well above 10~bar \citep{Selsis:2023}.

By comparison, emission spectra of CO$_2$-dominated atmospheres exhibit stronger variations when changing surface pressure, which confirms the results of \citet{Koll:2019b} (calculated between 5-12~$\mu$m) and \citet{Ih:2023} (for TRAPPIST-1b). This is due to the fact that CO$_2$-dominated atmospheres have many more spectral windows (compared to H$_2$O), where the planetary thermal emission is thus sensitive to the surface temperatures (which depends on the greenhouse effect of CO$_2$, and thus on the total pressure of CO$_2$). It is clearly visible in Fig.~\ref{fig:emission_spectra_T1bc} that CO$_2$ spectra look similar in the CO$_2$ absorption bands (e.g., near 15~$\mu$m), but look different in the CO$_2$ spectral windows.

Our emission spectra reveal that for the two planets, CO$_2$, H$_2$O and airless atmospheres leave distinct signatures, which are detectable by the MIRI instrument onboard JWST (see Table~\ref{table:num_eclipses}). The differences between our simulated cases are particularly strong near 15$\mu$m, where JWST Cycle 1 observations have been obtained for the two planets \citep{Greene:2023,Zieba:2023}. Our GCM-based calculations confirm that thick CO$_2$-dominated atmospheres are unlikely for TRAPPIST-1b and c (rejected at $>$~6~$\sigma$ for TRAPPIST-1b ; rejected at $>$~3~$\sigma$ for TRAPPIST-1c). While a thick H$_2$O-dominated atmosphere is also unlikely for TRAPPIST-1b (rejected at $>$~3~$\sigma$), it is compatible within $\sim$1~$\sigma$ with the F1500W MIRI observations of TRAPPIST-1c. This makes TRAPPIST-1c a promising candidate to test the existence of water vapor and clouds using transit spectroscopy as well as secondary eclipses in other MIRI filters.

\begin{figure*}
\subfloat[TRAPPIST-1b, brightness temperature]{
\label{subfig:Tbright_T1b}
\includegraphics[width=17cm]{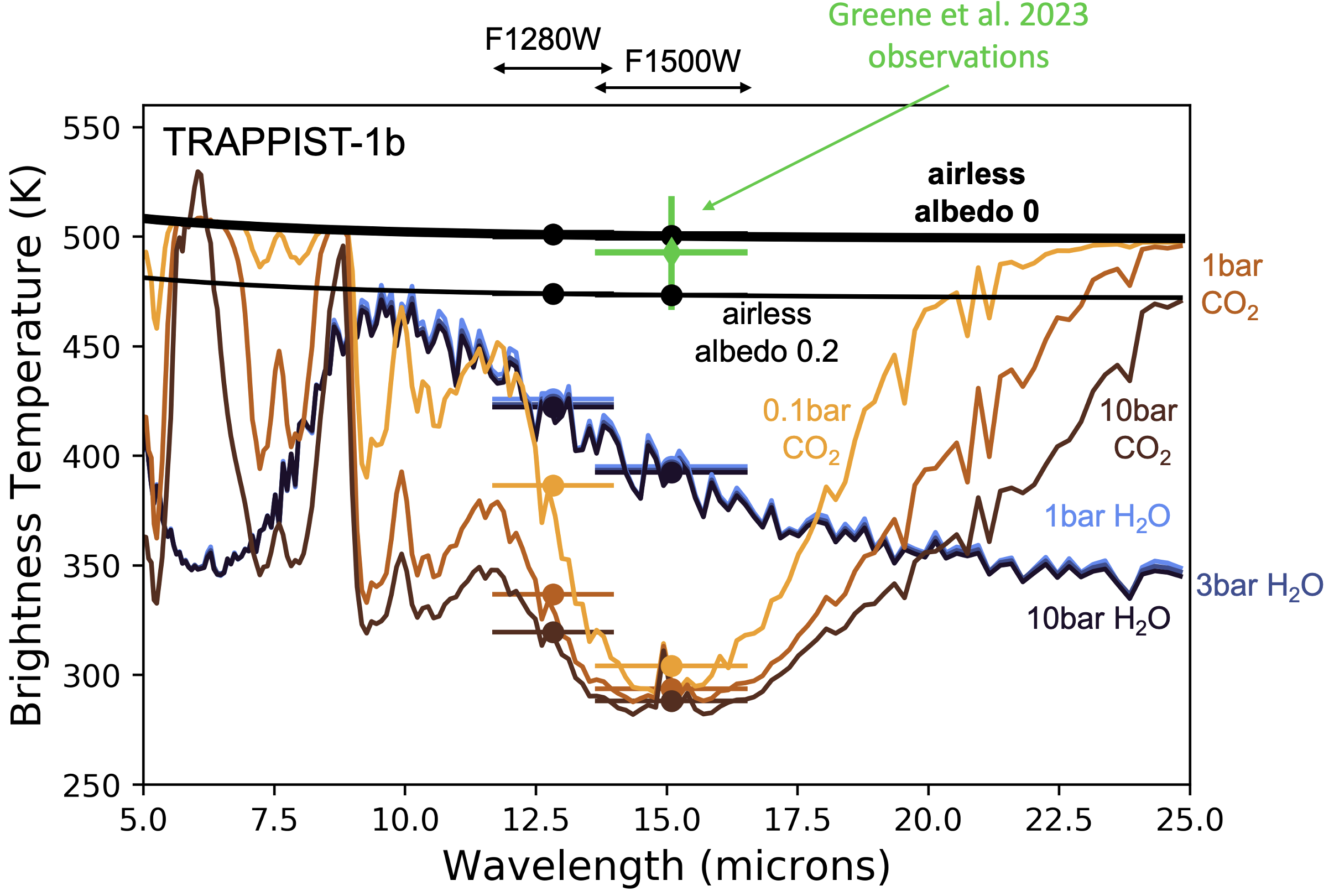}}
\\
\subfloat[TRAPPIST-1c, brightness temperature]{
\label{subfig:Tbright_T1c}
\includegraphics[width=15.7cm]{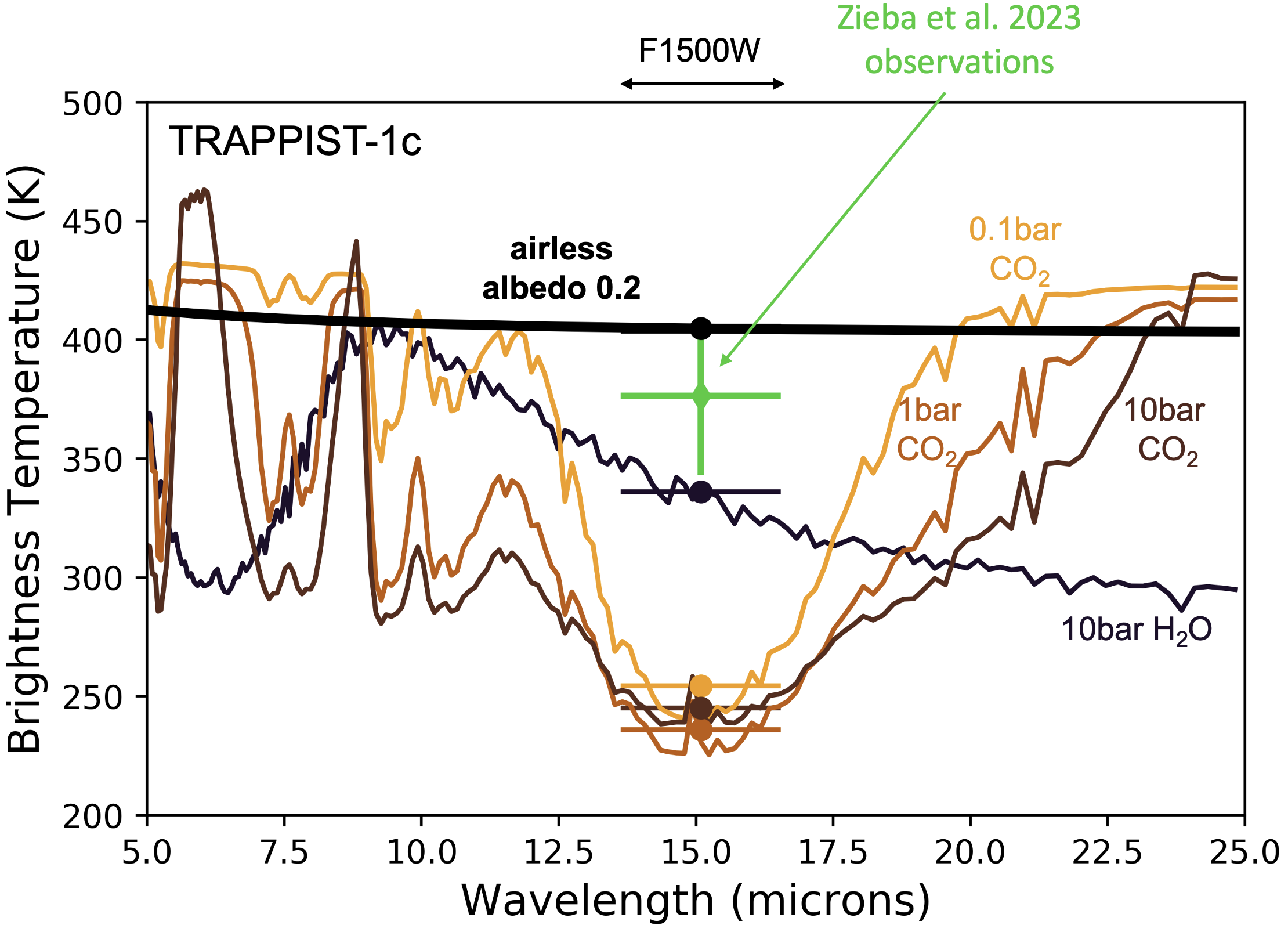}}
\caption{Same as Fig.~\ref{fig:emission_spectra_T1bc}, with the emission converted in brightness temperature units. 
\label{fig:Tbright_T1bc}}
\end{figure*}

To isolate the contribution of the planet (and thus get rid of corrections on the stellar flux), and facilitate comparison with previous works \citep{Lustig-Yaeger:2019}, we have converted the eclipse depths into brightness temperatures. This was done by first (1) computing a grid of secondary eclipse spectra at fixed brightness temperatures (to get the correspondence between eclipse depth and brightness temperature), and then (2) interpolating our thermal emission spectra (based on GCM simulations) using this grid. Fig.~\ref{fig:Tbright_T1bc} summarizes our results for TRAPPIST-1b and c, along with published observations \citep{Greene:2023,Zieba:2023}. We note that we have recalculated the observed brightness temperatures and found values a few degrees lower than in \citet{Greene:2023} and \citet{Zieba:2023} ($493^{+26}_{-26}$~K instead of $503^{+26}_{-27}$~K for TRAPPIST-1b ; $376^{+31}_{-33}$~K instead of $380^{+31}_{-31}$~K for TRAPPIST-1c). In our brightness temperature calculations, we aimed at treating observations and models using the same procedure in order to make comparisons as fair as possible. We thus computed first, for a grid of brightness temperatures, a grid of eclipse depth spectra. We then converted the grid of spectra into a grid of F1500W MIRI eclipse depths by integrating the fluxes in the F1500W MIRI filter. Finally, we interpolated in this grid using the measured F1500W eclipse depth values (of \citealt{Greene:2023} for the TRAPPIST-1b grid ; of \citet{Zieba:2023} for the TRAPPIST-1c grid) to derive the observed brightness temperatures. For CO$_2$-dominated atmospheres (at least the 10~bar simulation), we find results that are qualitatively similar to \citet{Lustig-Yaeger:2019}. Most differences are likely due to different atmospheric compositions used (e.g., \citealt{Lustig-Yaeger:2019} have lower brightness temperature near 8.5~$\mu$m likely due to SO$_2$ absorption included), which makes it difficult to compare other effects (3D vs 1D for example). We note that although the measurements at 15~$\mu$m are incompatible for TRAPPIST-1b with almost all our models of CO$_2$ and H$_2$O atmospheres, this does not demonstrate the absence of an atmosphere that could have a composition very different from that explored in the present manuscript.

\subsubsection{Phase Curves}
\label{subsubsection_phase_curves}

\begin{figure*}
\centering 
\subfloat[TRAPPIST-1b, thermal phase curves]{
\label{subfig:PC_T1b}
\includegraphics[width=17cm]{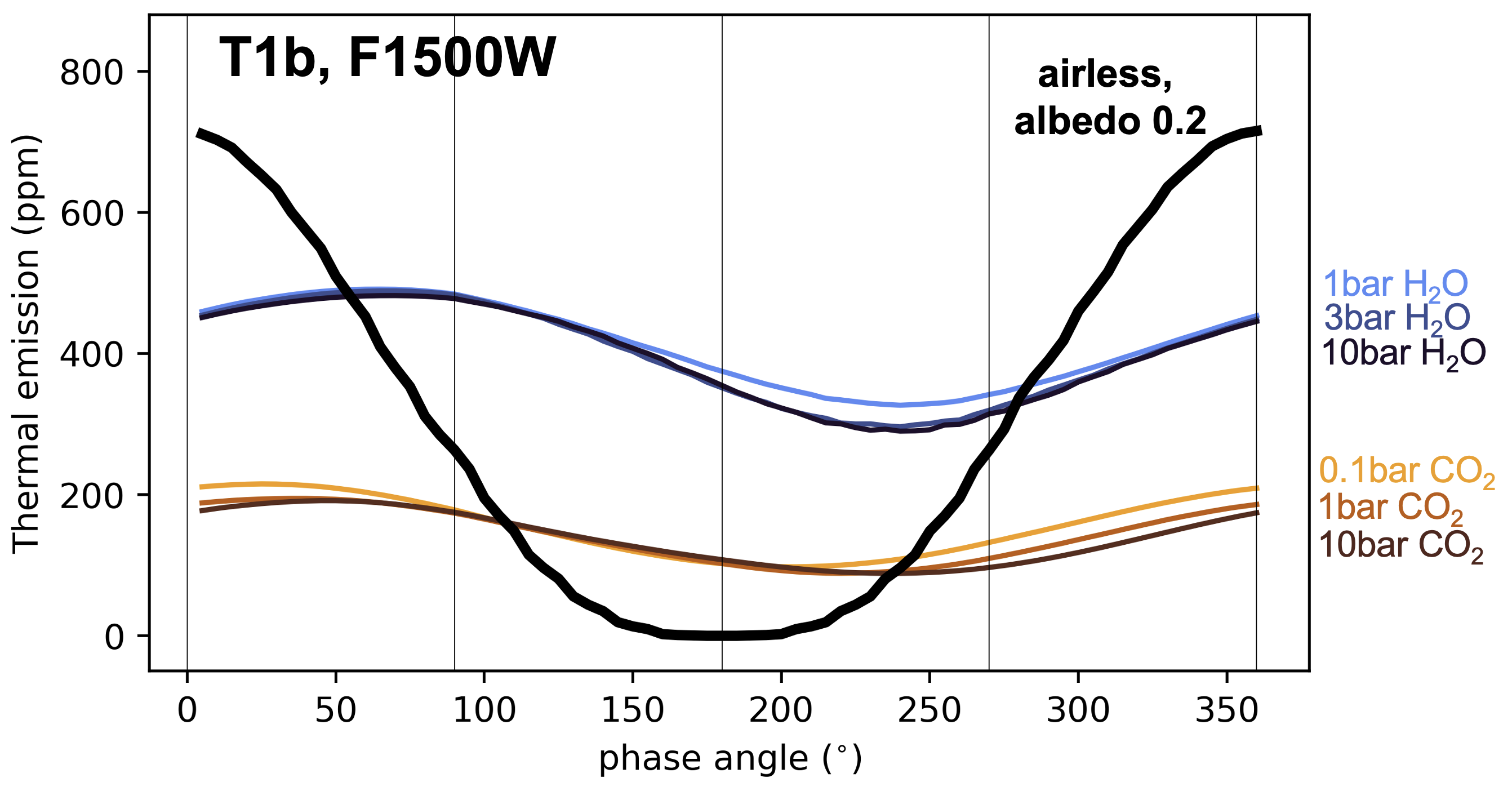}}
\\
\subfloat[TRAPPIST-1c, thermal phase curves]{
\label{subfig:PC_T1c}
\includegraphics[width=17cm]{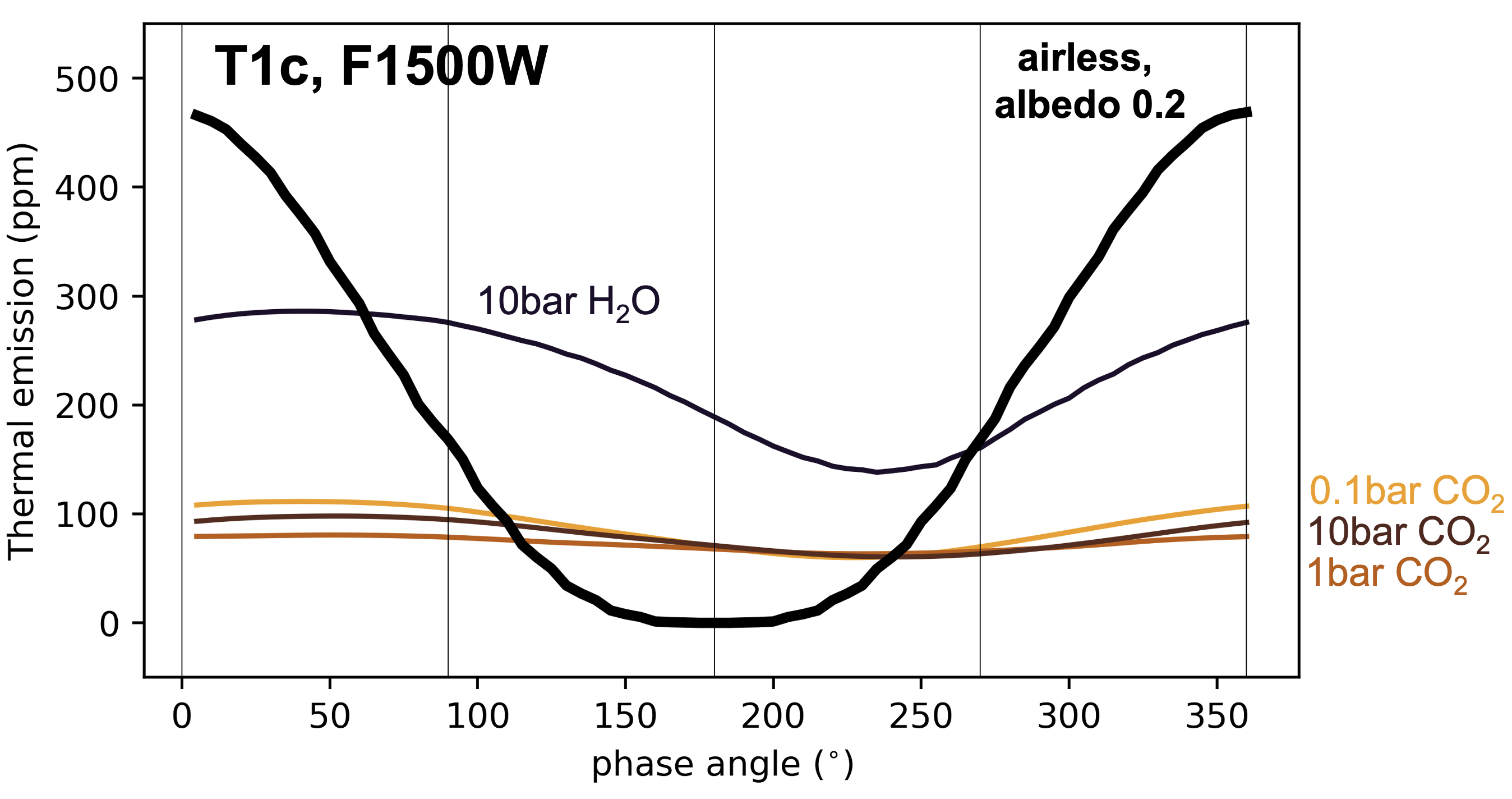}}
\caption{Thermal phase curves of TRAPPIST-1b (upper panel) and c (lower panel), integrated in the F1500W filter (centered around 15$\mu$m), for various atmospheric scenarios: H$_2$O-rich (blue to grey colors), CO$_2$-rich atmospheres (red to orange colors), airless planet (bold black color). H$_2$O-rich atmospheres also include 1~bar of N$_2$.
\label{fig:phase_curves_T1bc}}
\end{figure*}

Last but not least, using again the same GCM simulations, we computed their expected signals in thermal phase curves. Such observations are planned for the Cycle 2 of JWST (Program 3077, 55~hours of observations ; i.e. about 1 orbit of TRAPPIST-1c or 1.5 orbit of TRAPPIST-1b) with the MIRI instrument using the F1500W filter. This observation mode is a promising avenue to characterize the atmospheres of TRAPPIST-1b and c. Fig.~\ref{fig:phase_curves_T1bc} summarizes our results by showing the thermal emission phase curves of H$_2$O-dominated atmospheres compared -- for the context -- with CO$_2$-dominated atmospheres, as well as airless planets. For this figure, we have integrated the thermal emission on the F1500W filter (centered around 15~$\mu$m) as it is the selected observing mode for the phase curve measurements to be obtained within JWST Cycle 2.

There are significant differences between the different types of atmospheres. The phase curves of H$_2$O-dominated atmospheres have relatively low day-to-night amplitudes. For TRAPPIST-1b, c and d, there is a factor of 1.7 to 2 (in the F1500W filter) between dayside and nightside emissions. We interpret this behaviour, which is similar to what is observed in \citet{Wolf:2019} for the incipient runaway case, as follows. At first order, the signal of the phase curve can be decomposed into two parts, (1) a continuous signal (which persists at all phase angles) corresponding to the thermal emission of the Simpson-Nakajima limit \citep{Nakajima:1992,Goldblatt:2012} ; (2) a second source of emission localized on the dayside, which is produced by the re-emission of the ISR absorbed high in the atmosphere \citep{Wolf:2019}. In fact, the first component is lower than the Simpson-Nakajima limit due to part of the emission being absorbed by nightside clouds. This is well visible by looking at the OTR maps in Fig.~\ref{fig_yang_P60days_2Rearth_cloud_alb_otr}, \ref{fig_map_CLOUD_ALB_OTR_stars} and \ref{fig_map_TRAPPIST-1bcd} (bottom rows). The presence of nightside water clouds reduces in fact the planet's thermal emission on the nightside, leaving a visible signature on the phase curve. The effect is qualitatively comparable to that of silicate clouds on the phase curves of hot giant exoplanets \citep{Parmentier:2021}. Clouds are more prominent and thus have a stronger impact on the phase curves for planets receiving an insolation close to the water condensation limit.

Again, we note that the phase curves look very similar for all the surface pressures explored (from 1 to 10~bar of H$_2$O). Interestingly, we observe a significant hotspot phase shift in the H$_2$O-rich simulations (which is less pronounced in the CO$_2$ phase curves). We interpret this hotspot shift as a consequence of the stellar flux being absorbed high in the atmosphere (combined with the effect of the mean molecular weight ; see \citealt{Zhang:2017}). This is because (1) the shortwave heating combined with rotation rate produce strong westerly winds transporting heat from the day to the nightside. This is well illustrated in Fig.~\ref{fig_TL_plots_yang2013} that shows the strong horizontal winds in the stratosphere. This is also because (2) the thermal emission comes from low pressure atmospheric layers, where horizontal winds are strong. Interestingly, CO$_2$-dominated cases have a much less pronounced hotspot offset (near 15~$\mu$m, which is the strongest CO$_2$ absorption band), mostly due to the fact that atmospheric winds are much weaker due to less shortwave flux being absorbed high in the atmosphere (where most of the OTR is emitted at 15~$\mu$m). The amplitude of the phase curve and hotspot offset depend of course on the wavelength considered. This is well illustrated in the Figs.~\ref{fig:phase_curves_T1bB}, \ref{fig:phase_curves_T1cB}, \ref{fig:phase_curves_T1dB}.

We note that the simulated emission spectra of TRAPPIST-1b, c and d are provided at different phase angles (including at secondary eclipse, discussed in the previous Section) in Fig. \ref{fig:phase_curves_T1bA} to \ref{fig:phase_curves_T1dB}).

\section{Discussions}
\label{section_discussions}

\subsection{Comparisons with other GCMs}
\label{comparison_GCM}

3D simulations in \citet{Turbet2021} and in this study were performed with the same GCM: The Generic PCM. This raises the question of whether nighside cloud formation in H$_2$O-dominated atmospheres is also obtained in other 3D climate models. The multi-model validation \citep{Fauchez:2020,Fauchez:2021} is indeed a crucial step to validate this cloud effect, which has potentially strong impacts on the habitability of planets \citep{Turbet2021,Kasting2021}. The best illustration of this multi-model approach is the demonstration that substellar convective clouds appear in all models that have simulated a synchronously rotating aquaplanet, giving great confidence in this cloud feedback (see GCM intercomparisons in \citealt{Yang:2019a,Sergeev:2022}).

We thus surveyed the literature to look for evidence (or absence of evidence) of a similar cloud feedback regarding the formation of nightside water clouds in other models. First of all, and to the best of our knowledge and except in \citet{Turbet2021}, there are no published results of 3D GCM simulations of post-runaway atmospheres (i.e. water-dominated atmospheres with an exhausted surface water reservoir). \citet{Boukrouche:2022} (PhD manuscript) recently performed cloud-free GCM simulations of post-runaway water-dominated atmospheres and confirmed -- using the ExoFMS-SOCRATES GCM -- the preferential formation of clouds on the nightside, based on distance to saturation. Apart from this work, the best analogues we have can be found in numerical studies that simulated the climate of planets in incipient runaway greenhouse \citep{Kopparapu:2017,Wolf:2017,Wolf:2019,Kane:2022,Chaverot:2023}. Using the ExoCAM GCM, \citet{Kopparapu:2017} first studied the climate and clouds of tidally-locked rocky exoplanets transitioning into the runaway greenhouse. They found that as soon as a planet reaches the runaway greenhouse transition, its substellar convective cloud layer dissipates, drastically reducing the albedo and accelerating the runaway greenhouse. The same effect is also seen with ExoCAM in \citet{Wolf:2017} with GCM simulations of TRAPPIST-1d. A snapshot of cloud distribution on TRAPPIST-1d as it evolves into the runaway greenhouse transition (see \citealt{Wolf:2017}, Fig.~4) not only reveal the absence of clouds on the dayside, but also a preferential accumulation of clouds on the nightside, with two local maxima near the wind gyre structures, similar to our findings. Similar cloud distributions are obtained again with ExoCAM (see \citealt{Wolf:2019}, Figs.~4 and 5), for a broader range of host star types (from T$_{\text{eff}}$~=~2600~K to 4500~K). Using ROCKE-3D GCM, \citet{Kane:2018} also performed GCM simulations of the exoplanet Kepler-1649b in incipient runaway greenhouse. The runaway greenhouse produces a sharp decrease of the bond albedo. We analyzed the netCDF simulation file, provided in \citet{Kane:2022}, and that reveals that water clouds are also preferentially located on the nightside in the upper part of the atmosphere.

In summary, our literature review reveals that the preferential formation of nightside clouds is also obtained in other GCMs (ExoCAM, ROCKE-3D, ExoFMS) as soon as they reach conditions where water vapor becomes a dominant gas in the atmosphere (in the so-called incipient runaway greenhouse). This cloud feedback of course deserves to be studied and validated in the context of post-runaway water-dominated atmospheres with other GCMs (including the effects of water clouds).

\subsection{Detecting the cloud feedback}

Our simulations of synthetic spectra (see Sections \ref{subsubsection_transit_spectra}, \ref{subsubsection_emission_spectra} and \ref{subsubsection_phase_curves}) reveal that JWST has the capability of probing water-rich atmospheres around TRAPPIST-1b, c and d either in transit spectra or in emission. Observations are particularly favorable for TRAPPIST-1b and c, because the ISR received on the planets is high enough (1) to suppress cloud at the terminator and thus strongly increase the amplitude of water bands in transit spectra ; and (2) to increase the temperatures, and thus the thermal emission of the planets. Measurements at 15~$\mu$m (MIRI F1500W filter) show however that a thick H$_2$O-dominated atmosphere is unlikely for TRAPPIST-1b (rejected at $>$~3~$\sigma$) leaving TRAPPIST-1c as the best candidate so far (compatible at $\sim$1~$\sigma$). For TRAPPIST-1d, the observations are more difficult not only because the planet is smaller and cooler, but also because water clouds blur the transit spectra. However, by accumulating enough transits (see Table~\ref{table:num_transits}), transit spectra could reveal the presence of water vapour. Together, comparative observations of TRAPPIST-1 inner planets (planet c versus planet d) could be used to reveal the presence of nightside clouds, in case their composition turn out to be dominated by water vapor.

Although we used the TRAPPIST-1 planets as targets to compute observables, more easily accessible targets such as water-rich mini-Neptunes \citep{Luque:2022} could be used to detect this cloud effect, in particular by comparing observations of planets at different values of ISR. We identified indeed a threshold ISR, around 2$\times$ the insolation on Earth (the exact value depends on many factors that deserve to be explored in more details), above which water clouds become significantly depleted at the terminator, significantly increasing the amplitude of absorption features in transit spectra. Based on the simulations we did, we also identified another threshold ISR, around 5$\times$ the insolation on Earth, above which water clouds entirely disappear. Although these ISR thresholds may vary from one planet to another, depending on its amount of water vapor, its size, its type of star, etc., we looked for planets in this ISR range with the best observability metrics, using the ESM (Emission Spectroscopy Metrics) and TSM (Transmission Spectroscopy Metrics) of \citet{Kempton:2018}. In addition to the three inner planets of the TRAPPIST-1 (which are the planets with the best TSM and amonst the best ESM), we have identified LP 791-18c \citep{Crossfield:2019,Peterson:2023}, TOI-270d \citep{Gunther:2019} and LTT 3780c \citep{Cloutier:2020}. These three planets have radii between 2.1 and 2.4$\times$ that of the Earth and their density is compatible with the presence of both water and hydrogen envelopes. If these planets happen to be enriched in water, then they could serve as a natural laboratory for testing the behavior of water clouds, using transit spectra, secondary eclipses and phase curves. Previous simulations of K2-18b with the Generic PCM \citep{Charnay:2021} show indeed qualitatively similar cloud behaviour for temperate mini-Neptune with atmospheric composition of very high metallicity ($>$300). At lower metallicity, clouds can form on the dayside essentially due to two factors \citep{Charnay:2015a,Charnay:2015b}: (1) less water vapour induces less radiative heating on the dayside and less radiative cooling on the nightside ; (2) the specific heat capacity decreases due to higher mixing ratios of hydrogen and helium, which further reduces the day-to-night temperature contrast.

Detecting the water cloud feedback -- using TRAPPIST-1 planets or any population of planets endowed with water-rich atmospheres -- would have profound consequences for our understanding of the habitability of exoplanets, but also of that of the solar system planets (Earth vs Venus).

\subsection{Caveats}

In this study, we explored the ability of exoplanets to form their first oceans by simulating the condensation of the primordial water reservoir at the end of the magma ocean phase \citep{Hamano:2013,Lichtenberg:2023}. 

In \citet{Turbet2021} and this new study, we have performed several sensitivity studies, including: varying the amount of CCN \citep{Turbet2021} ; varying the amount of water (this work ; \citealt{Turbet2021}), of carbon dioxide \citep{Turbet2021}. However, to build confidence in these simulations, more sensitivity studies are needed. Firstly, one can question the choice of parameterizations used in this work. Previous works \citep{Yang:2019a,Sergeev:2020,Sergeev:2022} have shown, for example, that the choice of moist convection scheme has strong consequences on the properties of the substellar cloud layer predicted in the ‘cold start’ simulations. Cloud formation in our ‘hot start’ Generic PCM simulations is largely dominated by large-scale condensation (that is, water vapour condensation driven by large-scale air movements) and not by subgrid-scale condensation (driven by small convective cells, and parameterized using our moist convection scheme). This indicates that our results should be weakly sensitive to the choice of the subgrid moist convection parameterization, which is a strong indication that preferential nightside cloud formation is likely to be a robust mechanism across GCMs (see Section~\ref{comparison_GCM}). However, this deserves to be tested in detail with other GCMs  as well as with cloud-resolving simulations \citep{Tan:2021}. Secondly, one can question the choice of atmospheric compositions used in this work. It has been shown that various atmospheric compositions could be expected at the end of the magma ocean phase, depending on the redox state \citep{Bower:2022}. Modeling a wider range of atmospheric compositions, more representative of all the possible post-magma scenarios, is therefore a necessary step to validate our results.

Moreover, even if the primordial water reservoir never condensed, it is possible that late volcanic degassing, or volatile delivery by cometary impacts, could bring water contributing to form surface oceans \citep{Moore:2020}. This would probably require the planet to have lost its atmosphere at some point -- temporarily transforming the planet into a so-called land planet \citep{Kodama:2015} -- otherwise the delivered water would probably remain vaporized. This scenario also deserves to be explored in details, at least in part with the help of dedicated 3D numerical climate simulations.

\section{Conclusions}
\label{section_conclusions}

In this study, we used the 3D Generic Planetary Climate Model to perform a large grid of simulations for H$_2$O-dominated atmospheres planets orbiting around a wide variety of Main-Sequence stars. The simulations were designed to reproduce the conditions of early ocean formation on rocky planets due to the condensation of the primordial water reservoir at the end of the magma ocean phase \citep{Hamano:2013,Lichtenberg:2023}. 

So far, all Global Climate Model (GCM) simulations -- assuming the entire water reservoir is initially fully condensed on the surface -- have shown that low-altitude dayside convective clouds stabilize oceans even at high incoming stellar radiation (ISR) \citep{Yang:2013,Yang:2014b,Kopparapu:2016,Kopparapu:2017,Way:2016,Yang:2019a,Yang:2019b,Sergeev:2020,Sergeev:2022}. In contrast, our new GCM simulations -- assuming the entire water reservoir is initially fully vaporized in the atmosphere --
show that high-altitude nightside stratospheric clouds prevent surface water condensation from happening even at low ISR. This behaviour was already observed by \citet{Turbet2021} in the case of early Earth and Venus, and we thus extend it here to synchronously-rotating planets around all types of Main-Sequence stars (M, K and G types). More generally, this confirms and extends the results of \citet{Turbet2021} that the formation of nightside clouds seem to occur independently of the rotation rate assumed. Future work could test up to what rotation rate this result holds, given that there should be a threshold rotation rate above which Coriolis forces inhibit day-night stratospheric transport. This is particularly relevant for young planets, which are likely to have a much faster rotation rate than that determined by synchronous rotation. We propose to name this family of planets "mochi planets", with the planetary surface represented by red bean paste, and the cloud layer represented by the sticky rice in which someone has taken a bite. Yummy!

Based on these results, we introduced a Water Condensation Zone (WCZ), whose inner edge is located at lower ISR than the inner edge of the Habitable Zone (HZ). In other words, we found that the ISR required to form oceans by condensation is always significantly lower than that required to vaporize oceans. The contrast is particularly striking with all the recent 3D GCM results that assume most water is initially on the planetary surface \citep{Yang:2013,Yang:2014b,Kopparapu:2016,Kopparapu:2017,Way:2016}. Interestingly, the inner edge of the WCZ lies quite close (but still at lower ISR, due to the greenhouse effect of nightside clouds) to the -- widely used -- inner edge of the HZ of \citet{Kopparapu:2013}, based on 1D cloud-free calculations. This is mainly because (1) water vapor profiles are similar between 1D and 3D climate models, and (2) the absence of dayside clouds makes the atmosphere behave as if it were cloud-free from the point of view of shortwave heating.

We confirm with 3D GCM simulations the results of \citet{Selsis:2023} that in thick H$_2$O-dominated atmospheres, the vertical thermal profile diverges from that calculated using 1D inverse radiative-convective models \citep{Kopparapu:2013,Turbet2019aa,Boukrouche:2021}. This has important consequences for the lifetime of the magma ocean period, as well as the calculations of mass-radius relationships for irradiated planets \citep{Turbet2020aa}.

Last but not least, we show -- taking the inner planets of the TRAPPIST-1 system as an example -- that the JWST has the capability to probe water-rich atmospheres. We identify a threshold insolation, around 2 times the insolation on Earth (the exact value depends on many factors), above which water clouds disappear from the terminator, significantly increasing the amplitude of absorption features in transit spectra. For the TRAPPIST-1 system, we predict the transition occurs between TRAPPIST-1c and d. JWST observations using transit spectroscopy, but also secondary eclipses and thermal phase curves, can be used to probe first the existence of water-rich atmospheres on TRAPPIST-1 inner planets, and eventually test the existence of the nightside cloud feedback. Although we used the TRAPPIST-1 planets as a testbed, other accessible targets such as water-rich mini-Neptunes \citep{Luque:2022} could be used to detect this cloud effect. This would have profound consequences for our understanding of the habitability of exoplanets, but also of that of the solar system planets (Earth vs Venus).

\section*{Acknowledgements}
This project has received funding from the European Union’s Horizon 2020 research and innovation program under the Marie Sklodowska-Curie Grant Agreement No. 832738/ESCAPE. This work has been carried out within the framework of the National Centre of Competence in Research PlanetS supported by the Swiss National Science Foundation. M.T. acknowledges the financial support of the SNSF. M.T. thanks the Gruber Foundation for its generous support to this research. MT acknowledges support from the Tremplin 2022 program of the Faculty of Science and Engineering of Sorbonne University. MG, MT and FS acknowledge support from the BELSPO program BRAIN-be 2.0 (Belgian Research Action through Interdisciplinary Networks) contract B2/212/B1/PORTAL.. This work was performed using the High-Performance Computing (HPC) resources of Centre Informatique National de l'Enseignement Supérieur (CINES) under the allocations No. A0060110391, A0080110391, A0100110391, A0120110391 made by Grand Équipement National de Calcul Intensif (GENCI). A total of $\sim$~1.3M CPU hours were used for this project on the OCCIGEN supercomputer, resulting in $\sim$~2.3~t eq. of CO$_2$ emissions. TJF acknowledges the support of the NASA GSFC Sellers Exoplanet Environments Collaboration (SEEC), which is funded by the NASA Planetary Science Divisions Internal Scientist Funding Model. JL and FS acknowledge funding from the European Research Council (ERC) under the European Union’s Horizon 2020 research and innovation programme (grant agreement No. 679030/WHIPLASH), and from the french state: CNES, Programme National de Planétologie (PNP), the ANR (ANR-20-CE49-0009: SOUND), and in the framework of the Investments for the Future programme IdEx, Université de Bordeaux/RRI ORIGINS. The authors thank the Generic PCM team for the teamwork development and improvement of the model. This research has made use of NASA's Astrophysics Data System. M.T. thanks the LMD Planeto team as well as Nadège Lagarde (LAB) and Xavier Delfosse (IPAG) for useful discussions. We thank the anonymous reviewer for the very insightful comments and suggestions, which improved the quality of the manuscript. The Generic-PCM (and documentation on how to use the model) can be downloaded from our SVN repository \url{https://svn.lmd.jussieu.fr/Planeto/trunk/LMDZ.GENERIC/}. More information and documentation on the model are available on \url{http://www-planets.lmd.jussieu.fr}. The GCM outputs of the TRAPPIST-1 planet simulations are provided here: \url{https://zenodo.org/record/5627945}. Other data underlying this article can be shared upon reasonable request to the corresponding author.

\bibliographystyle{aa}
\bibliography{Thesis}

\begin{thebibliography}{159}
\expandafter\ifx\csname natexlab\endcsname\relax\def\natexlab#1{#1}\fi

\bibitem[{{Abe}(1997)}]{Abe:1997}
{Abe}, Y. 1997, Physics of the Earth and Planetary Interiors, 100, 27

\bibitem[{{Abe} \& {Matsui}(1988)}]{Abe:1988}
{Abe}, Y. \& {Matsui}, T. 1988, Journal of Atmospheric Sciences, 45, 3081

\bibitem[{{Agol} {et~al.}(2021){Agol}, {Dorn}, {Grimm}, {Turbet}, {Ducrot},
  {Delrez}, {Gillon}, {Demory}, {Burdanov}, {Barkaoui}, {Benkhaldoun},
  {Bolmont}, {Burgasser}, {Carey}, {de Wit}, {Fabrycky}, {Foreman-Mackey},
  {Haldemann}, {Hernandez}, {Ingalls}, {Jehin}, {Langford}, {Leconte},
  {Lederer}, {Luger}, {Malhotra}, {Meadows}, {Morris}, {Pozuelos}, {Queloz},
  {Raymond}, {Selsis}, {Sestovic}, {Triaud}, \& {Van Grootel}}]{Agol2021}
{Agol}, E., {Dorn}, C., {Grimm}, S.~L., {et~al.} 2021, The Planetary Science
  Journal, 2, 1

\bibitem[{{Angelo} {et~al.}(2017){Angelo}, {Rowe}, {Howell}, {Quintana},
  {Still}, {Mann}, {Burningham}, {Barclay}, {Ciardi}, {Huber}, \&
  {Kane}}]{Angelo:2017}
{Angelo}, I., {Rowe}, J.~F., {Howell}, S.~B., {et~al.} 2017, \aj, 153, 162

\bibitem[{{Anglada-Escud{\'e}} {et~al.}(2016){Anglada-Escud{\'e}}, {Amado},
  {Barnes}, {Berdi{\~n}as}, {Butler}, {Coleman}, {de La Cueva}, {Dreizler},
  {Endl}, {Giesers}, {Jeffers}, {Jenkins}, {Jones}, {Kiraga}, {K{\"u}rster},
  {L{\'o}pez-Gonz{\'a}lez}, {Marvin}, {Morales}, {Morin}, {Nelson}, {Ortiz},
  {Ofir}, {Paardekooper}, {Reiners}, {Rodr{\'{\i}}guez},
  {Rodr{\'{\i}}guez-L{\'o}pez}, {Sarmiento}, {Strachan}, {Tsapras}, {Tuomi}, \&
  {Zechmeister}}]{Anglada:2016}
{Anglada-Escud{\'e}}, G., {Amado}, P.~J., {Barnes}, J., {et~al.} 2016, Nature,
  536, 437

\bibitem[{{Anglada-Escud{\'e}} {et~al.}(2013){Anglada-Escud{\'e}}, {Tuomi},
  {Gerlach}, {Barnes}, {Heller}, {Jenkins}, {Wende}, {Vogt}, {Butler},
  {Reiners}, \& {Jones}}]{Anglada-Escude:2013}
{Anglada-Escud{\'e}}, G., {Tuomi}, M., {Gerlach}, E., {et~al.} 2013, \aap, 556,
  A126

\bibitem[{{Astudillo-Defru} {et~al.}(2017){Astudillo-Defru}, {Forveille},
  {Bonfils}, {S{\'e}gransan}, {Bouchy}, {Delfosse}, {Lovis}, {Mayor}, {Murgas},
  {Pepe}, {Santos}, {Udry}, \& {W{\"u}nsche}}]{Astudillo-Defru:2017}
{Astudillo-Defru}, N., {Forveille}, T., {Bonfils}, X., {et~al.} 2017, \aap,
  602, A88

\bibitem[{{Baraffe} {et~al.}(2015){Baraffe}, {Homeier}, {Allard}, \&
  {Chabrier}}]{Baraffe2015}
{Baraffe}, I., {Homeier}, D., {Allard}, F., \& {Chabrier}, G. 2015, Astronomy
  $\&$ Astrophysics, 577, A42

\bibitem[{{Bolmont} {et~al.}(2016){Bolmont}, {Libert}, {Leconte}, \&
  {Selsis}}]{Bolmont:2016aa}
{Bolmont}, E., {Libert}, A.-S., {Leconte}, J., \& {Selsis}, F. 2016, Astronomy
  $\&$ Astrophysics, 591, A106

\bibitem[{{Bolmont} {et~al.}(2017){Bolmont}, {Selsis}, {Owen}, {Ribas},
  {Raymond}, {Leconte}, \& {Gillon}}]{Bolmont:2017}
{Bolmont}, E., {Selsis}, F., {Owen}, J.~E., {et~al.} 2017, Monthly Notices of
  the Royal Astronomical Society, 464, 3728

\bibitem[{{Bonfils} {et~al.}(2018{\natexlab{a}}){Bonfils}, {Almenara},
  {Cloutier}, {W{\"u}nsche}, {Astudillo-Defru}, {Berta-Thompson}, {Bouchy},
  {Charbonneau}, {Delfosse}, {D{\'\i}az}, {Dittmann}, {Doyon}, {Forveille},
  {Irwin}, {Lovis}, {Mayor}, {Menou}, {Murgas}, {Newton}, {Pepe}, {Santos}, \&
  {Udry}}]{Bonfils:2018b}
{Bonfils}, X., {Almenara}, J.~M., {Cloutier}, R., {et~al.} 2018{\natexlab{a}},
  \aap, 618, A142

\bibitem[{{Bonfils} {et~al.}(2018{\natexlab{b}}){Bonfils}, {Astudillo-Defru},
  {D{\'\i}az}, {Almenara}, {Forveille}, {Bouchy}, {Delfosse}, {Lovis}, {Mayor},
  {Murgas}, {Pepe}, {Santos}, {S{\'e}gransan}, {Udry}, \&
  {W{\"u}nsche}}]{Bonfils:2018}
{Bonfils}, X., {Astudillo-Defru}, N., {D{\'\i}az}, R., {et~al.}
  2018{\natexlab{b}}, \aap, 613, A25

\bibitem[{{Bonfils} {et~al.}(2013){Bonfils}, {Delfosse}, {Udry}, {Forveille},
  {Mayor}, {Perrier}, {Bouchy}, {Gillon}, {Lovis}, {Pepe}, {Queloz}, {Santos},
  {Segransan}, \& {Bertaux}}]{Bonfils:2013}
{Bonfils}, X., {Delfosse}, X., {Udry}, S., {et~al.} 2013, Astronomy $\&$
  Astrophysics, 549, A109

\bibitem[{{Borucki} {et~al.}(2011){Borucki}, {Koch}, {Basri}, {Batalha},
  {Brown}, {Bryson}, {Caldwell}, {Christensen-Dalsgaard}, {Cochran}, {DeVore},
  {Dunham}, {Gautier}, {Geary}, {Gilliland}, {Gould}, {Howell}, {Jenkins},
  {Latham}, {Lissauer}, {Marcy}, {Rowe}, {Sasselov}, {Boss}, {Charbonneau},
  {Ciardi}, {Doyle}, {Dupree}, {Ford}, {Fortney}, {Holman}, {Seager},
  {Steffen}, {Tarter}, {Welsh}, {Allen}, {Buchhave}, {Christiansen}, {Clarke},
  {Das}, {D{\'e}sert}, {Endl}, {Fabrycky}, {Fressin}, {Haas}, {Horch},
  {Howard}, {Isaacson}, {Kjeldsen}, {Kolodziejczak}, {Kulesa}, {Li}, {Lucas},
  {Machalek}, {McCarthy}, {MacQueen}, {Meibom}, {Miquel}, {Prsa}, {Quinn},
  {Quintana}, {Ragozzine}, {Sherry}, {Shporer}, {Tenenbaum}, {Torres},
  {Twicken}, {Van Cleve}, {Walkowicz}, {Witteborn}, \& {Still}}]{Borucki:2011}
{Borucki}, W.~J., {Koch}, D.~G., {Basri}, G., {et~al.} 2011, The Astrophysical
  Journal, 736, 19

\bibitem[{{Boucher} {et~al.}(1995){Boucher}, {Le Treut}, \&
  {Baker}}]{Boucher:1995}
{Boucher}, O., {Le Treut}, H., \& {Baker}, M.~B. 1995, Journal of Geophysical
  Research, 100, 16

\bibitem[{Boukrouche(2022)}]{Boukrouche:2022}
Boukrouche, R. 2022, PhD thesis, {University of Oxford, Hertford College}

\bibitem[{{Boukrouche} {et~al.}(2021){Boukrouche}, {Lichtenberg}, \&
  {Pierrehumbert}}]{Boukrouche:2021}
{Boukrouche}, R., {Lichtenberg}, T., \& {Pierrehumbert}, R.~T. 2021, The
  Astrophysical Journal, 919, 130

\bibitem[{{Boutle} {et~al.}(2020){Boutle}, {Joshi}, {Lambert}, {Mayne},
  {Lyster}, {Manners}, {Ridgway}, \& {Kohary}}]{Boutle:2020}
{Boutle}, I.~A., {Joshi}, M., {Lambert}, F.~H., {et~al.} 2020, Nature
  Communications, 11, 2731

\bibitem[{{Boutle} {et~al.}(2017){Boutle}, {Mayne}, {Drummond}, {Manners},
  {Goyal}, {Hugo Lambert}, {Acreman}, \& {Earnshaw}}]{Boutle:2017}
{Boutle}, I.~A., {Mayne}, N.~J., {Drummond}, B., {et~al.} 2017, Astronomy $\&$
  Astrophysics, 601, A120

\bibitem[{{Bower} {et~al.}(2022){Bower}, {Hakim}, {Sossi}, \&
  {Sanan}}]{Bower:2022}
{Bower}, D.~J., {Hakim}, K., {Sossi}, P.~A., \& {Sanan}, P. 2022, The Planetary
  Science Journal, 3, 93

\bibitem[{{Carone} {et~al.}(2014){Carone}, {Keppens}, \& {Decin}}]{Carone:2014}
{Carone}, L., {Keppens}, R., \& {Decin}, L. 2014, Monthly Notices of the Royal
  Astronomical Society, 445, 930

\bibitem[{{Carone} {et~al.}(2015){Carone}, {Keppens}, \& {Decin}}]{Carone:2015}
{Carone}, L., {Keppens}, R., \& {Decin}, L. 2015, Monthly Notices of the Royal
  Astronomical Society, 453, 2412

\bibitem[{{Carone} {et~al.}(2016){Carone}, {Keppens}, \& {Decin}}]{Carone:2016}
{Carone}, L., {Keppens}, R., \& {Decin}, L. 2016, Monthly Notices of the Royal
  Astronomical Society, 461, 1981

\bibitem[{{Charnay} {et~al.}(2021){Charnay}, {Blain}, {B{\'e}zard}, {Leconte},
  {Turbet}, \& {Falco}}]{Charnay:2021}
{Charnay}, B., {Blain}, D., {B{\'e}zard}, B., {et~al.} 2021, Astronomy $\&$
  Astrophysics, 646, A171

\bibitem[{{Charnay} {et~al.}(2015{\natexlab{a}}){Charnay}, {Meadows}, \&
  {Leconte}}]{Charnay:2015a}
{Charnay}, B., {Meadows}, V., \& {Leconte}, J. 2015{\natexlab{a}}, The
  Astrophysical Journal, 813, 15

\bibitem[{{Charnay} {et~al.}(2015{\natexlab{b}}){Charnay}, {Meadows}, {Misra},
  {Leconte}, \& {Arney}}]{Charnay:2015b}
{Charnay}, B., {Meadows}, V., {Misra}, A., {Leconte}, J., \& {Arney}, G.
  2015{\natexlab{b}}, The Astrophysical Journal Letters, 813, L1

\bibitem[{{Chaverot} {et~al.}(2023){Chaverot}, {Bolmont}, \&
  {Turbet}}]{Chaverot:2023}
{Chaverot}, G., {Bolmont}, E., \& {Turbet}, M. 2023, to be submitted to
  Astronomy \& Astrophysics

\bibitem[{{Chaverot} {et~al.}(2022){Chaverot}, {Turbet}, {Bolmont}, \&
  {Leconte}}]{Chaverot:2022}
{Chaverot}, G., {Turbet}, M., {Bolmont}, E., \& {Leconte}, J. 2022, Astronomy
  $\&$ Astrophysics, 658, A40

\bibitem[{{Cloutier} {et~al.}(2020){Cloutier}, {Eastman}, {Rodriguez},
  {Astudillo-Defru}, {Bonfils}, {Mortier}, {Watson}, {Stalport}, {Pinamonti},
  {Lienhard}, {Harutyunyan}, {Damasso}, {Latham}, {Collins}, {Massey}, {Irwin},
  {Winters}, {Charbonneau}, {Ziegler}, {Matthews}, {Crossfield}, {Kreidberg},
  {Quinn}, {Ricker}, {Vanderspek}, {Seager}, {Winn}, {Jenkins}, {Vezie},
  {Udry}, {Twicken}, {Tenenbaum}, {Sozzetti}, {S{\'e}gransan}, {Schlieder},
  {Sasselov}, {Santos}, {Rice}, {Rackham}, {Poretti}, {Piotto}, {Phillips},
  {Pepe}, {Molinari}, {Mignon}, {Micela}, {Melo}, {de Medeiros}, {Mayor},
  {Matson}, {Martinez Fiorenzano}, {Mann}, {Magazz{\'u}}, {Lovis},
  {L{\'o}pez-Morales}, {Lopez}, {Lissauer}, {L{\'e}pine}, {Law}, {Kielkopf},
  {Johnson}, {Jensen}, {Howell}, {Gonzales}, {Ghedina}, {Forveille},
  {Figueira}, {Dumusque}, {Dressing}, {Doyon}, {D{\'\i}az}, {Fabrizio},
  {Delfosse}, {Cosentino}, {Conti}, {Collins}, {Cameron}, {Ciardi}, {Caldwell},
  {Burke}, {Buchhave}, {Brice{\~n}o}, {Boyd}, {Bouchy}, {Beichman}, {Artigau},
  \& {Almenara}}]{Cloutier:2020}
{Cloutier}, R., {Eastman}, J.~D., {Rodriguez}, J.~E., {et~al.} 2020, \aj, 160,
  3

\bibitem[{{Colose} {et~al.}(2021){Colose}, {Haqq-Misra}, {Wolf}, {Del Genio},
  {Barnes}, {Way}, \& {Ruedy}}]{Colose:2021}
{Colose}, C.~M., {Haqq-Misra}, J., {Wolf}, E.~T., {et~al.} 2021, The
  Astrophysical Journal, 921, 25

\bibitem[{{Crossfield} {et~al.}(2015){Crossfield}, {Petigura}, {Schlieder},
  {Howard}, {Fulton}, {Aller}, {Ciardi}, {L{\'e}pine}, {Barclay}, {de Pater},
  {de Kleer}, {Quintana}, {Christiansen}, {Schlafly}, {Kaltenegger}, {Crepp},
  {Henning}, {Obermeier}, {Deacon}, {Weiss}, {Isaacson}, {Hansen}, {Liu},
  {Greene}, {Howell}, {Barman}, \& {Mordasini}}]{Crossfield:2015}
{Crossfield}, I. J.~M., {Petigura}, E., {Schlieder}, J.~E., {et~al.} 2015, The
  Astrophysical Journal, 804, 10

\bibitem[{{Crossfield} {et~al.}(2019){Crossfield}, {Waalkes}, {Newton},
  {Narita}, {Muirhead}, {Ment}, {Matthews}, {Kraus}, {Kostov}, {Kosiarek},
  {Kane}, {Isaacson}, {Halverson}, {Gonzales}, {Everett}, {Dragomir},
  {Collins}, {Chontos}, {Berardo}, {Winters}, {Winn}, {Scott}, {Rojas-Ayala},
  {Rizzuto}, {Petigura}, {Peterson}, {Mocnik}, {Mikal-Evans}, {Mehrle},
  {Matson}, {Kuzuhara}, {Irwin}, {Huber}, {Huang}, {Howell}, {Howard},
  {Hirano}, {Fulton}, {Dupuy}, {Dressing}, {Dalba}, {Charbonneau}, {Burt},
  {Berta-Thompson}, {Benneke}, {Watanabe}, {Twicken}, {Tamura}, {Schlieder},
  {Seager}, {Rose}, {Ricker}, {Quintana}, {L{\'e}pine}, {Latham}, {Kotani},
  {Jenkins}, {Hori}, {Colon}, \& {Caldwell}}]{Crossfield:2019}
{Crossfield}, I. J.~M., {Waalkes}, W., {Newton}, E.~R., {et~al.} 2019, \apjl,
  883, L16

\bibitem[{{Del Genio} {et~al.}(2019){Del Genio}, {Way}, {Amundsen}, {Aleinov},
  {Kelley}, {Kiang}, \& {Clune}}]{Delgenio2019}
{Del Genio}, A.~D., {Way}, M.~J., {Amundsen}, D.~S., {et~al.} 2019,
  Astrobiology, 19, 99

\bibitem[{{Delrez} {et~al.}(2022){Delrez}, {Murray}, {Pozuelos}, {Narita},
  {Ducrot}, {Timmermans}, {Watanabe}, {Burgasser}, {Hirano}, {Rackham},
  {Stassun}, {Van Grootel}, {Aganze}, {Cointepas}, {Howell}, {Kaltenegger},
  {Niraula}, {Sebastian}, {Almenara}, {Barkaoui}, {Baycroft}, {Bonfils},
  {Bouchy}, {Burdanov}, {Caldwell}, {Charbonneau}, {Ciardi}, {Collins},
  {Daylan}, {Demory}, {de Wit}, {Dransfield}, {Fajardo-Acosta}, {Fausnaugh},
  {Fukui}, {Furlan}, {Garcia}, {Gnilka}, {G{\'o}mez Maqueo Chew},
  {G{\'o}mez-Mu{\~n}oz}, {G{\"u}nther}, {Harakawa}, {Heng}, {Hooton}, {Hori},
  {Ikoma}, {Jehin}, {Jenkins}, {Kagetani}, {Kawauchi}, {Kimura}, {Kodama},
  {Kotani}, {Krishnamurthy}, {Kudo}, {Kunovac}, {Kusakabe}, {Latham},
  {Littlefield}, {McCormac}, {Melis}, {Mori}, {Murgas}, {Palle}, {Pedersen},
  {Queloz}, {Ricker}, {Sabin}, {Schanche}, {Schroffenegger}, {Seager}, {Shiao},
  {Sohy}, {Standing}, {Tamura}, {Theissen}, {Thompson}, {Triaud}, {Vanderspek},
  {Vievard}, {Wells}, {Winn}, {Zou}, {Z{\'u}{\~n}iga-Fern{\'a}ndez}, \&
  {Gillon}}]{Delrez:2022}
{Delrez}, L., {Murray}, C.~A., {Pozuelos}, F.~J., {et~al.} 2022, Astronomy \&
  Astrophysics, 667, A59

\bibitem[{{Demory} {et~al.}(2020){Demory}, {Pozuelos}, {G{\'o}mez Maqueo Chew},
  {Sabin}, {Petrucci}, {Schroffenegger}, {Grimm}, {Sestovic}, {Gillon},
  {McCormac}, {Barkaoui}, {Benz}, {Bieryla}, {Bouchy}, {Burdanov}, {Collins},
  {de Wit}, {Dressing}, {Garcia}, {Giacalone}, {Guerra}, {Haldemann}, {Heng},
  {Jehin}, {Jofr{\'e}}, {Kane}, {Lillo-Box}, {Maign{\'e}}, {Mordasini},
  {Morris}, {Niraula}, {Queloz}, {Rackham}, {Savel}, {Soubkiou}, {Srdoc},
  {Stassun}, {Triaud}, {Zambelli}, {Ricker}, {Latham}, {Seager}, {Winn},
  {Jenkins}, {Calvario-Vel{\'a}squez}, {Franco Herrera}, {Colorado}, {Cadena
  Zepeda}, {Figueroa}, {Watson}, {Lugo-Ibarra}, {Carigi}, {Guisa}, {Herrera},
  {Sierra D{\'\i}az}, {Su{\'a}rez}, {Barrado}, {Batalha}, {Benkhaldoun},
  {Chontos}, {Dai}, {Essack}, {Ghachoui}, {Huang}, {Huber}, {Isaacson},
  {Lissauer}, {Morales-Calder{\'o}n}, {Robertson}, {Roy}, {Twicken},
  {Vanderburg}, \& {Weiss}}]{Demory:2020}
{Demory}, B.~O., {Pozuelos}, F.~J., {G{\'o}mez Maqueo Chew}, Y., {et~al.} 2020,
  \aap, 642, A49

\bibitem[{{Ding} \& {Wordsworth}(2020)}]{Ding:2020}
{Ding}, F. \& {Wordsworth}, R.~D. 2020, The Astrophysical Journal Letters, 891,
  L18

\bibitem[{{Dransfield} {et~al.}(2023){Dransfield}, {Timmermans}, {Triaud},
  {D{\'e}vora-Pajares}, {Aganze}, {Barkaoui}, {Burgasser}, {Collins},
  {Cointepas}, {Ducrot}, {G{\"u}nther}, {Howell}, {Murray}, {Niraula},
  {Rackham}, {Sebastian}, {Stassun}, {Z{\'u}{\~n}iga-Fern{\'a}ndez},
  {Almenara}, {Bonfils}, {Bouchy}, {Burke}, {Charbonneau}, {Christiansen},
  {Delrez}, {Gan}, {Garc{\'\i}a}, {Gillon}, {Chew}, {Hesse}, {Hooton}, {Isopi},
  {Jehin}, {Jenkins}, {Latham}, {Mallia}, {Murgas}, {Pedersen}, {Pozuelos},
  {Queloz}, {Rodriguez}, {Schanche}, {Seager}, {Srdoc}, {Stockdale}, {Twicken},
  {Vanderspek}, {Wells}, {Winn}, {de Wit}, \& {Zapparata}}]{Dransfield:2023}
{Dransfield}, G., {Timmermans}, M., {Triaud}, A. H.~M.~J., {et~al.} 2023,
  \mnras [\eprint[arXiv]{2305.06206}]

\bibitem[{{Dreizler} {et~al.}(2020){Dreizler}, {Jeffers}, {Rodr{\'\i}guez},
  {Zechmeister}, {Barnes}, {Haswell}, {Coleman}, {Lalitha}, {Hidalgo Soto},
  {Strachan}, {Hambsch}, {L{\'o}pez-Gonz{\'a}lez}, {Morales}, {Rodr{\'\i}guez
  L{\'o}pez}, {Berdi{\~n}as}, {Ribas}, {Pall{\'e}}, {Reiners}, \&
  {Anglada-Escud{\'e}}}]{Dreizler:2020}
{Dreizler}, S., {Jeffers}, S.~V., {Rodr{\'\i}guez}, E., {et~al.} 2020, \mnras,
  493, 536

\bibitem[{{Dressing} {et~al.}(2017){Dressing}, {Vanderburg}, {Schlieder},
  {Crossfield}, {Knutson}, {Newton}, {Ciardi}, {Fulton}, {Gonzales}, {Howard},
  {Isaacson}, {Livingston}, {Petigura}, {Sinukoff}, {Everett}, {Horch}, \&
  {Howell}}]{Dressing:2017}
{Dressing}, C.~D., {Vanderburg}, A., {Schlieder}, J.~E., {et~al.} 2017, The
  Astronomical Journal, 154, 207

\bibitem[{{Ducrot} {et~al.}(2020){Ducrot}, {Gillon}, {Delrez}, {Agol},
  {Rimmer}, {Turbet}, {G{\"u}nther}, {Demory}, {Triaud}, {Bolmont},
  {Burgasser}, {Carey}, {Ingalls}, {Jehin}, {Leconte}, {Lederer}, {Queloz},
  {Raymond}, {Selsis}, {Van Grootel}, \& {de Wit}}]{Ducrot:2020}
{Ducrot}, E., {Gillon}, M., {Delrez}, L., {et~al.} 2020, Astronomy $\&$
  Astrophysics, 640, A112

\bibitem[{{Faria} {et~al.}(2022){Faria}, {Su{\'a}rez Mascare{\~n}o},
  {Figueira}, {Silva}, {Damasso}, {Demangeon}, {Pepe}, {Santos}, {Rebolo},
  {Cristiani}, {Adibekyan}, {Alibert}, {Allart}, {Barros}, {Cabral},
  {D'Odorico}, {Di Marcantonio}, {Dumusque}, {Ehrenreich}, {Gonz{\'a}lez
  Hern{\'a}ndez}, {Hara}, {Lillo-Box}, {Lo Curto}, {Lovis}, {Martins},
  {M{\'e}gevand}, {Mehner}, {Micela}, {Molaro}, {Nunes}, {Pall{\'e}},
  {Poretti}, {Sousa}, {Sozzetti}, {Tabernero}, {Udry}, \& {Zapatero
  Osorio}}]{Faria2022}
{Faria}, J.~P., {Su{\'a}rez Mascare{\~n}o}, A., {Figueira}, P., {et~al.} 2022,
  Astronomy \& Astrophysics, 658, A115

\bibitem[{{Fauchez} {et~al.}(2021){Fauchez}, {Turbet}, {Sergeev}, {Mayne},
  {Spiga}, {Sohl}, {Saxena}, {Deitrick}, {Gilli}, {Domagal-Goldman}, {Forget},
  {Consentino}, {Barnes}, {Haqq-Misra}, {Way}, {Wolf}, {Olson}, {Crouse},
  {Janin}, {Bolmont}, {Leconte}, {Chaverot}, {Jaziri}, {Tsigaridis}, {Yang},
  {Pidhorodetska}, {Kopparapu}, {Chen}, {Boutle}, {Lefevre}, {Charnay},
  {Burnett}, {Cabra}, \& {Bouldin}}]{Fauchez:2021}
{Fauchez}, T.~J., {Turbet}, M., {Sergeev}, D.~E., {et~al.} 2021, The Planetary
  Science Journal, 2, 106

\bibitem[{{Fauchez} {et~al.}(2019){Fauchez}, {Turbet}, {Villanueva}, {Wolf},
  {Arney}, {Kopparapu}, {Lincowski}, {Mandell}, {de Wit}, {Pidhorodetska},
  {Domagal-Goldman}, \& {Stevenson}}]{Fauchez:2019}
{Fauchez}, T.~J., {Turbet}, M., {Villanueva}, G.~L., {et~al.} 2019, The
  Astrophysical Journal, 887, 194

\bibitem[{{Fauchez} {et~al.}(2020){Fauchez}, {Turbet}, {Wolf}, {Boutle}, {Way},
  {Del Genio}, {Mayne}, {Tsigaridis}, {Kopparapu}, {Yang}, {Forget}, {Mandell},
  \& {Domagal Goldman}}]{Fauchez:2020}
{Fauchez}, T.~J., {Turbet}, M., {Wolf}, E.~T., {et~al.} 2020, Geoscientific
  Model Development, 13, 707

\bibitem[{{Forget} {et~al.}(2013){Forget}, {Wordsworth}, {Millour},
  {Madeleine}, {Kerber}, {Leconte}, {Marcq}, \& {Haberle}}]{Forget:2013}
{Forget}, F., {Wordsworth}, R., {Millour}, E., {et~al.} 2013, Icarus, 222, 81

\bibitem[{{Fu} \& {Liou}(1992)}]{Fu:1992}
{Fu}, Q. \& {Liou}, K.~N. 1992, Journal of Atmospheric Sciences, 49, 2139

\bibitem[{{Fujii} {et~al.}(2017){Fujii}, {Del Genio}, \&
  {Amundsen}}]{Fujii:2017}
{Fujii}, Y., {Del Genio}, A.~D., \& {Amundsen}, D.~S. 2017, The Astrophysical
  Journal, 848, 100

\bibitem[{{Gilbert} {et~al.}(2020){Gilbert}, {Barclay}, {Schlieder},
  {Quintana}, {Hord}, {Kostov}, {Lopez}, {Rowe}, {Hoffman}, {Walkowicz},
  {Silverstein}, {Rodriguez}, {Vanderburg}, {Suissa}, {Airapetian}, {Clement},
  {Raymond}, {Mann}, {Kruse}, {Lissauer}, {Col{\'o}n}, {Kopparapu},
  {Kreidberg}, {Zieba}, {Collins}, {Quinn}, {Howell}, {Ziegler}, {Vrijmoet},
  {Adams}, {Arney}, {Boyd}, {Brande}, {Burke}, {Cacciapuoti}, {Chance},
  {Christiansen}, {Covone}, {Daylan}, {Dineen}, {Dressing}, {Essack},
  {Fauchez}, {Galgano}, {Howe}, {Kaltenegger}, {Kane}, {Lam}, {Lee}, {Lewis},
  {Logsdon}, {Mandell}, {Monsue}, {Mullally}, {Mullally}, {Paudel},
  {Pidhorodetska}, {Plavchan}, {Reyes}, {Rinehart}, {Rojas-Ayala}, {Smith},
  {Stassun}, {Tenenbaum}, {Vega}, {Villanueva}, {Wolf}, {Youngblood}, {Ricker},
  {Vanderspek}, {Latham}, {Seager}, {Winn}, {Jenkins}, {Bakos}, {Brice{\~n}o},
  {Ciardi}, {Cloutier}, {Conti}, {Couperus}, {Di Sora}, {Eisner}, {Everett},
  {Gan}, {Hartman}, {Henry}, {Isopi}, {Jao}, {Jensen}, {Law}, {Mallia},
  {Matson}, {Shappee}, {Le Wood}, \& {Winters}}]{Gilbert:2020}
{Gilbert}, E.~A., {Barclay}, T., {Schlieder}, J.~E., {et~al.} 2020, The
  Astronomical Journal, 160, 116

\bibitem[{{Gilbert} {et~al.}(2023){Gilbert}, {Vanderburg}, {Rodriguez}, {Hord},
  {Clement}, {Barclay}, {Quintana}, {Schlieder}, {Kane}, {Jenkins}, {Twicken},
  {Kunimoto}, {Vanderspek}, {Arney}, {Charbonneau}, {G{\"u}nther}, {Huang},
  {Isopi}, {Kostov}, {Kristiansen}, {Latham}, {Mallia}, {Mamajek}, {Mireles},
  {Quinn}, {Ricker}, {Schulte}, {Seager}, {Suissa}, {Winn}, {Youngblood}, \&
  {Zapparata}}]{Gilbert:2023}
{Gilbert}, E.~A., {Vanderburg}, A., {Rodriguez}, J.~E., {et~al.} 2023, The
  Astrophysical Journal Letters, 944, L35

\bibitem[{{Gillon} {et~al.}(2017){Gillon}, {Triaud}, {Demory}, {Jehin}, {Agol},
  {Deck}, {Lederer}, {de Wit}, {Burdanov}, {Ingalls}, {Bolmont}, {Leconte},
  {Raymond}, {Selsis}, {Turbet}, {Barkaoui}, {Burgasser}, {Burleigh}, {Carey},
  {Chaushev}, {Copperwheat}, {Delrez}, {Fernandes}, {Holdsworth}, {Kotze}, {Van
  Grootel}, {Almleaky}, {Benkhaldoun}, {Magain}, \& {Queloz}}]{Gillon:2017}
{Gillon}, M., {Triaud}, A.~H.~M.~J., {Demory}, B.-O., {et~al.} 2017, Nature,
  542, 456

\bibitem[{{Godolt} {et~al.}(2015){Godolt}, {Grenfell}, {Hamann-Reinus},
  {Kitzmann}, {Kunze}, {Langematz}, {von Paris}, {Patzer}, {Rauer}, \&
  {Stracke}}]{Godolt:2015}
{Godolt}, M., {Grenfell}, J.~L., {Hamann-Reinus}, A., {et~al.} 2015, Planetary
  and Space Science, 111, 62

\bibitem[{{Goldblatt}(2015)}]{goldblatt:2015}
{Goldblatt}, C. 2015, Astrobiology, 15, 362

\bibitem[{{Goldblatt} {et~al.}(2013){Goldblatt}, {Robinson}, {Zahnle}, \&
  {Crisp}}]{Goldblatt:2013}
{Goldblatt}, C., {Robinson}, T.~D., {Zahnle}, K.~J., \& {Crisp}, D. 2013,
  Nature Geoscience, 6, 661

\bibitem[{{Goldblatt} \& {Watson}(2012)}]{Goldblatt:2012}
{Goldblatt}, C. \& {Watson}, A.~J. 2012, Philosophical Transactions of the
  Royal Society of London Series A, 370, 4197

\bibitem[{{Gordon} {et~al.}(2022){Gordon}, {Rothman}, {Hargreaves}, {Hashemi},
  {Karlovets}, {Skinner}, {Conway}, {Hill}, {Kochanov}, {Tan}, {Wcis{\l}o},
  {Finenko}, {Nelson}, {Bernath}, {Birk}, {Boudon}, {Campargue}, {Chance},
  {Coustenis}, {Drouin}, {Flaud}, {Gamache}, {Hodges}, {Jacquemart}, {Mlawer},
  {Nikitin}, {Perevalov}, {Rotger}, {Tennyson}, {Toon}, {Tran}, {Tyuterev},
  {Adkins}, {Baker}, {Barbe}, {Can{\`e}}, {Cs{\'a}sz{\'a}r}, {Dudaryonok},
  {Egorov}, {Fleisher}, {Fleurbaey}, {Foltynowicz}, {Furtenbacher}, {Harrison},
  {Hartmann}, {Horneman}, {Huang}, {Karman}, {Karns}, {Kassi}, {Kleiner},
  {Kofman}, {Kwabia-Tchana}, {Lavrentieva}, {Lee}, {Long}, {Lukashevskaya},
  {Lyulin}, {Makhnev}, {Matt}, {Massie}, {Melosso}, {Mikhailenko}, {Mondelain},
  {M{\"u}ller}, {Naumenko}, {Perrin}, {Polyansky}, {Raddaoui}, {Raston},
  {Reed}, {Rey}, {Richard}, {T{\'o}bi{\'a}s}, {Sadiek}, {Schwenke},
  {Starikova}, {Sung}, {Tamassia}, {Tashkun}, {Vander Auwera}, {Vasilenko},
  {Vigasin}, {Villanueva}, {Vispoel}, {Wagner}, {Yachmenev}, \&
  {Yurchenko}}]{Gordon:2022}
{Gordon}, I.~E., {Rothman}, L.~S., {Hargreaves}, R.~J., {et~al.} 2022, Journal
  of Quantitative Spectroscopy and Radiative Transfer, 277, 107949

\bibitem[{{Greene} {et~al.}(2023){Greene}, {Bell}, {Ducrot}, {Dyrek}, {Lagage},
  \& {Fortney}}]{Greene:2023}
{Greene}, T.~P., {Bell}, T.~J., {Ducrot}, E., {et~al.} 2023, \nat, 618, 39

\bibitem[{{Greene} {et~al.}(2017){Greene}, {Lagage}, {Rieke}, \&
  {Schlawin}}]{Greene:2017}
{Greene}, T.~P., {Lagage}, P.-O., {Rieke}, M.~J., \& {Schlawin}, E. 2017, {MIRI
  observations of transiting exoplanets}, JWST Proposal. Cycle 1, ID. \#1177

\bibitem[{{Gregory}(1995)}]{Gregory:1995}
{Gregory}, D. 1995, Monthly Weather Review, 123, 2716

\bibitem[{{Grimm} {et~al.}(2018){Grimm}, {Demory}, {Gillon}, {Dorn}, {Agol},
  {Burdanov}, {Delrez}, {Sestovic}, {Triaud}, {Turbet}, {Bolmont}, {Caldas},
  {de Wit}, {Jehin}, {Leconte}, {Raymond}, {Van Grootel}, {Burgasser}, {Carey},
  {Fabrycky}, {Heng}, {Hernandez}, {Ingalls}, {Lederer}, {Selsis}, \&
  {Queloz}}]{Grimm:2018}
{Grimm}, S.~L., {Demory}, B.-O., {Gillon}, M., {et~al.} 2018, \aap, 613, A68

\bibitem[{{G{\"u}nther} {et~al.}(2019){G{\"u}nther}, {Pozuelos}, {Dittmann},
  {Dragomir}, {Kane}, {Daylan}, {Feinstein}, {Huang}, {Morton}, {Bonfanti},
  {Bouma}, {Burt}, {Collins}, {Lissauer}, {Matthews}, {Montet}, {Vanderburg},
  {Wang}, {Winters}, {Ricker}, {Vanderspek}, {Latham}, {Seager}, {Winn},
  {Jenkins}, {Armstrong}, {Barkaoui}, {Batalha}, {Bean}, {Caldwell}, {Ciardi},
  {Collins}, {Crossfield}, {Fausnaugh}, {Furesz}, {Gan}, {Gillon}, {Guerrero},
  {Horne}, {Howell}, {Ireland}, {Isopi}, {Jehin}, {Kielkopf}, {Lepine},
  {Mallia}, {Matson}, {Myers}, {Palle}, {Quinn}, {Relles}, {Rojas-Ayala},
  {Schlieder}, {Sefako}, {Shporer}, {Su{\'a}rez}, {Tan}, {Ting}, {Twicken}, \&
  {Waite}}]{Gunther:2019}
{G{\"u}nther}, M.~N., {Pozuelos}, F.~J., {Dittmann}, J.~A., {et~al.} 2019,
  Nature Astronomy, 3, 1099

\bibitem[{{Hamano} {et~al.}(2013){Hamano}, {Abe}, \& {Genda}}]{Hamano:2013}
{Hamano}, K., {Abe}, Y., \& {Genda}, H. 2013, Nature, 497, 607

\bibitem[{{Hamano} {et~al.}(2015){Hamano}, {Kawahara}, {Abe}, {Onishi}, \&
  {Hashimoto}}]{Hamano:2015}
{Hamano}, K., {Kawahara}, H., {Abe}, Y., {Onishi}, M., \& {Hashimoto}, G.~L.
  2015, The Astrophysical Journal, 806, 216

\bibitem[{{Harakawa} {et~al.}(2022){Harakawa}, {Takarada}, {Kasagi}, {Hirano},
  {Kotani}, {Kuzuhara}, {Omiya}, {Kawahara}, {Fukui}, {Hori}, {Ishikawa},
  {Ogihara}, {Livingston}, {Brandt}, {Currie}, {Aoki}, {Beichman}, {Henning},
  {Hodapp}, {Ishizuka}, {Izumiura}, {Jacobson}, {Janson}, {Kambe}, {Kodama},
  {Kokubo}, {Konishi}, {Krishnamurthy}, {Kudo}, {Kurokawa}, {Kusakabe}, {Kwon},
  {Matsumoto}, {McElwain}, {Mitsui}, {Nakagawa}, {Narita}, {Nishikawa},
  {Nugroho}, {Serabyn}, {Serizawa}, {Takahashi}, {Ueda}, {Uyama}, {Vievard},
  {Wang}, {Wisniewski}, {Tamura}, \& {Sato}}]{Harakawa:2022}
{Harakawa}, H., {Takarada}, T., {Kasagi}, Y., {et~al.} 2022, \pasj, 74, 904

\bibitem[{{Ih} {et~al.}(2023){Ih}, {Kempton}, {Whittaker}, \&
  {Lessard}}]{Ih:2023}
{Ih}, J., {Kempton}, E. M.~R., {Whittaker}, E.~A., \& {Lessard}, M. 2023, The
  Astrophysical Journal Letters, 952, L4

\bibitem[{{Kaltenegger}(2017)}]{Kaltenegger:2017}
{Kaltenegger}, L. 2017, Annual Review of Astronomy and Astrophysics, 55, 433

\bibitem[{{Kane}(2022)}]{Kane:2022}
{Kane}, S.~R. 2022, Nature Astronomy, 6, 420

\bibitem[{{Kane} {et~al.}(2018){Kane}, {Ceja}, {Way}, \&
  {Quintana}}]{Kane:2018}
{Kane}, S.~R., {Ceja}, A.~Y., {Way}, M.~J., \& {Quintana}, E.~V. 2018, The
  Astrophysical Journal, 869, 46

\bibitem[{{Karman} {et~al.}(2019){Karman}, {Gordon}, {van der Avoird},
  {Baranov}, {Boulet}, {Drouin}, {Groenenboom}, {Gustafsson}, {Hartmann},
  {Kurucz}, {Rothman}, {Sun}, {Sung}, {Thalman}, {Tran}, {Wishnow},
  {Wordsworth}, {Vigasin}, {Volkamer}, \& {van der Zande}}]{Karman:2019}
{Karman}, T., {Gordon}, I.~E., {van der Avoird}, A., {et~al.} 2019, Icarus,
  328, 160

\bibitem[{{Kasting} \& {Harman}(2021)}]{Kasting2021}
{Kasting}, J.~F. \& {Harman}, C.~E. 2021, Nature, 598, 259

\bibitem[{{Kasting} {et~al.}(1993){Kasting}, {Whitmire}, \&
  {Reynolds}}]{Kasting:1993}
{Kasting}, J.~F., {Whitmire}, D.~P., \& {Reynolds}, R.~T. 1993, Icarus, 101,
  108

\bibitem[{{Kempton} {et~al.}(2018){Kempton}, {Bean}, {Louie}, {Deming}, {Koll},
  {Mansfield}, {Christiansen}, {L{\'o}pez-Morales}, {Swain}, {Zellem},
  {Ballard}, {Barclay}, {Barstow}, {Batalha}, {Beatty}, {Berta-Thompson},
  {Birkby}, {Buchhave}, {Charbonneau}, {Cowan}, {Crossfield}, {de Val-Borro},
  {Doyon}, {Dragomir}, {Gaidos}, {Heng}, {Hu}, {Kane}, {Kreidberg}, {Mallonn},
  {Morley}, {Narita}, {Nascimbeni}, {Pall{\'e}}, {Quintana}, {Rauscher},
  {Seager}, {Shkolnik}, {Sing}, {Sozzetti}, {Stassun}, {Valenti}, \& {von
  Essen}}]{Kempton:2018}
{Kempton}, E. M.~R., {Bean}, J.~L., {Louie}, D.~R., {et~al.} 2018, \pasp, 130,
  114401

\bibitem[{{Kodama} {et~al.}(2015){Kodama}, {Genda}, {Abe}, \&
  {Zahnle}}]{Kodama:2015}
{Kodama}, T., {Genda}, H., {Abe}, Y., \& {Zahnle}, K.~J. 2015, The
  Astrophysical Journal, 812, 165

\bibitem[{{Kodama} {et~al.}(2019){Kodama}, {Genda}, {O'ishi}, {Abe-Ouchi}, \&
  {Abe}}]{Kodama:2019}
{Kodama}, T., {Genda}, H., {O'ishi}, R., {Abe-Ouchi}, A., \& {Abe}, Y. 2019,
  Journal of Geophysical Research (Planets), 124, 2306

\bibitem[{{Koll}(2020)}]{Koll:2020}
{Koll}, D. 2020, {TLC: Tidally Locked Coordinates}, Astrophysics Source Code
  Library, record ascl:2011.013

\bibitem[{{Koll} \& {Abbot}(2015)}]{Koll:2015}
{Koll}, D. D.~B. \& {Abbot}, D.~S. 2015, The Astrophysical Journal, 802, 21

\bibitem[{{Koll} \& {Cronin}(2019)}]{Koll:2019}
{Koll}, D. D.~B. \& {Cronin}, T.~W. 2019, The Astrophysical Journal, 881, 120

\bibitem[{{Koll} {et~al.}(2019){Koll}, {Malik}, {Mansfield}, {Kempton}, {Kite},
  {Abbot}, \& {Bean}}]{Koll:2019b}
{Koll}, D. D.~B., {Malik}, M., {Mansfield}, M., {et~al.} 2019, The
  Astrophysical Journal, 886, 140

\bibitem[{{Kopparapu} {et~al.}(2013){Kopparapu}, {Ramirez}, {Kasting}, {Eymet},
  {Robinson}, {Mahadevan}, {Terrien}, {Domagal-Goldman}, {Meadows}, \&
  {Deshpande}}]{Kopparapu:2013}
{Kopparapu}, R.~K., {Ramirez}, R., {Kasting}, J.~F., {et~al.} 2013, The
  Astrophysical Journal, 765, 131

\bibitem[{{Kopparapu} {et~al.}(2014){Kopparapu}, {Ramirez}, {SchottelKotte},
  {Kasting}, {Domagal-Goldman}, \& {Eymet}}]{Kopparapu:2014}
{Kopparapu}, R.~K., {Ramirez}, R.~M., {SchottelKotte}, J., {et~al.} 2014, The
  Astrophysical Journal Letters, 787, L29

\bibitem[{{Kopparapu} {et~al.}(2017){Kopparapu}, {Wolf}, {Arney}, {Batalha},
  {Haqq-Misra}, {Grimm}, \& {Heng}}]{Kopparapu:2017}
{Kopparapu}, R.~K., {Wolf}, E.~T., {Arney}, G., {et~al.} 2017, The
  Astrophysical Journal, 845, 5

\bibitem[{{Kopparapu} {et~al.}(2016){Kopparapu}, {Wolf}, {Haqq-Misra}, {Yang},
  {Kasting}, {Meadows}, {Terrien}, \& {Mahadevan}}]{Kopparapu:2016}
{Kopparapu}, R.~K., {Wolf}, E.~T., {Haqq-Misra}, J., {et~al.} 2016, The
  Astrophysical Journal, 819, 84

\bibitem[{{Kossakowski} {et~al.}(2023){Kossakowski}, {K{\"u}rster}, {Trifonov},
  {Henning}, {Kemmer}, {Caballero}, {Burn}, {Sabotta}, {Crouse}, {Fauchez},
  {Nagel}, {Kaminski}, {Herrero}, {Rodr{\'\i}guez}, {Gonz{\'a}lez-{\'A}lvarez},
  {Quirrenbach}, {Amado}, {Ribas}, {Reiners}, {Aceituno}, {B{\'e}jar},
  {Baroch}, {Bastelberger}, {Chaturvedi}, {Cifuentes}, {Dreizler}, {Jeffers},
  {Kopparapu}, {Lafarga}, {L{\'o}pez-Gonz{\'a}lez}, {Mart{\'\i}n-Ruiz},
  {Montes}, {Morales}, {Pall{\'e}}, {Pavlov}, {Pedraz}, {Perdelwitz},
  {P{\'e}rez-Torres}, {Perger}, {Reffert}, {Rodr{\'\i}guez L{\'o}pez},
  {Schlecker}, {Sch{\"o}fer}, {Schweitzer}, {Shan}, {Shields}, {Stock}, {Wolf},
  {Zapatero Osorio}, \& {Zechmeister}}]{Kossakowski:2023}
{Kossakowski}, D., {K{\"u}rster}, M., {Trifonov}, T., {et~al.} 2023, \aap, 670,
  A84

\bibitem[{{Kostov} {et~al.}(2019){Kostov}, {Schlieder}, {Barclay}, {Quintana},
  {Col{\'o}n}, {Brande}, {Collins}, {Feinstein}, {Hadden}, {Kane}, {Kreidberg},
  {Kruse}, {Lam}, {Matthews}, {Montet}, {Pozuelos}, {Stassun}, {Winters},
  {Ricker}, {Vanderspek}, {Latham}, {Seager}, {Winn}, {Jenkins}, {Afanasev},
  {Armstrong}, {Arney}, {Boyd}, {Barentsen}, {Barkaoui}, {Batalha}, {Beichman},
  {Bayliss}, {Burke}, {Burdanov}, {Cacciapuoti}, {Carson}, {Charbonneau},
  {Christiansen}, {Ciardi}, {Clampin}, {Collins}, {Conti}, {Coughlin},
  {Covone}, {Crossfield}, {Delrez}, {Domagal-Goldman}, {Dressing}, {Ducrot},
  {Essack}, {Everett}, {Fauchez}, {Foreman-Mackey}, {Gan}, {Gilbert}, {Gillon},
  {Gonzales}, {Hamann}, {Hedges}, {Hocutt}, {Hoffman}, {Horch}, {Horne},
  {Howell}, {Hynes}, {Ireland}, {Irwin}, {Isopi}, {Jensen}, {Jehin},
  {Kaltenegger}, {Kielkopf}, {Kopparapu}, {Lewis}, {Lopez}, {Lissauer}, {Mann},
  {Mallia}, {Mandell}, {Matson}, {Mazeh}, {Monsue}, {Moran}, {Moran}, {Morley},
  {Morris}, {Muirhead}, {Mukai}, {Mullally}, {Mullally}, {Murray}, {Narita},
  {Palle}, {Pidhorodetska}, {Quinn}, {Relles}, {Rinehart}, {Ritsko},
  {Rodriguez}, {Rowden}, {Rowe}, {Sebastian}, {Sefako}, {Shahaf}, {Shporer},
  {Ta{\~n}{\'o}n Reyes}, {Tenenbaum}, {Ting}, {Twicken}, {van Belle}, {Vega},
  {Volosin}, {Walkowicz}, \& {Youngblood}}]{Kostov:2019}
{Kostov}, V.~B., {Schlieder}, J.~E., {Barclay}, T., {et~al.} 2019, \aj, 158, 32

\bibitem[{{Kreidberg} {et~al.}(2021){Kreidberg}, {Agol}, {Bolmont}, {Gillon},
  {Hu}, {Koll}, {Mandell}, {Meadows}, {Morley}, {Schaefer}, {Selsis}, \& {de
  Wit}}]{Kreidberg:2021}
{Kreidberg}, L., {Agol}, E., {Bolmont}, E., {et~al.} 2021, {Hot Take on a Cool
  World: Does Trappist-1c Have an Atmosphere?}, JWST Proposal. Cycle 1, ID.
  \#2304

\bibitem[{{Lafreniere}(2017)}]{Lafreniere:2017}
{Lafreniere}, D. 2017, {NIRISS Exploration of the Atmospheric diversity of
  Transiting exoplanets (NEAT)}, JWST Proposal. Cycle 1, ID. \#1201

\bibitem[{{Lagage} \& {Bouwman}(2017)}]{Lagage:2017}
{Lagage}, P.-O. \& {Bouwman}, J. 2017, {Thermal emission from Trappist1-b},
  JWST Proposal. Cycle 1, ID. \#1279

\bibitem[{{Lebrun} {et~al.}(2013){Lebrun}, {Massol}, {Chassefi{\`e}Re},
  {Davaille}, {Marcq}, {Sarda}, {Leblanc}, \& {Brandeis}}]{Lebrun:2013}
{Lebrun}, T., {Massol}, H., {Chassefi{\`e}Re}, E., {et~al.} 2013, Journal of
  Geophysical Research (Planets), 118, 1155

\bibitem[{{Leconte} {et~al.}(2013{\natexlab{a}}){Leconte}, {Forget}, {Charnay},
  {Wordsworth}, \& {Pottier}}]{Leconte:2013nat}
{Leconte}, J., {Forget}, F., {Charnay}, B., {Wordsworth}, R., \& {Pottier}, A.
  2013{\natexlab{a}}, Nature, 504, 268

\bibitem[{{Leconte} {et~al.}(2013{\natexlab{b}}){Leconte}, {Forget}, {Charnay},
  {Wordsworth}, {Selsis}, \& {Millour}}]{Leconte:2013aa}
{Leconte}, J., {Forget}, F., {Charnay}, B., {et~al.} 2013{\natexlab{b}},
  Astronomy.and Astrophysics, in press

\bibitem[{{Lichtenberg} {et~al.}(2023){Lichtenberg}, {Schaefer}, {Nakajima}, \&
  {Fischer}}]{Lichtenberg:2023}
{Lichtenberg}, T., {Schaefer}, L.~K., {Nakajima}, M., \& {Fischer}, R.~A. 2023,
  in Astronomical Society of the Pacific Conference Series, Vol. 534,
  Protostars and Planets VII, ed. S.~{Inutsuka}, Y.~{Aikawa}, T.~{Muto},
  K.~{Tomida}, \& M.~{Tamura}, 907

\bibitem[{{Lim} {et~al.}(2021){Lim}, {Albert}, {Artigau}, {Benneke}, {Cowan},
  {Doyon}, \& {Lafreniere}}]{Lim:2021}
{Lim}, O., {Albert}, L., {Artigau}, E., {et~al.} 2021, {Atmospheric
  reconnaissance of the TRAPPIST-1 planets}, JWST Proposal. Cycle 1, ID. \#2589

\bibitem[{{Lunine}(2017)}]{Lunine:2017}
{Lunine}, J.~I. 2017, Acta Astronautica, 131, 123

\bibitem[{{Luque} \& {Pall{\'e}}(2022)}]{Luque:2022}
{Luque}, R. \& {Pall{\'e}}, E. 2022, Science, 377, 1211

\bibitem[{{Lustig-Yaeger} {et~al.}(2019){Lustig-Yaeger}, {Meadows}, \&
  {Lincowski}}]{Lustig-Yaeger:2019}
{Lustig-Yaeger}, J., {Meadows}, V.~S., \& {Lincowski}, A.~P. 2019, The
  Astronomical Journal, 158, 27

\bibitem[{{Manabe} \& {Wetherald}(1967)}]{Manabe:1967}
{Manabe}, S. \& {Wetherald}, R.~T. 1967, Journal of Atmospheric Sciences, 24,
  241

\bibitem[{{Mann} {et~al.}(2019){Mann}, {Dupuy}, {Kraus}, {Gaidos}, {Ansdell},
  {Ireland}, {Rizzuto}, {Hung}, {Dittmann}, {Factor}, {Feiden}, {Martinez},
  {Ru{\'\i}z-Rodr{\'\i}guez}, \& {Thao}}]{Mann:2019}
{Mann}, A.~W., {Dupuy}, T., {Kraus}, A.~L., {et~al.} 2019, The Astrophysical
  Journal, 871, 63

\bibitem[{{Massie} \& {Hervig}(2013)}]{Massie:2023}
{Massie}, S.~T. \& {Hervig}, M. 2013, Journal of Quantitative Spectroscopy and
  Radiative Transfer, 130, 373

\bibitem[{{Mlawer} {et~al.}(2012){Mlawer}, {Payne}, {Moncet}, {Delamere},
  {Alvarado}, \& {Tobin}}]{Mlawer:2012}
{Mlawer}, E.~J., {Payne}, V.~H., {Moncet}, J.-L., {et~al.} 2012, Philosophical
  Transactions of the Royal Society of London Series A, 370, 2520

\bibitem[{{Moore} \& {Cowan}(2020)}]{Moore:2020}
{Moore}, K. \& {Cowan}, N.~B. 2020, Monthly Notices of the Royal Astronomical
  Society, 496, 3786

\bibitem[{{Morley} {et~al.}(2017){Morley}, {Kreidberg}, {Rustamkulov},
  {Robinson}, \& {Fortney}}]{Morley:2017}
{Morley}, C.~V., {Kreidberg}, L., {Rustamkulov}, Z., {Robinson}, T., \&
  {Fortney}, J.~J. 2017, The Astrophysical Journal, 850, 121

\bibitem[{{Muirhead} {et~al.}(2015){Muirhead}, {Mann}, {Vanderburg}, {Morton},
  {Kraus}, {Ireland}, {Swift}, {Feiden}, {Gaidos}, \& {Gazak}}]{Muirhead:2015}
{Muirhead}, P.~S., {Mann}, A.~W., {Vanderburg}, A., {et~al.} 2015, \apj, 801,
  18

\bibitem[{{Nakajima} {et~al.}(1992){Nakajima}, {Hayashi}, \&
  {Abe}}]{Nakajima:1992}
{Nakajima}, S., {Hayashi}, Y.-Y., \& {Abe}, Y. 1992, Journal of Atmospheric
  Sciences, 49, 2256

\bibitem[{{Parmentier} {et~al.}(2021){Parmentier}, {Showman}, \&
  {Fortney}}]{Parmentier:2021}
{Parmentier}, V., {Showman}, A.~P., \& {Fortney}, J.~J. 2021, \mnras, 501, 78

\bibitem[{{Pepe} {et~al.}(2011){Pepe}, {Lovis}, {S{\'e}gransan}, {Benz},
  {Bouchy}, {Dumusque}, {Mayor}, {Queloz}, {Santos}, \& {Udry}}]{Pepe:2011}
{Pepe}, F., {Lovis}, C., {S{\'e}gransan}, D., {et~al.} 2011, \aap, 534, A58

\bibitem[{{Peterson} {et~al.}(2023){Peterson}, {Benneke}, {Collins}, {Piaulet},
  {Crossfield}, {Ali-Dib}, {Christiansen}, {Gagn{\'e}}, {Faherty}, {Kite},
  {Dressing}, {Charbonneau}, {Murgas}, {Cointepas}, {Almenara}, {Bonfils},
  {Kane}, {Werner}, {Gorjian}, {Roy}, {Shporer}, {Pozuelos}, {Socia},
  {Cloutier}, {Dietrich}, {Irwin}, {Weiss}, {Waalkes}, {Berta-Thomson},
  {Evans}, {Apai}, {Parviainen}, {Pall{\'e}}, {Narita}, {Howard}, {Dragomir},
  {Barkaoui}, {Gillon}, {Jehin}, {Ducrot}, {Benkhaldoun}, {Fukui}, {Mori},
  {Nishiumi}, {Kawauchi}, {Ricker}, {Latham}, {Winn}, {Seager}, {Isaacson},
  {Bixel}, {Gibbs}, {Jenkins}, {Smith}, {Chavez}, {Rackham}, {Henning},
  {Gabor}, {Chen}, {Espinoza}, {Jensen}, {Collins}, {Schwarz}, {Conti}, {Wang},
  {Kielkopf}, {Mao}, {Horne}, {Sefako}, {Quinn}, {Moldovan}, {Fausnaugh},
  {F{\.z}{\.z}r{\'e}sz}, \& {Barclay}}]{Peterson:2023}
{Peterson}, M.~S., {Benneke}, B., {Collins}, K., {et~al.} 2023, \nat, 617, 701

\bibitem[{{Piaulet} {et~al.}(2023){Piaulet}, {Benneke}, {Almenara}, {Dragomir},
  {Knutson}, {Thorngren}, {Peterson}, {Crossfield}, {Kempton}, {Kubyshkina},
  {Howard}, {Angus}, {Isaacson}, {Weiss}, {Beichman}, {Fortney}, {Fossati},
  {Lammer}, {McCullough}, {Morley}, \& {Wong}}]{Piaulet:2023}
{Piaulet}, C., {Benneke}, B., {Almenara}, J.~M., {et~al.} 2023, Nature
  Astronomy, 7, 206

\bibitem[{{Rackham} {et~al.}(2018){Rackham}, {Apai}, \&
  {Giampapa}}]{Rackham:2018}
{Rackham}, B.~V., {Apai}, D., \& {Giampapa}, M.~S. 2018, The Astrophysical
  Journal, 853, 122

\bibitem[{{Rackham} {et~al.}(2023){Rackham}, {Espinoza}, {Berdyugina},
  {Korhonen}, {MacDonald}, {Montet}, {Morris}, {Oshagh}, {Shapiro}, {Unruh},
  {Quintana}, {Zellem}, {Apai}, {Barclay}, {Barstow}, {Bruno}, {Carone},
  {Casewell}, {Cegla}, {Criscuoli}, {Fischer}, {Fournier}, {Giampapa}, {Giles},
  {Iyer}, {Kopp}, {Kostogryz}, {Krivova}, {Mallonn}, {McGruder},
  {Molaverdikhani}, {Newton}, {Panja}, {Peacock}, {Reardon}, {Roettenbacher},
  {Scandariato}, {Solanki}, {Stassun}, {Steiner}, {Stevenson}, {Tregloan-Reed},
  {Valio}, {Wedemeyer}, {Welbanks}, {Yu}, {Alam}, {Davenport}, {Deming},
  {Dong}, {Ducrot}, {Fisher}, {Gilbert}, {Kostov}, {L{\'o}pez-Morales}, {Line},
  {Mo{\v{c}}nik}, {Mullally}, {Paudel}, {Ribas}, \& {Valenti}}]{Rackham:2023}
{Rackham}, B.~V., {Espinoza}, N., {Berdyugina}, S.~V., {et~al.} 2023, RAS
  Techniques and Instruments, 2, 148

\bibitem[{{Ramirez} \& {Kaltenegger}(2014)}]{Ramirez:2014c}
{Ramirez}, R.~M. \& {Kaltenegger}, L. 2014, The Astrophysical Journal Letters,
  797, L25

\bibitem[{{Ramirez} \& {Kaltenegger}(2016)}]{Ramirez:2016}
{Ramirez}, R.~M. \& {Kaltenegger}, L. 2016, The Astrophysical Journal, 823, 6

\bibitem[{{Ramirez} \& {Kaltenegger}(2017)}]{Ramirez:2017b}
{Ramirez}, R.~M. \& {Kaltenegger}, L. 2017, The Astrophysical Journal Letters,
  837, L4

\bibitem[{Ramirez \& Kaltenegger(2018)}]{Ramirez:2018}
Ramirez, R.~M. \& Kaltenegger, L. 2018, The Astrophysical Journal, 858, 72

\bibitem[{{Rathcke} {et~al.}(2021){Rathcke}, {Bello-Arufe}, {Buchhave},
  {Burgasser}, {Diamond-Lowe}, {Espinoza}, {Fisher}, {Gibson}, {Guzman Mesa},
  {Heng}, {Hoeijmakers}, {Hooton}, {Kitzmann}, {Kozakis}, {Lopez-Morales},
  {Mendonca}, \& {Morris}}]{Rathcke:2021}
{Rathcke}, A., {Bello-Arufe}, A., {Buchhave}, L.~A., {et~al.} 2021, {Probing
  the Terrestrial Planet TRAPPIST-1c for the Presence of an Atmosphere}, JWST
  Proposal. Cycle 1, ID. \#2420

\bibitem[{{Rossow}(1978)}]{Rossow:1978}
{Rossow}, W.~B. 1978, Icarus, 36, 1

\bibitem[{{Rowe} {et~al.}(2014){Rowe}, {Bryson}, {Marcy}, {Lissauer},
  {Jontof-Hutter}, {Mullally}, {Gilliland}, {Issacson}, {Ford}, {Howell},
  {Borucki}, {Haas}, {Huber}, {Steffen}, {Thompson}, {Quintana}, {Barclay},
  {Still}, {Fortney}, {Gautier}, {Hunter}, {Caldwell}, {Ciardi}, {Devore},
  {Cochran}, {Jenkins}, {Agol}, {Carter}, \& {Geary}}]{Rowe:2014}
{Rowe}, J.~F., {Bryson}, S.~T., {Marcy}, G.~W., {et~al.} 2014, \apj, 784, 45

\bibitem[{{Salvador} {et~al.}(2017){Salvador}, {Massol}, {Davaille}, {Marcq},
  {Sarda}, \& {Chassefi{\`e}re}}]{Salvador:2017}
{Salvador}, A., {Massol}, H., {Davaille}, A., {et~al.} 2017, Journal of
  Geophysical Research (Planets), 122, 1458

\bibitem[{{Selsis} {et~al.}(2023){Selsis}, {Leconte}, {Turbet}, {Chaverot}, \&
  {Bolmont}}]{Selsis:2023}
{Selsis}, F., {Leconte}, J., {Turbet}, M., {Chaverot}, G., \& {Bolmont}, E.
  2023, Nature, 620, 287–291

\bibitem[{{Sergeev} {et~al.}(2022){Sergeev}, {Fauchez}, {Turbet}, {Boutle},
  {Tsigaridis}, {Way}, {Wolf}, {Domagal-Goldman}, {Forget}, {Haqq-Misra},
  {Kopparapu}, {Lambert}, {Manners}, \& {Mayne}}]{Sergeev:2022}
{Sergeev}, D.~E., {Fauchez}, T.~J., {Turbet}, M., {et~al.} 2022, The Planetary
  Science Journal, 3, 212

\bibitem[{{Sergeev} {et~al.}(2020){Sergeev}, {Lambert}, {Mayne}, {Boutle},
  {Manners}, \& {Kohary}}]{Sergeev:2020}
{Sergeev}, D.~E., {Lambert}, F.~H., {Mayne}, N.~J., {et~al.} 2020, The
  Astrophysical Journal, 894, 84

\bibitem[{{Shields} {et~al.}(2016){Shields}, {Ballard}, \&
  {Johnson}}]{Shields:2016}
{Shields}, A.~L., {Ballard}, S., \& {Johnson}, J.~A. 2016, Physics Reports,
  663, 1

\bibitem[{{Shields} {et~al.}(2013){Shields}, {Meadows}, {Bitz},
  {Pierrehumbert}, {Joshi}, \& {Robinson}}]{Shields:2013}
{Shields}, A.~L., {Meadows}, V.~S., {Bitz}, C.~M., {et~al.} 2013, Astrobiology,
  13, 715

\bibitem[{{Su{\'a}rez Mascare{\~n}o} {et~al.}(2023){Su{\'a}rez Mascare{\~n}o},
  {Gonz{\'a}lez-{\'A}lvarez}, {Zapatero Osorio}, {Lillo-Box}, {Faria},
  {Passegger}, {Gonz{\'a}lez Hern{\'a}ndez}, {Figueira}, {Sozzetti}, {Rebolo},
  {Pepe}, {Santos}, {Cristiani}, {Lovis}, {Silva}, {Ribas}, {Amado},
  {Caballero}, {Quirrenbach}, {Reiners}, {Zechmeister}, {Adibekyan}, {Alibert},
  {B{\'e}jar}, {Benatti}, {D'Odorico}, {Damasso}, {Delisle}, {Di Marcantonio},
  {Dreizler}, {Ehrenreich}, {Hatzes}, {Hara}, {Henning}, {Kaminski},
  {L{\'o}pez-Gonz{\'a}lez}, {Martins}, {Micela}, {Montes}, {Pall{\'e}},
  {Pedraz}, {Rodr{\'\i}guez}, {Rodr{\'\i}guez-L{\'o}pez}, {Tal-Or}, {Sousa}, \&
  {Udry}}]{Suarez-Mascareno:2023}
{Su{\'a}rez Mascare{\~n}o}, A., {Gonz{\'a}lez-{\'A}lvarez}, E., {Zapatero
  Osorio}, M.~R., {et~al.} 2023, \aap, 670, A5

\bibitem[{{Tan} {et~al.}(2021){Tan}, {Lef{\`e}vre}, \&
  {Pierrehumbert}}]{Tan:2021}
{Tan}, X., {Lef{\`e}vre}, M., \& {Pierrehumbert}, R.~T. 2021, The Astrophysical
  Journal Letters, 923, L15

\bibitem[{{Torres} {et~al.}(2017){Torres}, {Kane}, {Rowe}, {Batalha}, {Henze},
  {Ciardi}, {Barclay}, {Borucki}, {Buchhave}, {Crepp}, {Everett}, {Horch},
  {Howard}, {Howell}, {Isaacson}, {Jenkins}, {Latham}, {Petigura}, \&
  {Quintana}}]{Torres:2017}
{Torres}, G., {Kane}, S.~R., {Rowe}, J.~F., {et~al.} 2017, \aj, 154, 264

\bibitem[{{Torres} {et~al.}(2015){Torres}, {Kipping}, {Fressin}, {Caldwell},
  {Twicken}, {Ballard}, {Batalha}, {Bryson}, {Ciardi}, {Henze}, {Howell},
  {Isaacson}, {Jenkins}, {Muirhead}, {Newton}, {Petigura}, {Barclay},
  {Borucki}, {Crepp}, {Everett}, {Horch}, {Howard}, {Kolbl}, {Marcy},
  {McCauliff}, \& {Quintana}}]{Torres:2015}
{Torres}, G., {Kipping}, D.~M., {Fressin}, F., {et~al.} 2015, \apj, 800, 99

\bibitem[{{Tuomi} {et~al.}(2013){Tuomi}, {Jones}, {Jenkins}, {Tinney},
  {Butler}, {Vogt}, {Barnes}, {Wittenmyer}, {O'Toole}, {Horner}, {Bailey},
  {Carter}, {Wright}, {Salter}, \& {Pinfield}}]{Tuomi:2013}
{Tuomi}, M., {Jones}, H.~R.~A., {Jenkins}, J.~S., {et~al.} 2013, \aap, 551, A79

\bibitem[{Turbet(2018)}]{Turbet:2018}
Turbet, M. 2018, Theses, {Sorbonne Universit{\'e} / Universit{\'e} Pierre et
  Marie Curie - Paris VI}

\bibitem[{{Turbet} {et~al.}(2020{\natexlab{a}}){Turbet}, {Bolmont}, {Bourrier},
  {Demory}, {Leconte}, {Owen}, \& {Wolf}}]{Turbet:2020review}
{Turbet}, M., {Bolmont}, E., {Bourrier}, V., {et~al.} 2020{\natexlab{a}}, Space
  Science Reviews, 216, 100

\bibitem[{{Turbet} {et~al.}(2021){Turbet}, {Bolmont}, {Chaverot}, {Ehrenreich},
  {Leconte}, \& {Marcq}}]{Turbet2021}
{Turbet}, M., {Bolmont}, E., {Chaverot}, G., {et~al.} 2021, Nature, 598, 276

\bibitem[{{Turbet} {et~al.}(2020{\natexlab{b}}){Turbet}, {Bolmont},
  {Ehrenreich}, {Gratier}, {Leconte}, {Selsis}, {Hara}, \&
  {Lovis}}]{Turbet2020aa}
{Turbet}, M., {Bolmont}, E., {Ehrenreich}, D., {et~al.} 2020{\natexlab{b}},
  Astronomy \& Astrophysics, 638, A41

\bibitem[{{Turbet} {et~al.}(2018){Turbet}, {Bolmont}, {Leconte}, {Forget},
  {Selsis}, {Tobie}, {Caldas}, {Naar}, \& {Gillon}}]{Turbet:2018aa}
{Turbet}, M., {Bolmont}, E., {Leconte}, J., {et~al.} 2018, Astronomy $\&$
  Astrophysics, 612, A86

\bibitem[{{Turbet} {et~al.}(2019){Turbet}, {Ehrenreich}, {Lovis}, {Bolmont}, \&
  {Fauchez}}]{Turbet2019aa}
{Turbet}, M., {Ehrenreich}, D., {Lovis}, C., {Bolmont}, E., \& {Fauchez}, T.
  2019, Astronomy \& Astrophysics, 628, A12

\bibitem[{{Turbet} {et~al.}(2017){Turbet}, {Forget}, {Leconte}, {Charnay}, \&
  {Tobie}}]{Turbet:2017epsl}
{Turbet}, M., {Forget}, F., {Leconte}, J., {Charnay}, B., \& {Tobie}, G. 2017,
  Earth and Planetary Science Letters, 476, 11

\bibitem[{{Turbet} {et~al.}(2020{\natexlab{c}}){Turbet}, {Gillmann}, {Forget},
  {Baudin}, {Palumbo}, {Head}, \& {Karatekin}}]{Turbet:2020impact}
{Turbet}, M., {Gillmann}, C., {Forget}, F., {et~al.} 2020{\natexlab{c}},
  Icarus, 335, 113419

\bibitem[{{Turbet} {et~al.}(2016){Turbet}, {Leconte}, {Selsis}, {Bolmont},
  {Forget}, {Ribas}, {Raymond}, \& {Anglada-Escud{\'e}}}]{Turbet:2016}
{Turbet}, M., {Leconte}, J., {Selsis}, F., {et~al.} 2016, Astronomy $\&$
  Astrophysics, 596, A112

\bibitem[{{Turbet} \& {Selsis}(2023)}]{TurbetSelsis2021}
{Turbet}, M. \& {Selsis}, F. 2023, in Star-Planet Interactions, ed. L.~{Bigot},
  J.~{Bouvier}, Y.~{Lebreton}, A.~{Chiavassa}, \& A.~{L{\`e}bre}, 202

\bibitem[{{Vanderburg} {et~al.}(2020){Vanderburg}, {Rowden}, {Bryson},
  {Coughlin}, {Batalha}, {Collins}, {Latham}, {Mullally}, {Col{\'o}n}, {Henze},
  {Huang}, \& {Quinn}}]{Vanderburg:2020}
{Vanderburg}, A., {Rowden}, P., {Bryson}, S., {et~al.} 2020, \apjl, 893, L27

\bibitem[{{Villanueva} {et~al.}(2022){Villanueva}, {Liuzzi}, {Faggi},
  {Protopapa}, {Kofman}, {Fauchez}, {Stone}, \& {Mandell}}]{Villanueva:2022}
{Villanueva}, G.~L., {Liuzzi}, G., {Faggi}, S., {et~al.} 2022, {Fundamentals of
  the Planetary Spectrum Generator}

\bibitem[{{Villanueva} {et~al.}(2018){Villanueva}, {Smith}, {Protopapa},
  {Faggi}, \& {Mandell}}]{Villanueva:2018}
{Villanueva}, G.~L., {Smith}, M.~D., {Protopapa}, S., {Faggi}, S., \&
  {Mandell}, A.~M. 2018, Journal of Quantitative Spectroscopy and Radiative
  Transfer, 217, 86

\bibitem[{{Way} \& {Del Genio}(2020)}]{Way:2020}
{Way}, M.~J. \& {Del Genio}, A.~D. 2020, Journal of Geophysical Research
  (Planets), 125, e06276

\bibitem[{{Way} {et~al.}(2018){Way}, {Del Genio}, {Aleinov}, {Clune}, {Kelley},
  \& {Kiang}}]{Way:2018}
{Way}, M.~J., {Del Genio}, A.~D., {Aleinov}, I., {et~al.} 2018, The
  Astrophysical Journal Supplement Series, 239, 24

\bibitem[{{Way} {et~al.}(2016){Way}, {Del Genio}, {Kiang}, {Sohl}, {Grinspoon},
  {Aleinov}, {Kelley}, \& {Clune}}]{Way:2016}
{Way}, M.~J., {Del Genio}, A.~D., {Kiang}, N.~Y., {et~al.} 2016, Geophysical
  Research Letters, 43, 8376

\bibitem[{{Wolf}(2017)}]{Wolf:2017}
{Wolf}, E.~T. 2017, The Astrophysical Journal Letters, 839, L1

\bibitem[{{Wolf} {et~al.}(2020){Wolf}, {Haqq-Misra}, {Kopparapu}, {Fauchez},
  {Welsh}, {Kane}, {Kostov}, \& {Eggl}}]{Wolf:2020}
{Wolf}, E.~T., {Haqq-Misra}, J., {Kopparapu}, R., {et~al.} 2020, Journal of
  Geophysical Research (Planets), 125, e06576

\bibitem[{{Wolf} {et~al.}(2019){Wolf}, {Kopparapu}, \&
  {Haqq-Misra}}]{Wolf:2019}
{Wolf}, E.~T., {Kopparapu}, R.~K., \& {Haqq-Misra}, J. 2019, The Astrophysical
  Journal, 877, 35

\bibitem[{{Wolf} {et~al.}(2017){Wolf}, {Shields}, {Kopparapu}, {Haqq-Misra}, \&
  {Toon}}]{Wolf:2017b}
{Wolf}, E.~T., {Shields}, A.~L., {Kopparapu}, R.~K., {Haqq-Misra}, J., \&
  {Toon}, O.~B. 2017, The Astrophysical Journal, 837, 107

\bibitem[{{Wolf} \& {Toon}(2015)}]{Wolf:2015}
{Wolf}, E.~T. \& {Toon}, O.~B. 2015, Journal of Geophysical Research
  (Atmospheres), 120, 5775

\bibitem[{{Wordsworth} {et~al.}(2011){Wordsworth}, {Forget}, {Selsis},
  {Millour}, {Charnay}, \& {Madeleine}}]{Wordsworth:2011ajl}
{Wordsworth}, R.~D., {Forget}, F., {Selsis}, F., {et~al.} 2011, The
  Astrophysical Journal Letters, 733, L48

\bibitem[{{Wunderlich} {et~al.}(2019){Wunderlich}, {Godolt}, {Grenfell},
  {St{\"a}dt}, {Smith}, {Gebauer}, {Schreier}, {Hedelt}, \&
  {Rauer}}]{Wunderlich2019}
{Wunderlich}, F., {Godolt}, M., {Grenfell}, J.~L., {et~al.} 2019, Astronomy \&
  Astrophysics, 624, A49

\bibitem[{{Yang} {et~al.}(2019{\natexlab{a}}){Yang}, {Abbot}, {Koll}, {Hu}, \&
  {Showman}}]{Yang:2019b}
{Yang}, J., {Abbot}, D.~S., {Koll}, D. D.~B., {Hu}, Y., \& {Showman}, A.~P.
  2019{\natexlab{a}}, The Astrophysical Journal, 871, 29

\bibitem[{{Yang} {et~al.}(2014){Yang}, {Bou{\'e}}, {Fabrycky}, \&
  {Abbot}}]{Yang:2014b}
{Yang}, J., {Bou{\'e}}, G., {Fabrycky}, D.~C., \& {Abbot}, D.~S. 2014, The
  Astrophysical Journal Letters, 787, L2

\bibitem[{{Yang} {et~al.}(2013){Yang}, {Cowan}, \& {Abbot}}]{Yang:2013}
{Yang}, J., {Cowan}, N.~B., \& {Abbot}, D.~S. 2013, The Astrophysical Journal
  Letters, 771, L45

\bibitem[{{Yang} {et~al.}(2016){Yang}, {Leconte}, {Wolf}, {Goldblatt}, {Feldl},
  {Merlis}, {Wang}, {Koll}, {Ding}, {Forget}, \& {Abbot}}]{Yang:2016}
{Yang}, J., {Leconte}, J., {Wolf}, E.~T., {et~al.} 2016, The Astrophysical
  Journal, 826, 222

\bibitem[{{Yang} {et~al.}(2019{\natexlab{b}}){Yang}, {Leconte}, {Wolf},
  {Merlis}, {Koll}, {Forget}, \& {Abbot}}]{Yang:2019a}
{Yang}, J., {Leconte}, J., {Wolf}, E.~T., {et~al.} 2019{\natexlab{b}}, The
  Astrophysical Journal, 875, 46

\bibitem[{{Zahnle} \& {Catling}(2017)}]{Zahnle:2017}
{Zahnle}, K.~J. \& {Catling}, D.~C. 2017, The Astrophysical Journal, 843, 122

\bibitem[{{Zahnle} {et~al.}(1988){Zahnle}, {Kasting}, \&
  {Pollack}}]{Zahnle:1988}
{Zahnle}, K.~J., {Kasting}, J.~F., \& {Pollack}, J.~B. 1988, Icarus, 74, 62

\bibitem[{{Zechmeister} {et~al.}(2019){Zechmeister}, {Dreizler}, {Ribas},
  {Reiners}, {Caballero}, {Bauer}, {B{\'e}jar}, {Gonz{\'a}lez-Cuesta},
  {Herrero}, {Lalitha}, {L{\'o}pez-Gonz{\'a}lez}, {Luque}, {Morales},
  {Pall{\'e}}, {Rodr{\'\i}guez}, {Rodr{\'\i}guez L{\'o}pez}, {Tal-Or},
  {Anglada-Escud{\'e}}, {Quirrenbach}, {Amado}, {Abril}, {Aceituno},
  {Aceituno}, {Alonso-Floriano}, {Ammler-von Eiff}, {Antona Jim{\'e}nez},
  {Anwand-Heerwart}, {Arroyo-Torres}, {Azzaro}, {Baroch}, {Barrado},
  {Becerril}, {Ben{\'\i}tez}, {Berdi{\~n}as}, {Bergond}, {Bluhm},
  {Brinkm{\"o}ller}, {del Burgo}, {Calvo Ortega}, {Cano}, {Cardona
  Guill{\'e}n}, {Carro}, {C{\'a}rdenas V{\'a}zquez}, {Casal},
  {Casasayas-Barris}, {Casanova}, {Chaturvedi}, {Cifuentes}, {Claret},
  {Colom{\'e}}, {Cort{\'e}s-Contreras}, {Czesla}, {D{\'\i}ez-Alonso}, {Dorda},
  {Fern{\'a}ndez}, {Fern{\'a}ndez-Mart{\'\i}n}, {Fuhrmeister}, {Fukui},
  {Galad{\'\i}-Enr{\'\i}quez}, {Gallardo Cava}, {Garcia de la Fuente},
  {Garcia-Piquer}, {Garc{\'\i}a Vargas}, {Gesa}, {G{\'o}ngora Rueda},
  {Gonz{\'a}lez-{\'A}lvarez}, {Gonz{\'a}lez Hern{\'a}ndez},
  {Gonz{\'a}lez-Peinado}, {Gr{\"o}zinger}, {Gu{\`a}rdia}, {Guijarro}, {de
  Guindos}, {Hatzes}, {Hauschildt}, {Hedrosa}, {Helmling}, {Henning},
  {Hermelo}, {Hern{\'a}ndez Arabi}, {Hern{\'a}ndez Casta{\~n}o}, {Hern{\'a}ndez
  Otero}, {Hintz}, {Huke}, {Huber}, {Jeffers}, {Johnson}, {de Juan},
  {Kaminski}, {Kemmer}, {Kim}, {Klahr}, {Klein}, {Kl{\"u}ter}, {Klutsch},
  {Kossakowski}, {K{\"u}rster}, {Labarga}, {Lafarga}, {Llamas}, {Lamp{\'o}n},
  {Lara}, {Launhardt}, {L{\'a}zaro}, {Lodieu}, {L{\'o}pez del Fresno},
  {L{\'o}pez-Puertas}, {L{\'o}pez Salas}, {L{\'o}pez-Santiago}, {Mag{\'a}n
  Madinabeitia}, {Mall}, {Mancini}, {Mandel}, {Marfil}, {Mar{\'\i}n Molina},
  {Maroto Fern{\'a}ndez}, {Mart{\'\i}n}, {Mart{\'\i}n-Fern{\'a}ndez},
  {Mart{\'\i}n-Ruiz}, {Marvin}, {Mirabet}, {Monta{\~n}{\'e}s-Rodr{\'\i}guez},
  {Montes}, {Moreno-Raya}, {Nagel}, {Naranjo}, {Narita}, {Nortmann}, {Nowak},
  {Ofir}, {Oshagh}, {Panduro}, {Parviainen}, {Pascual}, {Passegger}, {Pavlov},
  {Pedraz}, {P{\'e}rez-Calpena}, {P{\'e}rez Medialdea}, {Perger}, {Perryman},
  {Rabaza}, {Ram{\'o}n Ballesta}, {Rebolo}, {Redondo}, {Reffert}, {Reinhardt},
  {Rhode}, {Rix}, {Rodler}, {Rodr{\'\i}guez Trinidad}, {Rosich}, {Sadegi},
  {S{\'a}nchez-Blanco}, {S{\'a}nchez Carrasco}, {S{\'a}nchez-L{\'o}pez},
  {Sanz-Forcada}, {Sarkis}, {Sarmiento}, {Sch{\"a}fer}, {Schmitt},
  {Sch{\"o}fer}, {Schweitzer}, {Seifert}, {Shulyak}, {Solano}, {Sota}, {Stahl},
  {Stock}, {Strachan}, {Stuber}, {St{\"u}rmer}, {Su{\'a}rez}, {Tabernero},
  {Tala Pinto}, {Trifonov}, {Veredas}, {Vico Linares}, {Vilardell}, {Wagner},
  {Wolthoff}, {Xu}, {Yan}, \& {Zapatero Osorio}}]{Zechmeister:2019}
{Zechmeister}, M., {Dreizler}, S., {Ribas}, I., {et~al.} 2019, Astronomy \&
  Astrophysics, 627, A49

\bibitem[{{Zhang} \& {Showman}(2017)}]{Zhang:2017}
{Zhang}, X. \& {Showman}, A.~P. 2017, The Astrophysical Journal, 836, 73

\bibitem[{{Zieba} {et~al.}(2023){Zieba}, L., {Ducrot}, {Gillon}, {Morley},
  {Schaefer}, {Tamburo}, {Lyu}, {Acuna}, {Agol}, {Iyer}, {Hu}, {Lincowski},
  {Meadows}, {Selsis}, {Bolmont}, {Mandell}, \& {Suissa}}]{Zieba:2023}
{Zieba}, S., L., K., {Ducrot}, E., {et~al.} 2023, Nature

\end{thebibliography}

\begin{appendix}

\section{Additional Tables}

We provide here -- in Tables~\ref{table_simu_1} and \ref{table_simu_2} -- the main list of 3D Global Climate Model (GCM) simulations performed with the Generic PCM in this study.

\begin{table*}
\centering
\begin{tabular}{lccccccc}
\hline
Simulation & Star Type (T$_{\text{eff}}$)  & Flux & Rotation Period & Mass & Radius & P$_{\rm H_2O}$ & Internal heat flux \\
Name & (K) & (F$_{\oplus}$) & (Earth days) & (M$_{\oplus}$) & (R$_{\oplus}$) & (bar) & (W~m$^{-2}$) \\
\hline
SUN-1 & Sun (5780) & 1.47 & 1 (non-TL) & 1 & 1 & 10 & 0\\
SUN-2 & Sun (5780) & 1.10 & 1 (non-TL) & 1 & 1 & 10 & 0\\
SUN-3 & Sun (5780) & 1.00 & 1 (non-TL) & 1 & 1 & 10 & 0\\
SUN-4 & Sun (5780) & 0.98 & 1 (non-TL) & 1 & 1 & 10 & 0\\
SUN-5 & Sun (5780) & 0.95 & 1 (non-TL) & 1 & 1 & 10 & 0\\
SUN-6 & Sun (5780) & 0.92 & 1 (non-TL) & 1 & 1 & 10 & 0\\
\hline
K5-1 & K5 (4400) & 1.47 & 87.0 (TL) & 1 & 1 & 10 & 0\\
K5-2 & K5 (4400) & 1.10 & 108.0 (TL) & 1 & 1 & 10 & 0\\
K5-3 & K5 (4400) & 1.03 & 113.7 (TL) & 1 & 1 & 10 & 0\\
K5-4 & K5 (4400) & 0.95 & 120.2 (TL) & 1 & 1 & 10 & 0\\
K5-5 & K5 (4400) & 0.88 & 127.6 (TL) & 1 & 1 & 10 & 0\\
K5-6 & K5 (4400) & 0.84 & 131.8 (TL) & 1 & 1 & 10 & 0\\
\hline
M3-1 & M3 (3400) & 1.47 & 18.3 (TL) & 1 & 1 & 10 & 0\\
M3-2 & M3 (3400) & 1.10 & 22.7 (TL) & 1 & 1 & 10 & 0\\
M3-3 & M3 (3400) & 1.03 & 23.9 (TL) & 1 & 1 & 10 & 0\\
M3-4 & M3 (3400) & 0.95 & 25.2 (TL) & 1 & 1 & 10 & 0\\
M3-5 & M3 (3400) & 0.88 & 26.8 (TL) & 1 & 1 & 10 & 0\\
M3-6 & M3 (3400) & 0.84 & 27.6 (TL) & 1 & 1 & 10 & 0\\
M3-7 & M3 (3400) & 0.81 & 28.6 (TL) & 1 & 1 & 10 & 0\\
\hline
Pcen-1 & M5.5 (3050) & 1.47 & 6.0 (TL) & 1 & 1 & 10 & 0\\
Pcen-2 & M5.5 (3050) & 1.10 & 7.5 (TL) & 1 & 1 & 10 & 0\\
Pcen-3 & M5.5 (3050) & 1.03 & 7.9 (TL) & 1 & 1 & 10 & 0\\
Pcen-4 & M5.5 (3050) & 0.95 & 8.3 (TL) & 1 & 1 & 10 & 0\\
Pcen-5 & M5.5 (3050) & 0.88 & 8.8 (TL) & 1 & 1 & 10 & 0\\
Pcen-6 & M5.5 (3050) & 0.81 & 9.4 (TL) & 1 & 1 & 10 & 0\\
Pcen-7 & M5.5 (3050) & 0.77 & 9.8 (TL) & 1 & 1 & 10 & 0\\
\hline
T1-1 & M8 (2600) & 1.47 & 3.3 (TL) & 1 & 1 & 10 & 0\\
T1-2 & M8 (2600) & 1.10 & 4.1 (TL) & 1 & 1 & 10 & 0\\
T1-3 & M8 (2600) & 1.03 & 4.3 (TL) & 1 & 1 & 10 & 0\\
T1-4 & M8 (2600) & 0.95 & 4.6 (TL) & 1 & 1 & 10 & 0\\
T1-5 & M8 (2600) & 0.88 & 4.8 (TL) & 1 & 1 & 10 & 0\\
T1-6 & M8 (2600) & 0.81 & 5.2 (TL) & 1 & 1 & 10 & 0\\
T1-7 & M8 (2600) & 0.77 & 5.3 (TL) & 1 & 1 & 10 & 0\\
T1-8 & M8 (2600) & 0.73 & 5.5 (TL) & 1 & 1 & 10 & 0\\
\hline
\end{tabular} 
\caption{List of the numerical experiments performed with the Generic PCM in this study. P$_{\rm H_2O}$ is the partial pressure of H$_2$O. All simulations assume a partial pressure P$_{\rm N_2}$ equal to 1~bar. TL means Tidally-Locked. Simulations with the Sun as the host star were directly taken from \cite{Turbet2021}. The names of the simulations always start with the name/type of the host star.}
\label{table_simu_1} 
\end{table*}

\begin{table*}
\centering
\begin{tabular}{lccccccc}
\hline
Simulation & Star Type & FluxT$_{\text{eff}}$ & Rotation Period & Mass & Radius & P$_{\rm H_2O}$ & Internal heat flux \\
Name & (K) & (F$_{\oplus}$) & (Earth days) & (M$_{\oplus}$) & (R$_{\oplus}$) & (bar) & (W~m$^{-2}$) \\
\hline
M3-YANG-1 & M3 (3400) & 1.47 & 60 (TL) & 5.6 & 2 & 10 & 0\\
M3-YANG-2 & M3 (3400) & 1.10 & 60 (TL) & 5.6 & 2 & 10 & 0\\
M3-YANG-3 & M3 (3400) & 1.03 & 60 (TL) & 5.6 & 2 & 10 & 0\\
M3-YANG-4 & M3 (3400) & 0.95 & 60 (TL) & 5.6 & 2 & 10 & 0\\
M3-YANG-5 & M3 (3400) & 0.88 & 60 (TL) & 5.6 & 2 & 10 & 0\\
M3-YANG-6 & M3 (3400) & 0.84 & 60 (TL) & 5.6 & 2 & 10 & 0\\
\hline
M3-3bar-1 & M3 (3400) & 1.47 & 18.3 (TL) & 1 & 1 & 3 & 0\\
M3-3bar-2 & M3 (3400) & 1.10 & 22.7 (TL) & 1 & 1 & 3 & 0\\
M3-3bar-3 & M3 (3400) & 1.03 & 23.9 (TL) & 1 & 1 & 3 & 0\\
M3-3bar-4 & M3 (3400) & 0.95 & 25.2 (TL) & 1 & 1 & 3 & 0\\
M3-3bar-5 & M3 (3400) & 0.88 & 26.8 (TL) & 1 & 1 & 3 & 0\\
M3-3bar-6 & M3 (3400) & 0.84 & 27.6 (TL) & 1 & 1 & 3 & 0\\
M3-3bar-7 & M3 (3400) & 0.81 & 28.6 (TL) & 1 & 1 & 3 & 0\\
\hline
M3-1bar-1 & M3 (3400) & 1.47 & 18.3 (TL) & 1 & 1 & 1 & 0\\
M3-1bar-2 & M3 (3400) & 1.10 & 22.7 (TL) & 1 & 1 & 1 & 0\\
M3-1bar-3 & M3 (3400) & 1.03 & 23.9 (TL) & 1 & 1 & 1 & 0\\
M3-1bar-4 & M3 (3400) & 0.95 & 25.2 (TL) & 1 & 1 & 1 & 0\\
M3-1bar-5 & M3 (3400) & 0.88 & 26.8 (TL) & 1 & 1 & 1 & 0\\
M3-1bar-6 & M3 (3400) & 0.84 & 27.6 (TL) & 1 & 1 & 1 & 0\\
M3-1bar-7 & M3 (3400) & 0.81 & 28.6 (TL) & 1 & 1 & 1 & 0\\
\hline
M3-0.3bar-1 & M3 (3400) & 1.47 & 18.3 (TL) & 1 & 1 & 0.3 & 0\\
M3-0.3bar-2 & M3 (3400) & 1.10 & 22.7 (TL) & 1 & 1 & 0.3 & 0\\
M3-0.3bar-3 & M3 (3400) & 1.03 & 23.9 (TL) & 1 & 1 & 0.3 & 0\\
M3-0.3bar-4 & M3 (3400) & 0.95 & 25.2 (TL) & 1 & 1 & 0.3 & 0\\
M3-0.3bar-5 & M3 (3400) & 0.88 & 26.8 (TL) & 1 & 1 & 0.3 & 0\\
M3-0.3bar-6 & M3 (3400) & 0.84 & 27.6 (TL) & 1 & 1 & 0.3 & 0\\
\hline
T1b & M8 (2600) & 4.15 & 1.51 (TL) & 1.37 & 1.12 & 10 & 0\\
T1b-3bar & M8 (2600) & 4.15 & 1.51 (TL) & 1.37 & 1.12 & 3 & 0\\
T1b-1bar & M8 (2600) & 4.15 & 1.51 (TL) & 1.37 & 1.12 & 1 & 0\\
T1b-Fgeo1 & M8 (2600) & 4.15 & 1.51 (TL) & 1.37 & 1.12 & 10 & 1\\
T1b-Fgeo5 & M8 (2600) & 4.15 & 1.51 (TL) & 1.37 & 1.12 & 10 & 5\\
T1b-Fgeo25 & M8 (2600) & 4.15 & 1.51 (TL) & 1.37 & 1.12 & 10 & 25\\
\hline
T1c & M8 (2600) & 2.21 & 2.42 (TL) & 1.31 & 1.10 & 10 & 0\\
T1d & M8 (2600) & 1.12 & 4.05 (TL) & 0.39 & 0.79 & 10 & 0\\
T1e & M8 (2600) & 0.65 & 6.10 (TL) & 0.69 & 0.92 & 10 & 0\\
Pcen-b & M5.5 (3050) & 0.68 & 11.2 (TL) & 1.35 & 1.09 & 10 & 0\\
\end{tabular} 
\caption{Continuation of Table~\ref{table_simu_1}.}
\label{table_simu_2} 
\end{table*}

\section{Additional observables}

\subsection{Transit spectra of TRAPPIST-1b, c and d}

\begin{figure*}
\centering 
\includegraphics[width=0.85\linewidth]{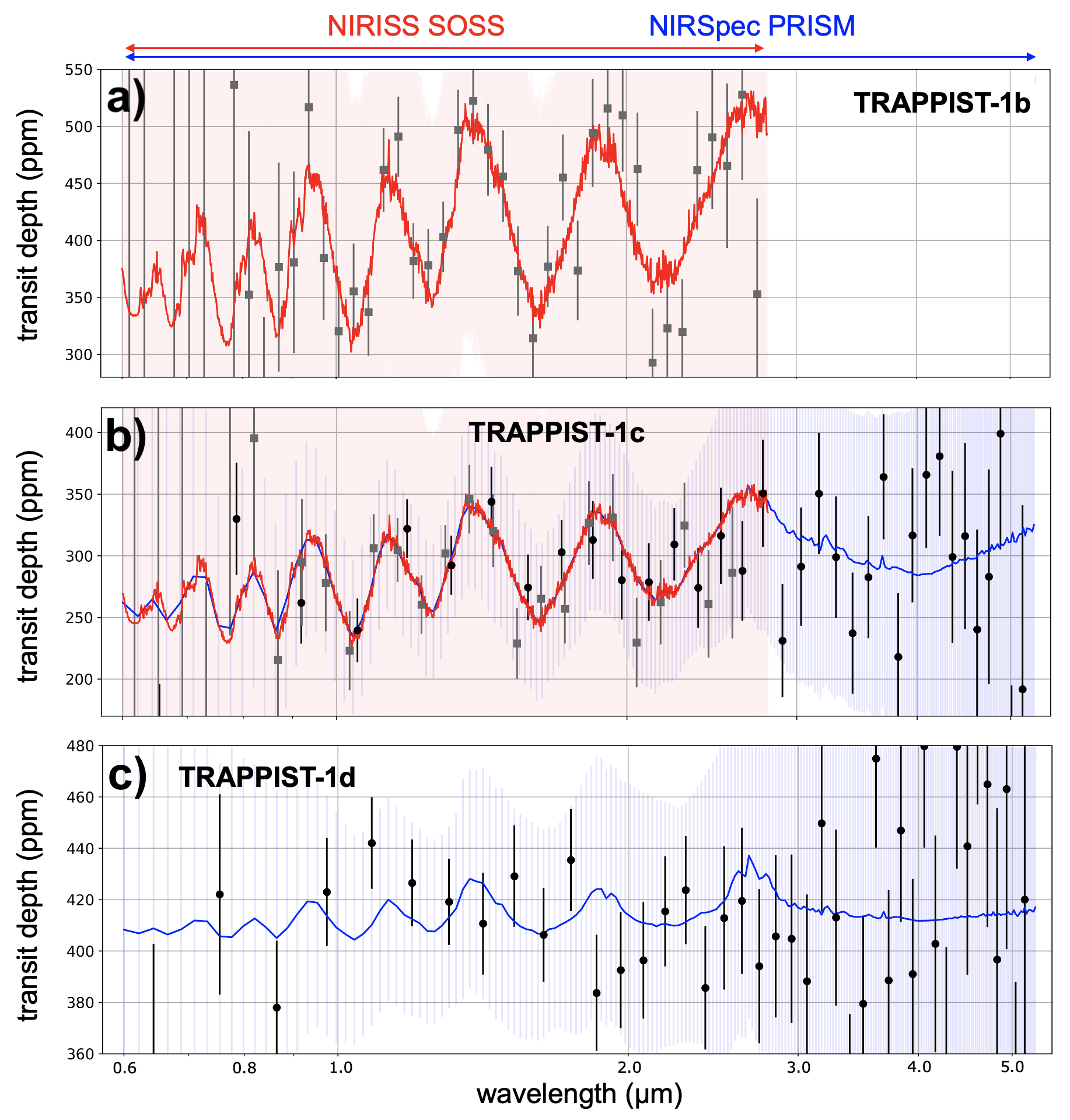}
\caption{Transit spectra computed for the three innermost planets of the TRAPPIST-1 system calculated from 3D GCM simulations, assuming a 10~bar H$_2$O atmosphere (+1~bar N$_2$). Red and blue spectra are calculated for the wavelength coverage of NIRISS SOSS and NIRSpec PRISM, respectively. 1-$\sigma$ error bars are calculated using the JWST Cycle 1 observation program (more details in Section-\ref{subsubsection_transit_spectra}). The transit spectra are in transit depth units (the transit depth is equal to zero at the surface).}
\label{transit_spectra_T1bcd_H2O}
\end{figure*}

\begin{figure*}
\centering 
\includegraphics[width=0.85\linewidth]{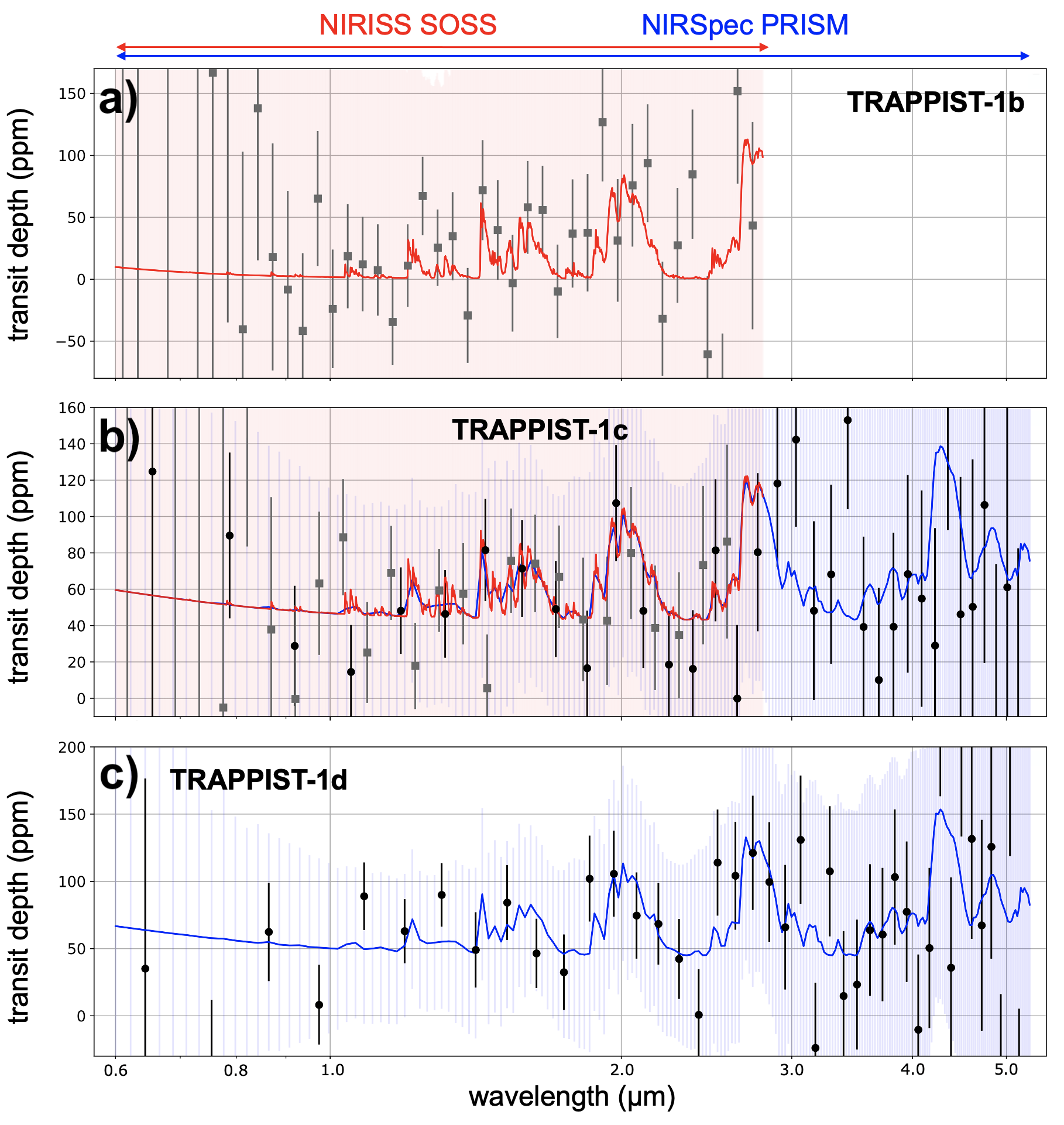}
\caption{Same as Fig.~\ref{transit_spectra_T1bcd_H2O}, for 10~bar CO$_2$ atmospheres.}
\label{transit_spectra_T1bcd_CO2}
\end{figure*}

We provide here additional figures of transit spectra computed for TRAPPIST-1b, c and d, for H$_2$O-rich atmospheres (see Fig.~\ref{transit_spectra_T1bcd_H2O}) and CO$_2$-rich atmospheres (see Fig.~\ref{transit_spectra_T1bcd_CO2}). Transit spectra include realistic error bars calculated based on JWST Cycle 1 planned observations with NIRSpec and NIRISS.

\subsection{Emission spectra and thermal phase curves of TRAPPIST-1b, c and d}

We provide here additional figures of emission spectra and thermal phase curves, for each individual GCM simulation of TRAPPIST-1b, c and d. Each plot shows (on the left) the emission spectra at all phase angles ; and (on the right) the associated thermal phase curve, integrated over all wavelengths. 

\begin{figure*}
\centering 
\subfloat[TRAPPIST-1b, 1~bar N$_2$, 1~bar H$_2$O]{
\label{subfig:T1b1}
\includegraphics[width=0.85\linewidth]{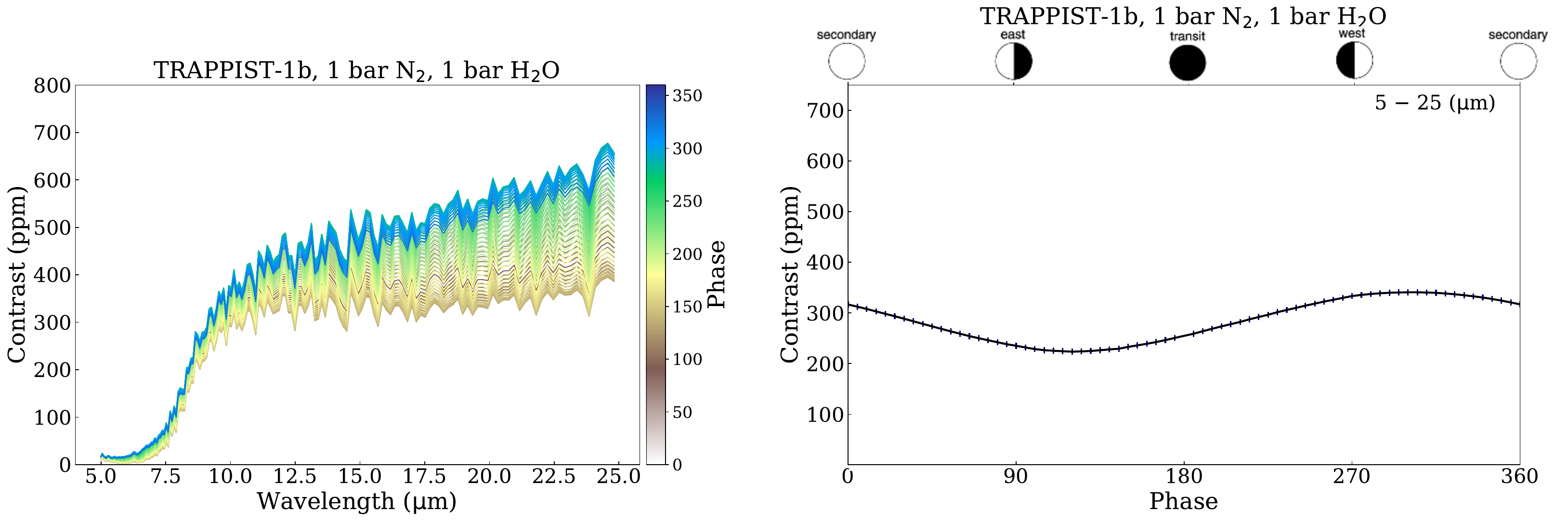}}
\\
\subfloat[TRAPPIST-1b, 1~bar N$_2$, 3~bar H$_2$O]{
\label{subfig:T1b2}
\includegraphics[width=0.85\linewidth]{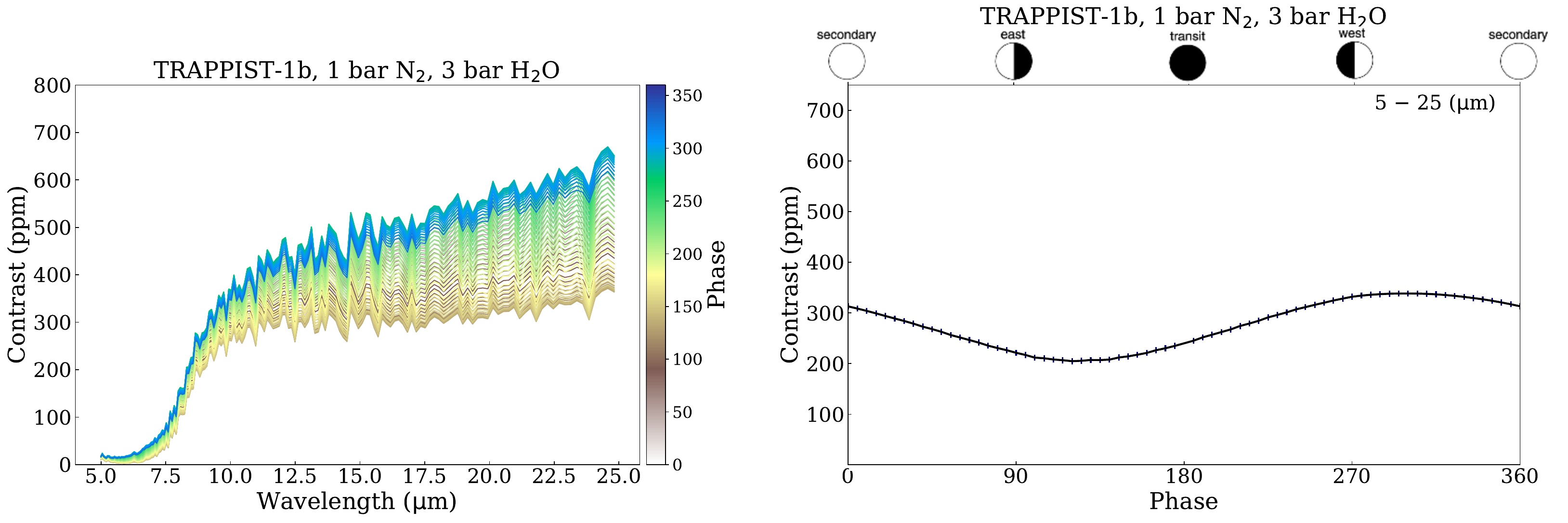}}
\\
\subfloat[TRAPPIST-1b, 1~bar N$_2$, 10~bar H$_2$O]{
\label{subfig:T1b3}
\includegraphics[width=0.85\linewidth]{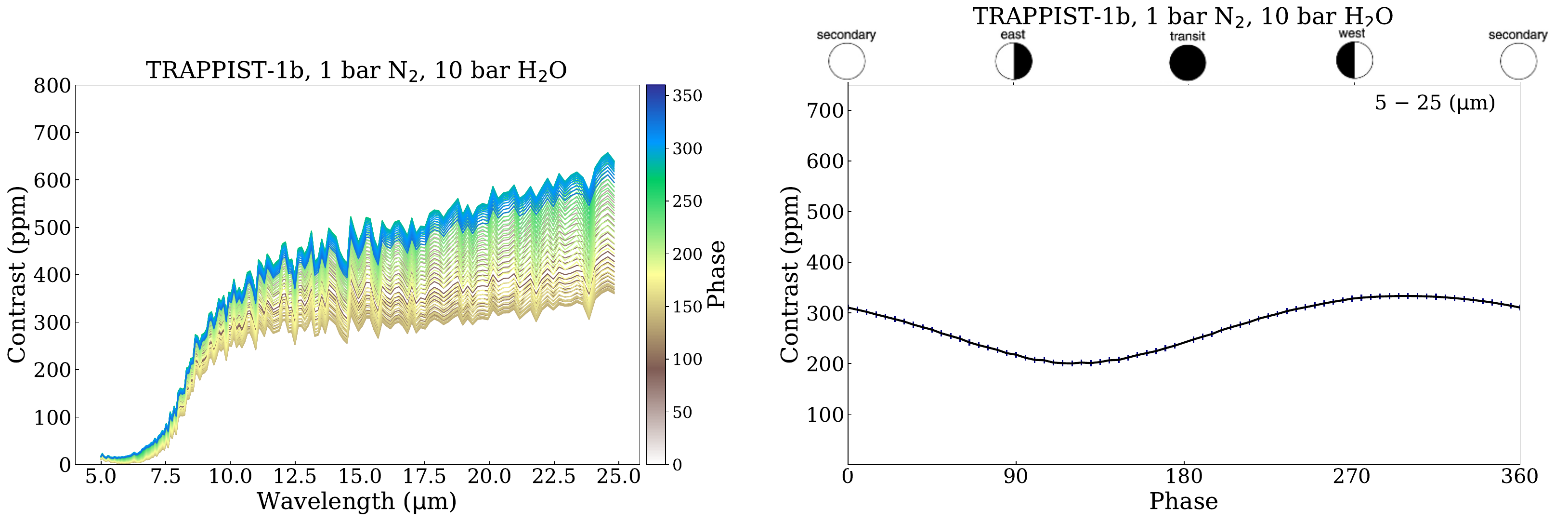}}
\caption{Thermal emission spectra at various phase angles (left). Thermal phase curve (right) integrated over all wavelengths. 
\label{fig:phase_curves_T1bA}}
\end{figure*}

\begin{figure*}
\centering 
\subfloat[TRAPPIST-1b, 10~bar CO$_2$]{
\label{subfig:T1b4}
\includegraphics[width=0.85\linewidth]{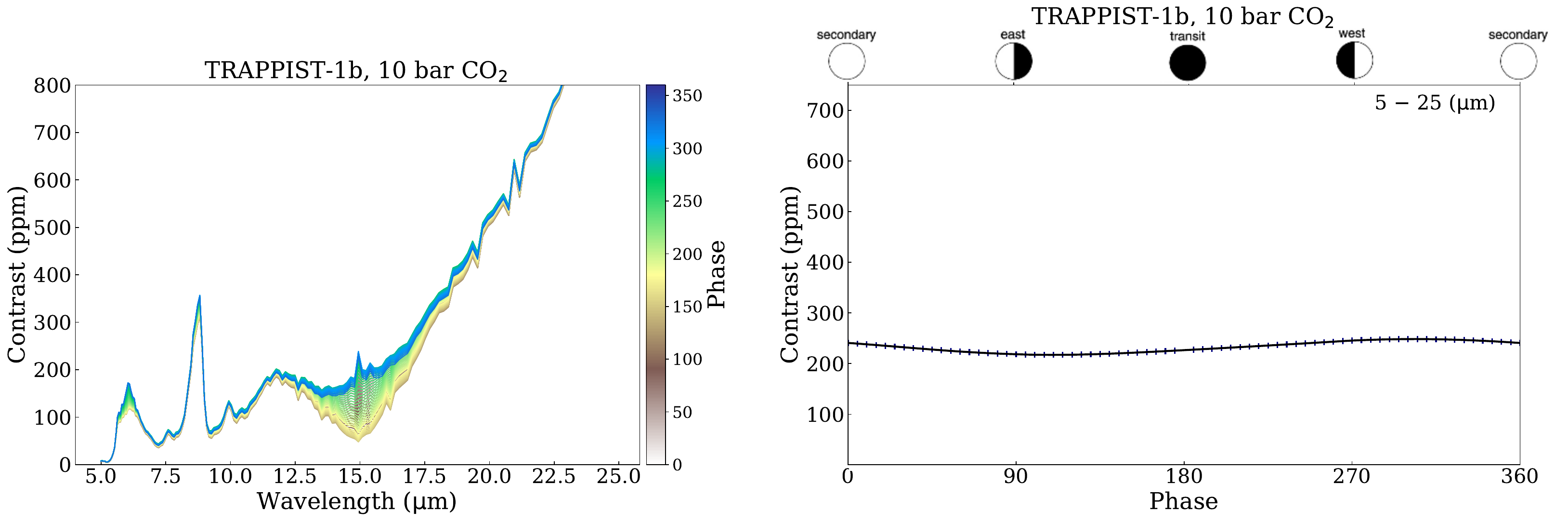}}
\\
\subfloat[TRAPPIST-1b, 1~bar CO$_2$]{
\label{subfig:T1b5}
\includegraphics[width=0.85\linewidth]{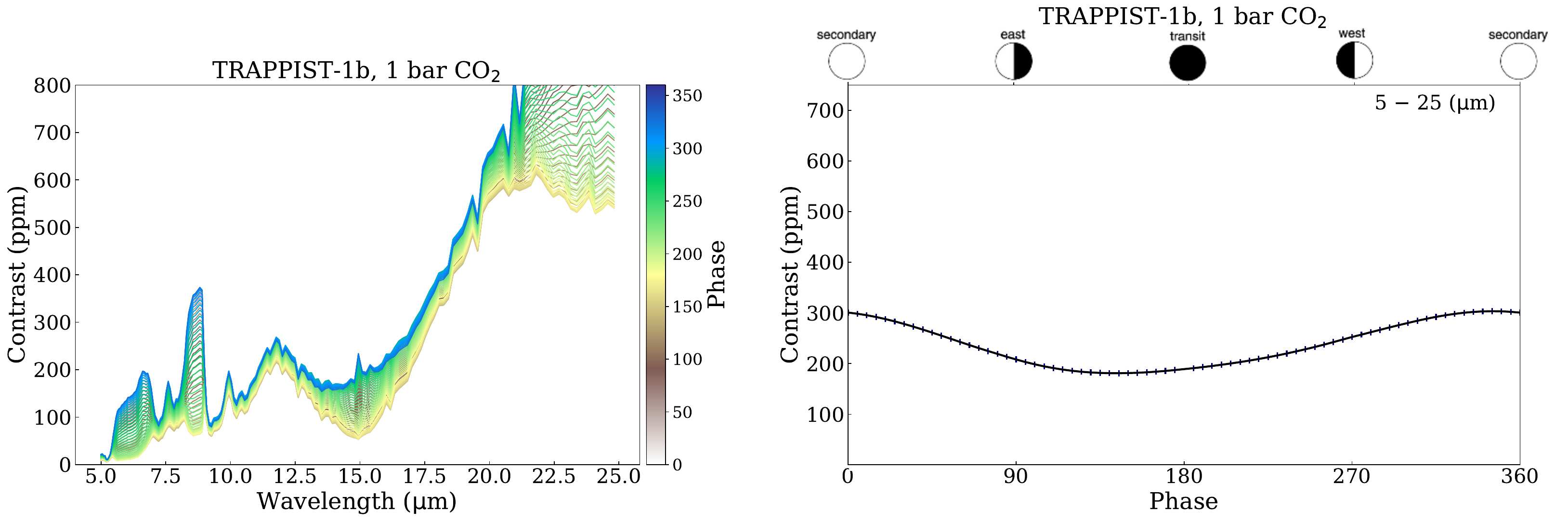}}
\\
\subfloat[TRAPPIST-1b, 100mbar CO$_2$]{
\label{subfig:T1b6}
\includegraphics[width=0.85\linewidth]{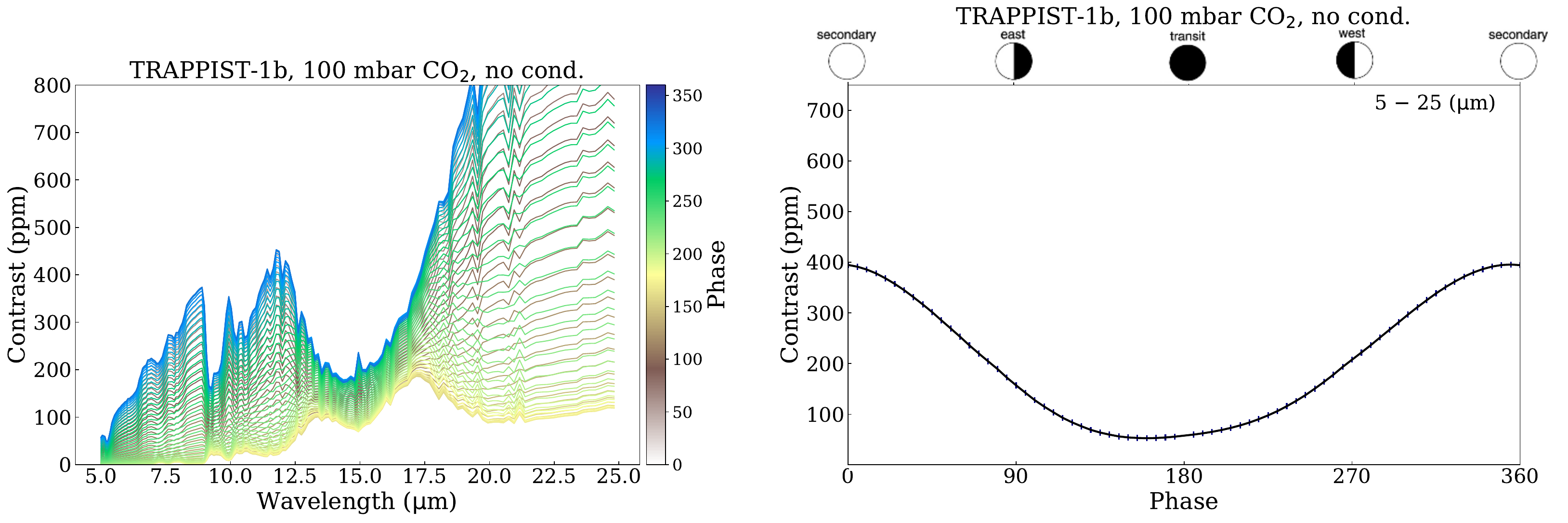}}
\caption{Thermal emission spectra at various phase angles (left). Thermal phase curve (right) integrated over all wavelengths.
\label{fig:phase_curves_T1bB}}
\end{figure*}

\begin{figure*}
\centering 
\subfloat[TRAPPIST-1b, airless]{
\label{subfig:T1b7}
\includegraphics[width=0.85\linewidth]{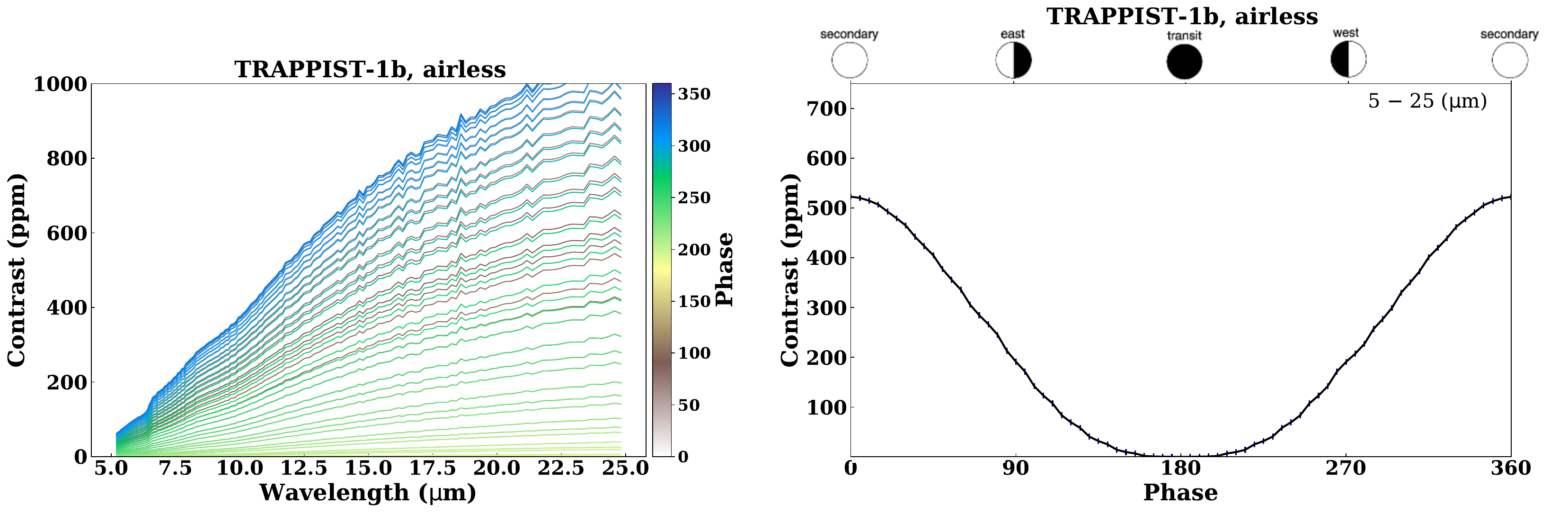}}
\caption{Thermal emission spectra at various phase angles (left). Thermal phase curve (right) integrated over all wavelengths. 
\label{fig:phase_curves_T1bC}}
\end{figure*}

\begin{figure*}
\centering 
\subfloat[TRAPPIST-1c, 1~bar N$_2$, 10~bar H$_2$O]{
\label{subfig:T1c4}
\includegraphics[width=0.85\linewidth]{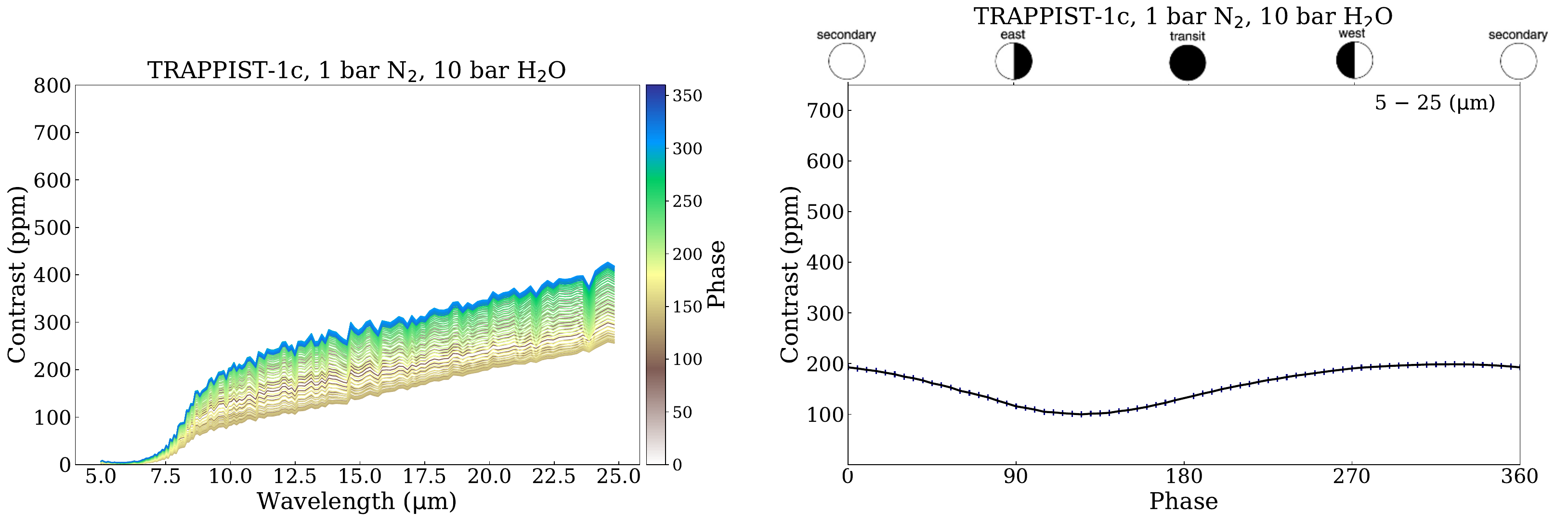}}
\\
\subfloat[TRAPPIST-1c, airless]{
\label{subfig:T1c5}
\includegraphics[width=0.85\linewidth]{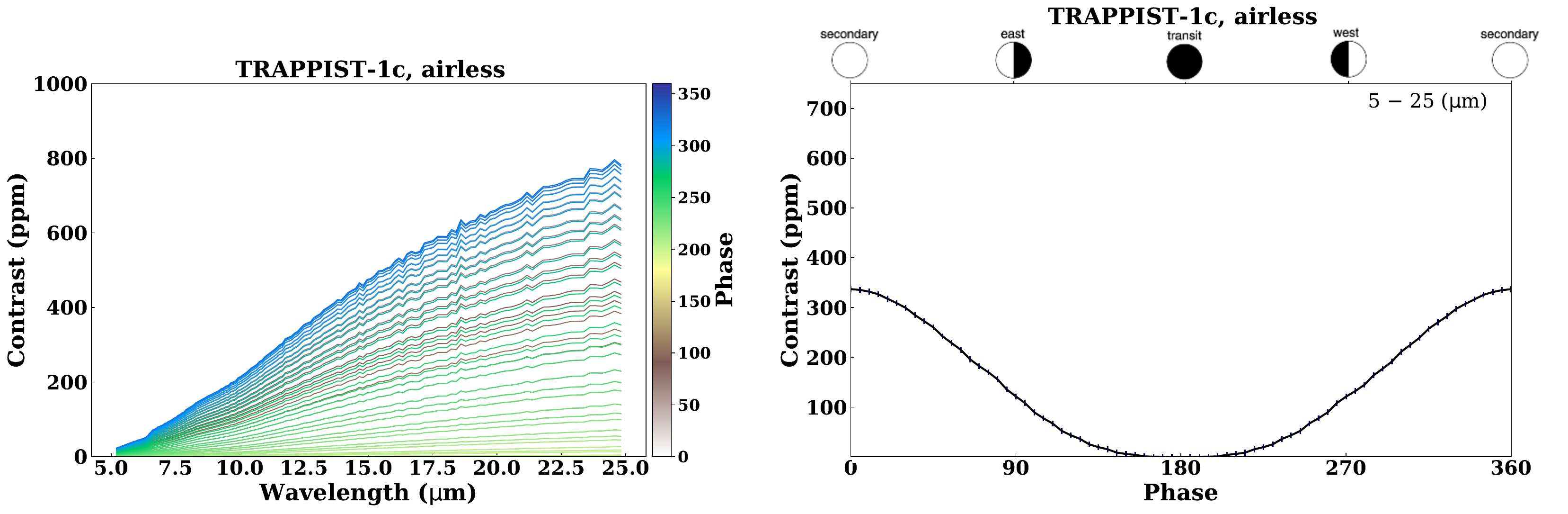}}
\caption{Thermal emission spectra at various phase angles (left). Thermal phase curve (right) integrated over all wavelengths.
\label{fig:phase_curves_T1cA}}
\end{figure*}

\begin{figure*}
\centering 
\subfloat[TRAPPIST-1c, 10~bar CO$_2$]{
\label{subfig:T1c1}
\includegraphics[width=0.85\linewidth]{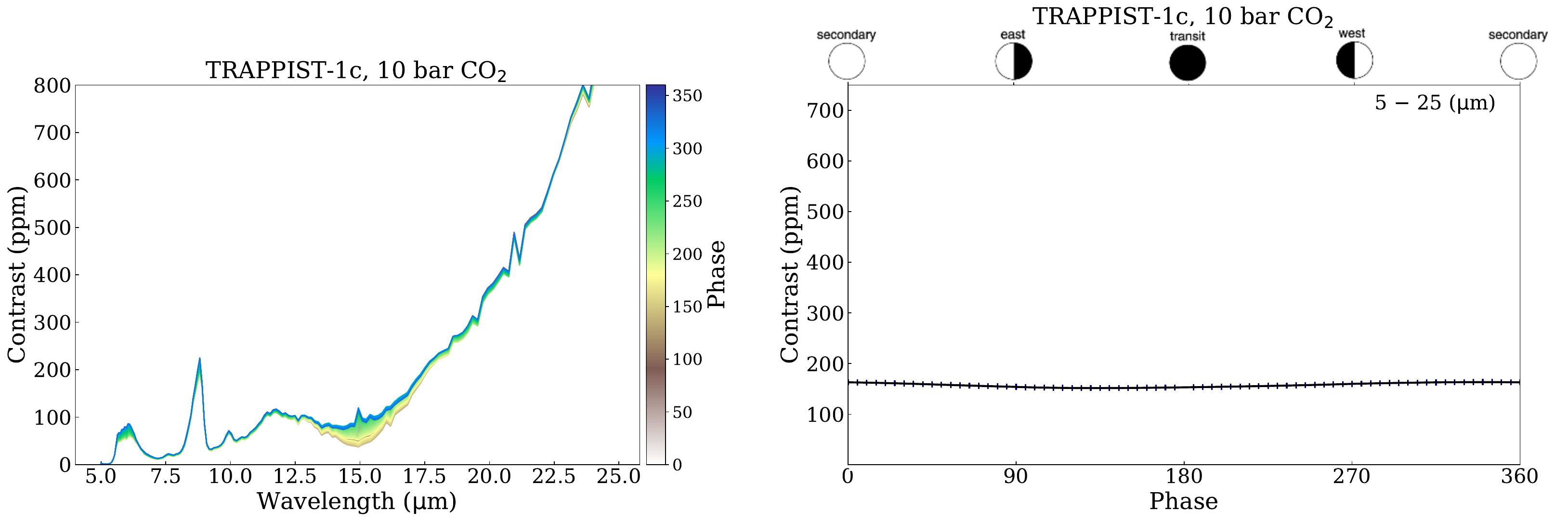}}
\\
\subfloat[TRAPPIST-1c, 1~bar CO$_2$]{
\label{subfig:T1c2}
\includegraphics[width=0.85\linewidth]{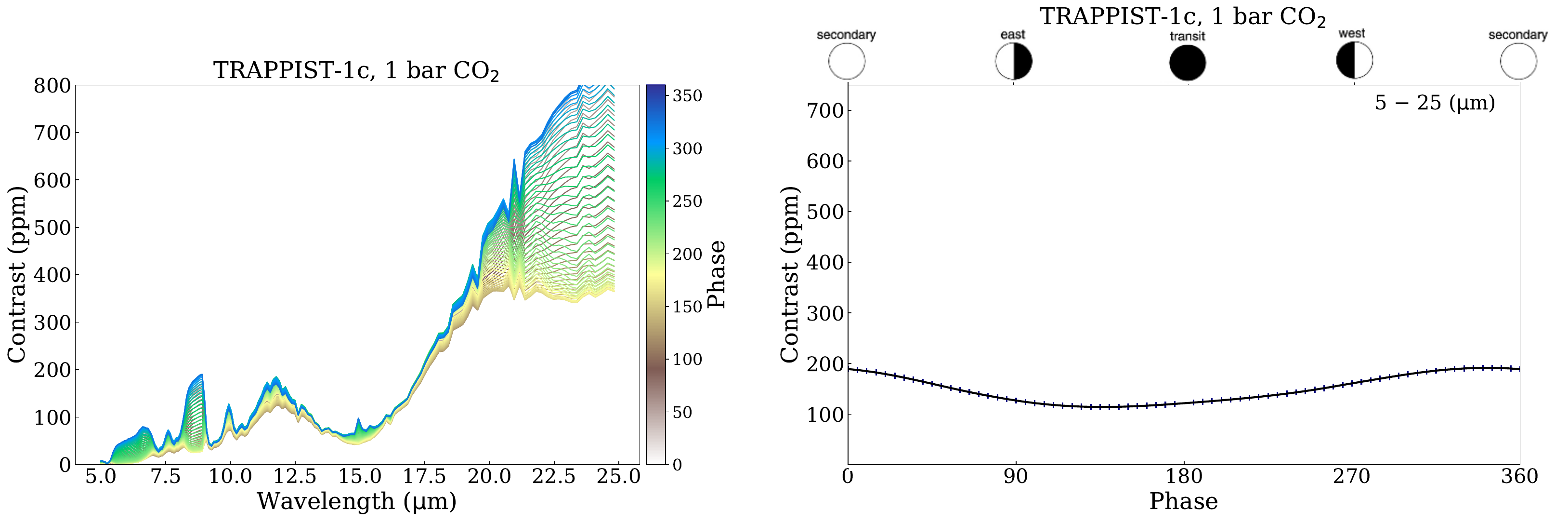}}
\\
\subfloat[TRAPPIST-1c, 100mbar CO$_2$]{
\label{subfig:T1c3}
\includegraphics[width=0.85\linewidth]{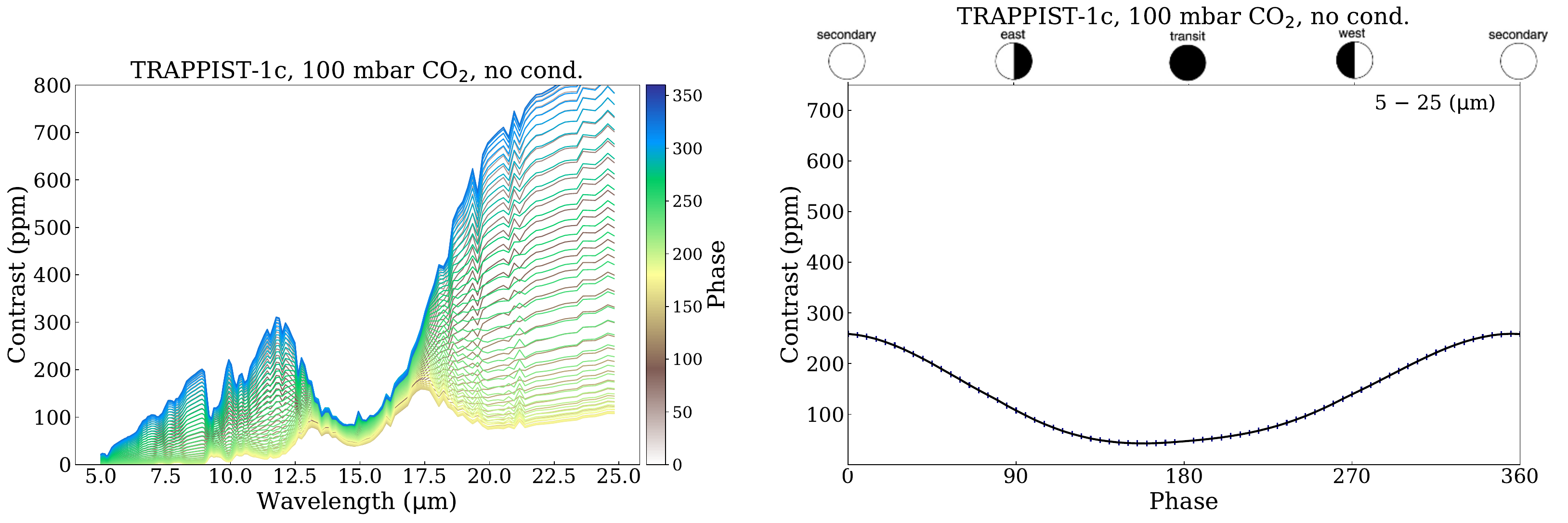}}
\caption{Thermal emission spectra at various phase angles (left). Thermal phase curve (right) integrated over all wavelengths. 
\label{fig:phase_curves_T1cB}}
\end{figure*}

\begin{figure*}
\centering 
\subfloat[TRAPPIST-1d, 1~bar N$_2$, 10~bar H$_2$O]{
\label{subfig:T1d4}
\includegraphics[width=0.85\linewidth]{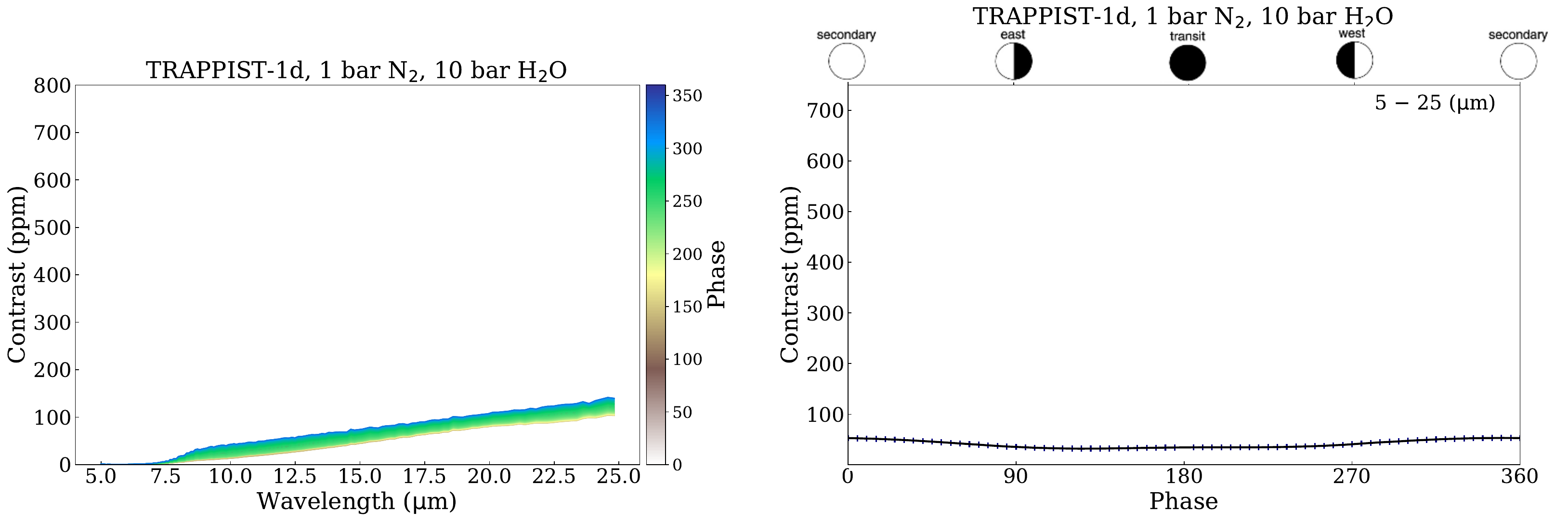}}
\\
\subfloat[TRAPPIST-1d, airless]{
\label{subfig:T1d5}
\includegraphics[width=0.85\linewidth]{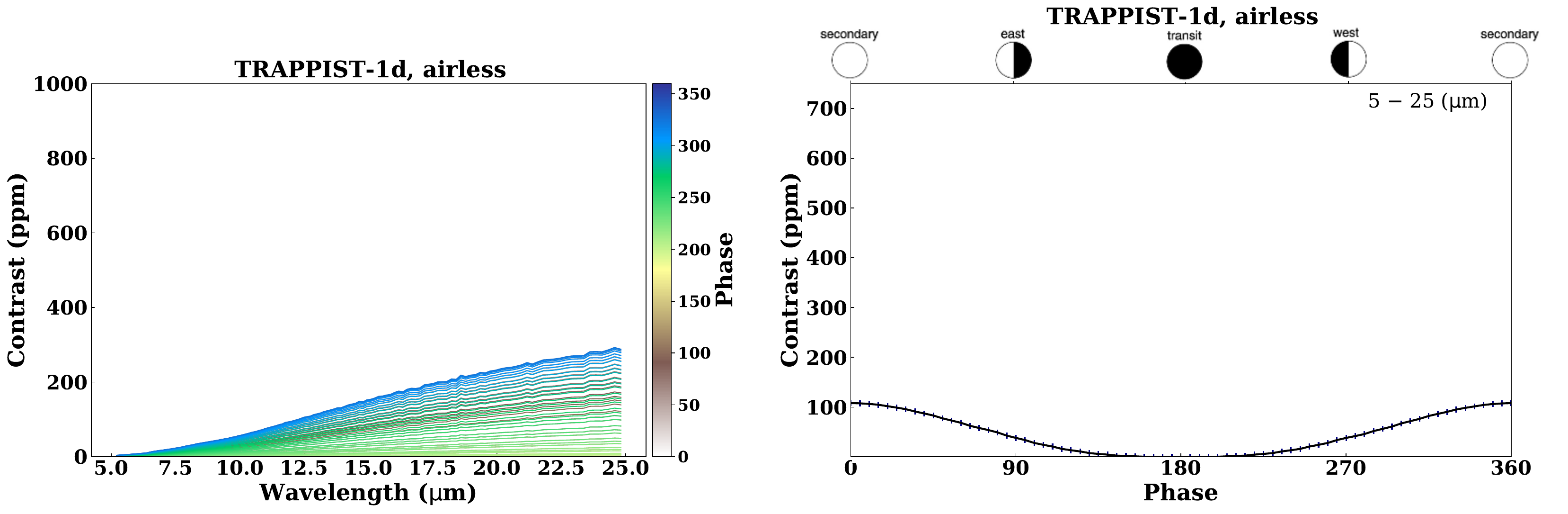}}
\caption{Thermal emission spectra at various phase angles (left). Thermal phase curve (right) integrated over all wavelengths.
\label{fig:phase_curves_T1dA}}
\end{figure*}

\begin{figure*}
\centering 
\subfloat[TRAPPIST-1d, 10~bar CO$_2$]{
\label{subfig:T1d1}
\includegraphics[width=0.85\linewidth]{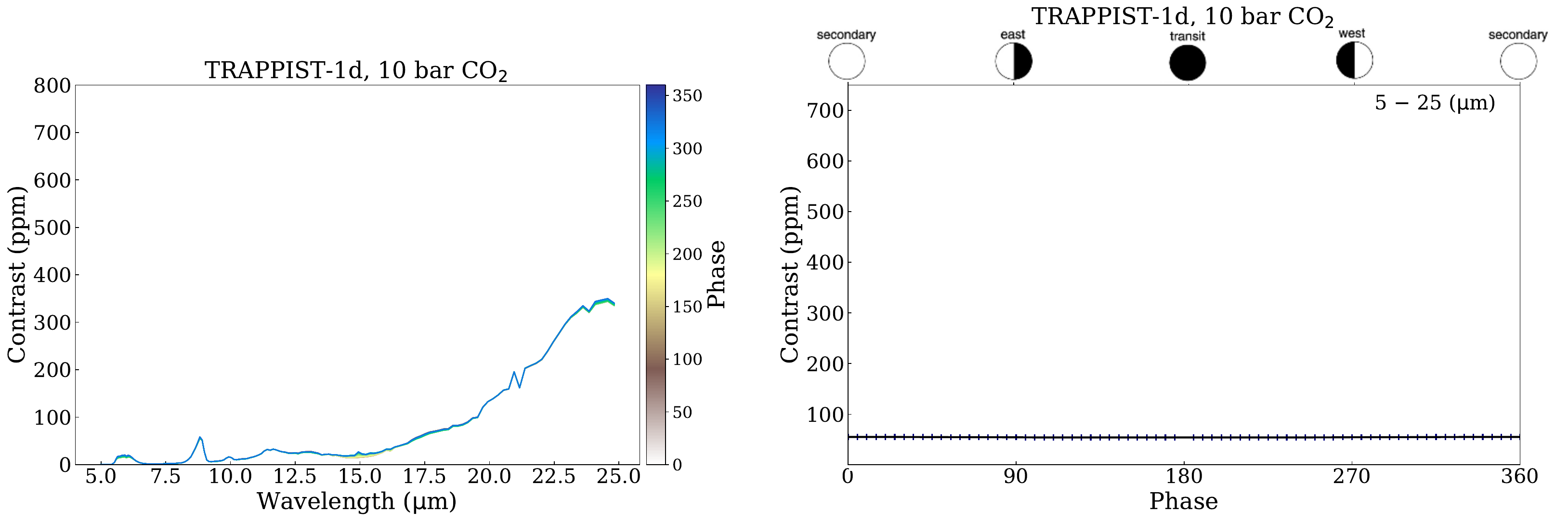}}
\\
\subfloat[TRAPPIST-1d, 1~bar CO$_2$]{
\label{subfig:T1d2}
\includegraphics[width=0.85\linewidth]{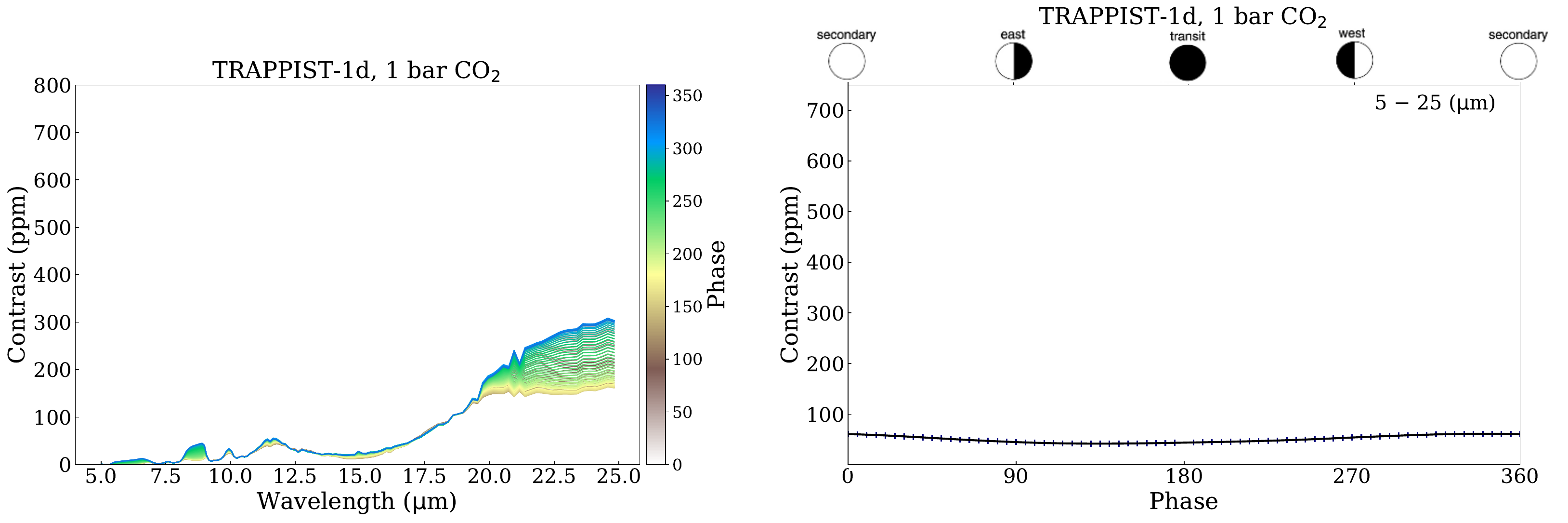}}
\\
\subfloat[TRAPPIST-1d, 100mbar CO$_2$]{
\label{subfig:T1d3}
\includegraphics[width=0.85\linewidth]{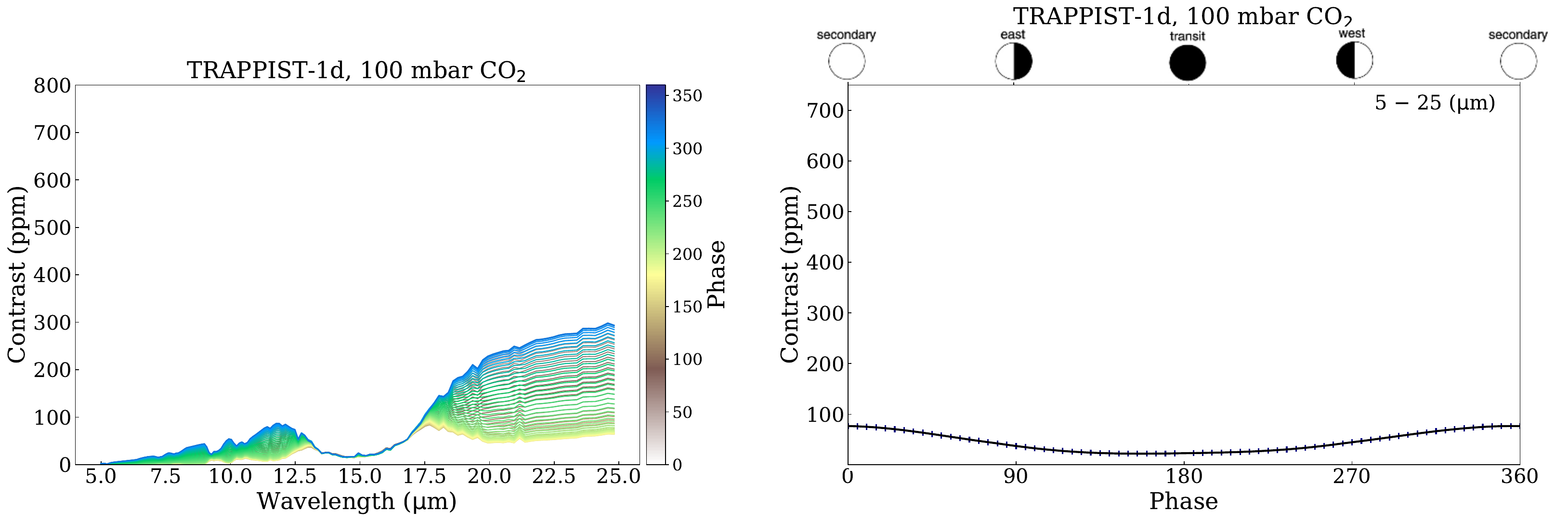}}
\caption{Thermal emission spectra at various phase angles (left). Thermal phase curve (right) integrated over all wavelengths. 
\label{fig:phase_curves_T1dB}}
\end{figure*}

\section{The effect of atmospheric expansion}

\begin{figure*}
\centering 
\includegraphics[width=\linewidth]{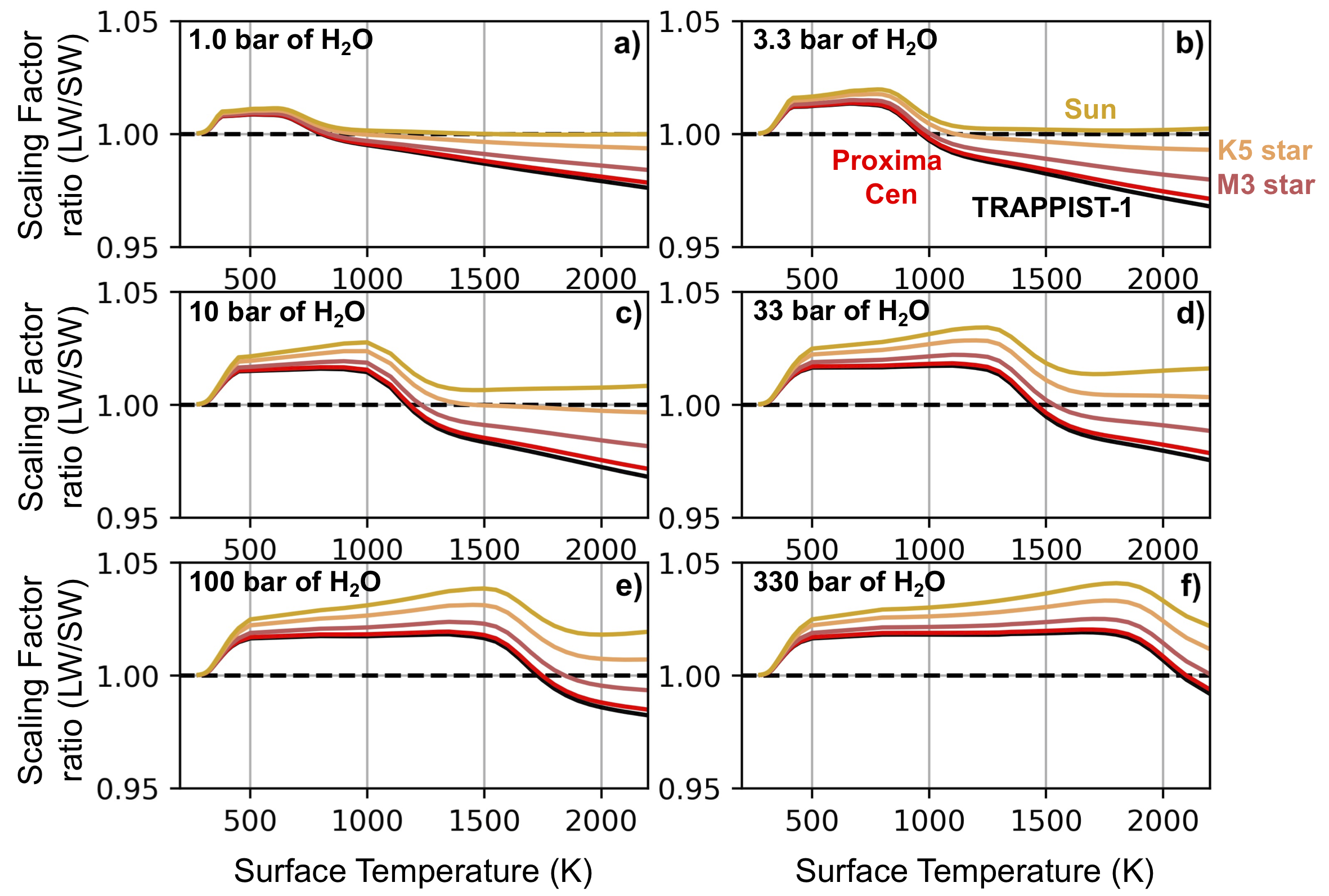}
\caption{Scaling factor ratio $(\frac{R_{\text{LW}}}{R_{\text{SW}}})^2$ as a function of surface temperature, with $R_{\text{LW}}$ the mean planetary radius at which the OTR is emitted ; and $R_{\text{SW}}$ the mean planetary radius at which the ISR is absorbed. Each panel corresponds to a given water content (from 1 to 330~bar of H$_2$O). The five colors correspond to the five different host stars.
\label{fig_scaling_factor}}
\end{figure*}

Post-runaway H$_2$O-dominated atmospheres have been shown to be significantly expanded \citep{Turbet2019aa,Turbet2020aa}. \citet{goldblatt:2015} showed that in extended atmospheres such as H$_2$O-rich atmospheres, the shortwave (i.e. emitted by the star) and longwave (i.e. emitted by the planet) fluxes can be significantly modified compared to the usual approximation. The absorbed and emitted fluxes are indeed usually calculated by taking the observed radius of the planet as the reference cross-section of the absorbed and emitted in fluxes. To evaluate the error made by this approximation, we have used 1D inverse radiative-convective calculations of H$_2$O-rich atmospheres \citep{Turbet2019aa,Turbet2020aa} and calculated for -- each spectral band in the radiative transfer -- the altitude and therefore the radius at which the opacity is equal to 1. We calculated a shortwave radius $R_{\text{SW}}$ (and a longwave radius $R_{\text{LW}}$) by weighting the spectral radius by the stellar spectrum (and the OTR spectrum, respectively). Fig.~\ref{fig_scaling_factor} shows the scaling factor ratio as a function of surface temperature of the H$_2$O-rich atmosphere. The scaling factor ratio is defined as $(\frac{R_{\text{LW}}}{R_{\text{SW}}})^2$ and provides an estimate of the error made on the TOA radiative budget. There are three important points to notice:
\begin{enumerate}
    \item The scaling factor ratio is always close (and within a few $\%$) to 1. This indicates that this expansion effect affects the water condensation limit to a few $\%$ as well. The scaling factor ratio being almost always $\ge$~1, this indicates that the ISR of the water condensation limit was slightly underestimated in our calculations.
    \item The evolution of the scaling factor as a function of surface temperature is more sophisticated here than what was calculated in \citet{goldblatt:2015}. This is because we calculated the $\tau$~=~1 radius for each spectral band of the model, while \citet{goldblatt:2015} calculated the radius of a wavelength-integrated $\tau$~=~1. As a result, and unlike what \citet{goldblatt:2015} calculated, the scaling factor ratio evolves non-monotonically as a function of surface temperature. At low surface temperatures, the scaling factor ratio increases because $R_{\text{LW}}$ increases much faster (the emission comes from a given atmospheric pressure as shown in \citealt{Boukrouche:2021}, so increasing the total atmospheric pressure increases $R_{\text{LW}}$) than $R_{\text{SW}}$ (the stellar flux always reaches the surface). At intermediate surface temperatures, the scaling factor ratio reaches a plateau, because most of the shortwave flux (and longwave flux) are directly absorbed (and emitted, respectively) by the atmosphere. Increasing the total surface pressure shifts equally $R_{\text{LW}}$) and $R_{\text{SW}}$). At high surface temperatures, the scaling factor ratio decreases. This is because the surface is hot enough that it starts to emit in the visible atmospheric windows of H$_2$O.
    \item The scaling factor varies significantly depending on the total water vapor pressure assumed, as well as the type of host star. This is well illustrated in Fig.~\ref{fig_scaling_factor}. Note however that the behaviour of the scaling factor ratio as a function of surface temperature is qualitatively similar for all the cases we explored.
\end{enumerate}

It is important to note however that Fig.~\ref{fig_scaling_factor} likely overestimates the effect of the atmospheric expansion on the TOA radiative budget. This is because we used in this calculation vertical thermal profiles from 1D inverse radiative-convective models, which overestimate temperatures in the low atmosphere, atmospheric scale height and thus atmospheric expansion \citep{Selsis:2023}. When high water contents are reached, the temperature profile in the lower atmosphere reaches an isotherm (see e.g. Fig.~\ref{vertical_profiles_T1bcd_2}c) thus limiting the atmospheric expansion and the effect on the TOA radiative budget.

Overall, the effect of atmospheric expansion is expected to play a minor role on our calculations of the Water Condensation Zone.

\end{appendix}
\end{document}